\documentclass[sigconf,10pt]{acmart}

\usepackage{array}
\usepackage{hyperref}
\usepackage{multirow}
\usepackage{graphicx}
\usepackage[utf8]{inputenc}
\usepackage[ruled, lined, longend, linesnumbered]{algorithm2e}
\usepackage{xcolor}
\usepackage{amsmath,multicol}

\usepackage{amssymb,textcomp,comment}
\usepackage{makecell}
\usepackage{ulem}

\usepackage{subcaption}

\usepackage{booktabs}
\usepackage{color}
\usepackage{colortbl}

\usepackage{listings}
\usepackage{enumitem}
\usepackage{cleveref}

\usepackage{pifont}

\usepackage{wasysym}

\crefname{section}{§}{§§}
\Crefname{section}{§}{§§}

\definecolor{raana}{HTML}{77DD77}

\definecolor{dkgreen}{rgb}{0,0.6,0}
\definecolor{gray}{rgb}{0.5,0.5,0.5}
\definecolor{mauve}{rgb}{0.58,0,0.82}

\newcommand{\yws}[1]{{\color{black}#1}}

\newcommand{\revision}[1]{{\color{black}#1}}
\usepackage{array}
\newcolumntype{L}[1]{>{\raggedright\let\newline\\\arraybackslash\hspace{0pt}}m{#1}}
\newcolumntype{C}[1]{>{\centering\let\newline\\\arraybackslash\hspace{0pt}}m{#1}}
\newcolumntype{R}[1]{>{\raggedleft\let\newline\\\arraybackslash\hspace{0pt}}m{#1}}

\newcommand{\sys}{\texttt{ElastiLM}\xspace}

\copyrightyear{2025}
\acmYear{2025}
\setcopyright{acmlicensed}\acmConference[ACM MOBICOM '25]{The 31st Annual International Conference on Mobile Computing and Networking}{November 4--8, 2025}{Hong Kong, China}
\acmBooktitle{The 31st Annual International Conference on Mobile Computing and Networking (ACM MOBICOM '25), November 4--8, 2025, Hong Kong, China}
\acmDOI{10.1145/3680207.3765259}
\acmISBN{979-8-4007-1129-9/2025/11}

\begin{document}

\title{Elastic On-Device LLM Service}

\author{Wangsong Yin$^{1}$, Rongjie Yi$^{2}$, Daliang Xu$^{2}$,
Gang Huang$^{1}$,\\ Mengwei Xu$^{2,3\#}$, Xuanzhe Liu$^{1\#}$}
\affiliation {
	\institution{$^1$Peking University, Beijing, China}
	\country{}
}

\affiliation {
	\institution{$^2$Beijing University of Posts and Telecommunications, Beijing, China}
	\country{}
}

\affiliation {
	\institution{$^3$Beiyou Shenzhen Institute, Shenzhen, China}
	\country{}
}

\email{yws@stu.pku.edu.cn}
\email{
    {hg, liuxuanzhe}@pku.edu.cn
}
\email{
    {mwx, xudaliang}@bupt.edu.cn}
\thanks{$^\#$Corresponding author.}

\renewcommand{\shortauthors}{Wangsong Yin et al.}

\begin{abstract}
On-device Large Language Models (LLMs) are transforming mobile AI, catalyzing applications like UI automation without privacy concerns.
Nowadays the common practice is to deploy a single yet powerful LLM as a general task solver for multiple requests.
We identify a key system challenge in this paradigm: current LLMs lack the elasticity to serve requests that have diversified Service-Level Objectives (SLOs) on inference latency.
To tackle this, we present \sys, an on-device LLM service that elasticizes both the model and the prompt dimension of a full LLM.
It incorporates 
(1) a one-shot neuron-reordering method, which leverages the intrinsic permutation consistency in transformer models to generate high-quality elasticized sub-models with minimal runtime switching overhead;
(2) a dual-head tiny language model, which efficiently and effectively refines the prompt and orchestrates the elastification between model and prompt.
We implement such an elastic on-device LLM service on multiple COTS smartphones, and evaluate \sys on both standalone NLP/mobile-agent datasets and end-to-end synthesized traces.
On diverse SLOs, \sys outperforms 7 strong baselines in (absolute) accuracy by up to 14.83\% and 10.45\% on average, with <1\% TTFT switching overhead, on-par memory consumption and <100 offline GPU hours.

\end{abstract}



\maketitle

\section{Introduction}

Large Language Models (LLMs) are ushering in a transformative era for mobile AI.
A multitude of killer apps are built on top of LLMs, encompassing mobile UI automation~\cite{autodriod}, API-calling~\cite{chen2024octopus,xie2024droidcall}, and screen content comprehension~\cite{rewind}.
For instance, one can easily place an order by simply saying ``Order a pizza now from Pizza Hut'' in smartphone.

\revision{
With ever-growing privacy concerns, deploying LLMs on local devices~\cite{yin2024llmservicemobiledevices,AICore, mei2024aiosllmagentoperating, packer2024memgptllmsoperatingsystems} is attracting attentions increasingly.
For instance, mobile GUI agent~\cite{li2024personal_llm_agents} handles users' requests by accessing the device screen information, which may contain highly private photos or chat history.
To this end, Google has developed \texttt{Android AICore}~\cite{AICore}, a built-in on-device LLM in Android OS that has been used by apps like GBoard smart reply and Pixel voice recorder.
}


\revision{
\noindent \textbf{On-device LLM needs elasticity.}
LLM can handle ubiquitous language tasks, at the cost of large parameter size, making it neither ``necessary'' nor ``practical'' to deploy separated LLMs on a device.
Instead, on-device LLM is often \textit{shared} across tasks and apps, e.g., LLM-as-a-Service in OS~\cite{AICore}.
However, a single static LLM cannot meet the diversified Service-Level Objective (SLO) demanded by different LLM tasks.
Consequently, on-device LLM needs elasticity.
Demand of elasticity is further exaggerated and complicated since LLM inference consists of two distinct stages: prefill (prompt processing speed) and decode (token generation speed), whose latencies are measured by Time-To-First-Token (TTFT) and Time-Per-Output-Token (TPOT), respectively.
For instance, a chatbot~\cite{llama.cpp} must behave both low TTFT and low TPOT to match human reading speed;
a UI-automation agent~\cite{zhang2024llamatouch, autodriod} typically requires a low TTFT and an acceptable TPOT, as TPOT can be overlapped with UI manipulations; a screen-event recorder~\cite{rewind} running in background only needs a tolerable TTFT/TPOT.
Even in a single app, the SLOs maybe also diversified due to the resource/task variation.
Failing to meet an SLO leads to serious consequences: a significant degradation of user experience, or failure in the interactions between LLM agents and the environments/tools~\cite{zhang2024llamatouch, chen2024octopus, Apple_Intelligence}.
}



\revision{
Ideally, a request is sent to the elastic LLM with its prompt as well as the SLO expected, and the LLM needs to provide the highest text generation quality without failing the SLO.
Fundamentally differing from prior work that elasticize CNNs in pre-LLM era~\cite{wen2023adaptivenetpostdeploymentneuralarchitecture, Han_2021, Fang_2018, fang2023depgraph},
a key opportunity in our system model is that, \textit{both model and input (prompts) dimensions of the LLM can be elasticized}.
Through multi-scale LLM pruning technique~\cite{xia2024shearedllamaacceleratinglanguage, ma2023llmpruner, sun2024simpleeffectivepruningapproach}, one can get a crucial subset of weights that running at different speed; similarly, through scoring the importance of each input token~\cite{jiang-etal-2024-longllmlingua, wu2024llmlingua2, modarressi-etal-2022-adapler}, one can prune the prompt into different lengths on demand.
In both ways, LLM accuracy is sacrificed for faster text generation in a flexible manner.
Our pilot experiments in $\S$\ref{subsec: opportunity} explore how each dimension of elasticity impacts the LLM inference latency:
TTFT (often more time consuming and requires higher elasticity) is proportional to both the prompt length and the model size; TPOT is mainly proportional to the model size with the help of KV cache~\cite{kwon2023efficientmemorymanagementlarge}.


\noindent \textbf{Challenges.}
However, elasticizing an on-device LLM faces the following unique challenges.
}

$\bullet$\textit{Costly runtime switching between elasticized models.}
An elastic LLM service has to frequently switch between the sub-models at request level, yet traditional model elastification methods (e.g. structural pruning) often ignore this switching overhead 
~\cite{hermes, wen2023adaptivenetpostdeploymentneuralarchitecture, ma2023llmpruner, cai2020onceforalltrainnetworkspecialize}.
For instance, generated by SoTA structural pruning~\cite{ma2023llmpruner}, a sub-model of LLaMA-7B with 20\% parameters takes 8.2s to switch to 30\% on Redmi K60 Champion.
The root cause is that, to utilize the deeply optimized on-device NN inference libraries and hardware throughput, the interleaved and overlapping sub-models‘ weights must be re-layouted to contiguous format in memory before inference (or each sub-model must be maintained an unacceptable standalone copy in memory).
This switching overhead is billed to TTFT since the switching can only be performed right after the LLM receiving a request.

$\bullet$ \textit{Sensitive prompt-model orchestration strategy.}
There exist multiple elastic strategies to meet an SLO, yet their output text quality could differ significantly.
Exemplified with a real prompt from \texttt{ARC\_E} dataset in Figure~\ref{fig:decision_example}, although both elastic strategies (50\%/20\% prompt/model pruning vs. 20\%/50\% prompt/model pruning) can meet the SLO, only the first strategy leads to a correct generated answer.
Another instance is that with randomized strategy, the top5 API selection of \texttt{Octopus} dataset exhibits a 15.2\% accuracy loss to the oracle strategy on an SLO with 50\% TTFT and 80\% TPOT of the full LLaMA-7B model.
How to orchestrate the two dimensions of elastification to maximize the LLM output quality at request level has not been touched in prior study.

\noindent \textbf{Our solution: \sys.}
We present \sys, a system for elasticizing on-device LLMs. It tackles the above challenges through the following novel techniques.

\noindent \textbf{One-shot reordering of permutation consistent units}  ($\S$\ref{subsec: model_elastification}).
This technique performs pruning on a fine granularity of  \textit{permutation consistent units} in Transformers.
Identified by \sys, these units can be offline arbitrarily layouted in a block (e.g., Attention or MLP) while guarantee the equivalent input/output as original block, thereby fundamentally avoiding the runtime switching overhead.
Specifically, such a unit is an entire attention head (i.e., columns in $W_{Q}/W_{K}/W_{V}$ and rows in $W_{O}$ with the same indices), or an entire MLP neuron (i.e., a column in $W_{up}$ and a row in $W_{down}$).
In the one-shot reordering, \sys first profiles the importance of the units through an eXplainable-AI (XAI)~\cite{huang2023elastictrainer, Selvaraju_2019, sundararajan2017axiomaticattributiondeepnetworks} guided method, which measures unit importance via gradients. 
Then, \sys reorders the units in memory based on their importance, making each contiguous memory segment (starting from base address to a memory pointer) a sub-model.
After reordering by importance, the pruning always extends from the periphery inward, thus requiring no online reordering.
The profiling and reordering are all done offline, incurring no online overhead.
The online switching of sub-models is zero-cost by only moving the pointer.
\sys further incorporates other optimizations such as LoRA recovery and anchor layers locking to improve the sub-models quality.

\noindent \textbf{Dual-head Tiny Language Model (TLM) for prompt-model elastification orchestration} ($\S$\ref{subsec:prompt_elastification}).
Different from prior work~\cite{wu2024llmlingua2, jiang-etal-2024-longllmlingua} that only focus on identifying important tokens, \sys designs a dual-head TLM as an end-to-end solution for determining an optimal prompt-model elastification strategy.
Specifically, the TLM features two heads, namely score-head and decision-head.
During inference,
the score-head estimates the token importance,
while the decision-head chooses the proper prompt- and model- elastification ratios (mapped to the real latency by a one-shot profiling on various SoCs).
At offline, the decision-head learns from the prompts and groundtruth of a generic corpora.
As a self-induced labelling process, all possible strategies of a prompt with its SLOs are traversed to get the inference result, and an optimal strategy is recorded as the label.
Both the training and inference of TLM are cost-effective with careful optimizations like reusing the compact MobileBert and bottom layers sharing.


\noindent \textbf{Evaluation.}
We have fully implemented \sys on 3 COTS smartphones and 5 base/instruction-tuned LLMs with 3B--7B parameters.
The datasets encompass fundamental language tasks \texttt{ARC\_E}/\texttt{OBQA}/\texttt{PIQA}/\texttt{SCIQ}~\cite{allenai:arc, OpenBookQA2018, PIQA, SciQ}, mobile UI automation tasks \texttt{LlamaTouch}~\cite{zhang2024llamatouch}, and mobile system API-calling tasks \texttt{Octopus}~\cite{chen2024octopus}.
Evaluation shows that \sys achieves up to 14.83\% and on average 10.45\% higher absolute accuracy on end-to-end traces, and up to 40\% on standalone datasets, when meeting all request SLOs.
Within 2\% absolute accuracy loss, \sys can speed up TTFT by up to 5$\times$ and TPOT by up to 2$\times$.
The runtime memory consumption is also on-par with non-elastic LLM service.
The entire elastification of \sys only takes 68.3 GPU hours (within \$100 GPU renting cost~\cite{hfgpurenting}) when elastictizing LLaMA-7B for MI14 smartphone, being affordable for most LLM service developers. 

\noindent \textbf{Contributions} are listed as follows:

\begin{itemize}[leftmargin=10pt]
    \item We highlight the strong motivation and key challenges of elastic on-device LLM service.

    \item \revision{We present \sys, a system that fully exploits the space of model and prompt elastification through two novel techniques/designs, one-shot reordering of permutation consistent units and dual-head tiny language model.}

    \item We conduct comprehensive experiments on \sys that demonstrate its superior performance over competitive baselines.
\end{itemize}



\section{Background and Motivations}


\subsection{Elastic on-device LLM service}
\label{subsec: bkg_llmservice}




\begin{figure}[t]
    \centering
    \includegraphics[width=0.46\textwidth]{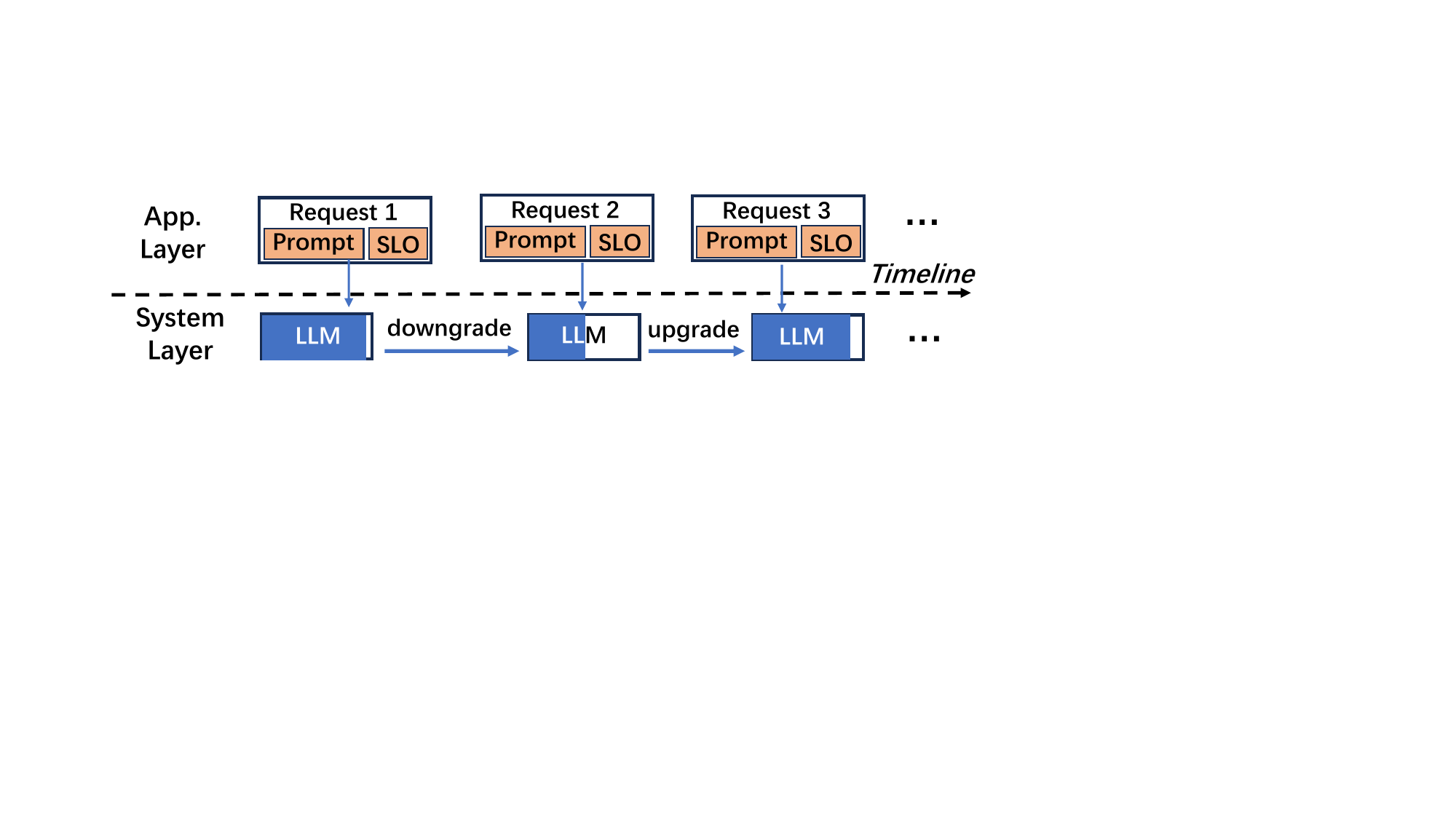}
    \vspace{-12pt}
    \caption{\revision{The system model. The service can be upgraded/downgraded at runtime to adapt to various SLOs.}}
    \label{fig:system_model}
    \vspace{-12pt}
\end{figure}




\revision{
\noindent \textbf{On-device LLM needs elasticity.}
As listed in Table~\ref{table-llmaas_apps}, a chatbot must behave both low TTFT and TPOT in order to match human reading speed. A UI-automation agent requires a relatively low TTFT to generate the first action and an acceptable TPOT, since the following latency can be overlapped with the manipulation of UI elements and thus transparent to users.
Besides, these mobile-agents typically only decode few tokens compared to the prompt length, making a lower TTFT more important.
Failing to provide satisfactory latency for a request leads to serious consequences: a significant degradation in user experience, or failure in the interactions between LLM agents and the environment/tools.
}

\begin{table}[t]
\footnotesize
\begin{tabular}{@{}c|c@{}}
\toprule

\textbf{Mobile LLM App.}         & \textbf{Service-Level Objective} \\ \midrule
Chatbot~\cite{llama.cpp}                          & Readable TTFT/TPOT               \\
Always-on Voice Assistant~\cite{siri, xiaoai}        & Very-Low TTFT, {\scriptsize medium TPOT}       \\
Background Screen-Event Recorder~\cite{rewind} & Tolerable TTFT/TPOT              \\
Smart Message Reply~\cite{gboard}              & Low TTFT, low TPOT               \\
API-Calling Agent~\cite{chen2024octopus}                & Low TTFT, acceptable TPOT        \\
UI-Automation Agent~\cite{zhang2024llamatouch, autodriod}              & Low TTFT, acceptable TPOT        \\ \bottomrule
\end{tabular}
\caption{SLOs of various mobile LLM applications.}
\vspace{-22pt}
\label{table-llmaas_apps}
\end{table}



\revision{
How to satisfy the heterogeneous demands of different LLM requests, while not degrade the LLM output quality significantly?
One plausible solution is to deploy a dedicated-sized LLM for each SLO\footnote{In cloud datacenters, a tighter SLO can be achieved by scaling up hardware resources, e.g., number of GPUs/TPUs. However, hardware resource of mobile devices is limited and not scalable.}.
This is unfriendly (and even infeasible) to both the LLM service developers and users. 
On one hand, costly GPU resources are required for pretraining multiple LLMs; on the other hand, memory consumption rises dramatically in order to manage these LLMs --- running counter to the motivation behind a single LLM service.
}

\revision{
\noindent \textbf{The system model.}
Thereby, we propose our system model.
As illustrated in Figure~\ref{fig:system_model}, at system layer, there is one running LLM that serves requests from the application layer (e.g., apps/agents).
Each request consists of a prompt (a text sequence as LLM input) and an SLO (inference latency constraint).
This single LLM can rapidly upgrade/downgrade itself to a more bulky/swift one at runtime to adapt to a specific SLO.
}

\subsection{Opportunities and challenges}
\label{subsec: opportunity}

\begin{figure}[t]
	\centering
	\begin{minipage}[b]{0.22\textwidth}
		\includegraphics[width=1\textwidth]{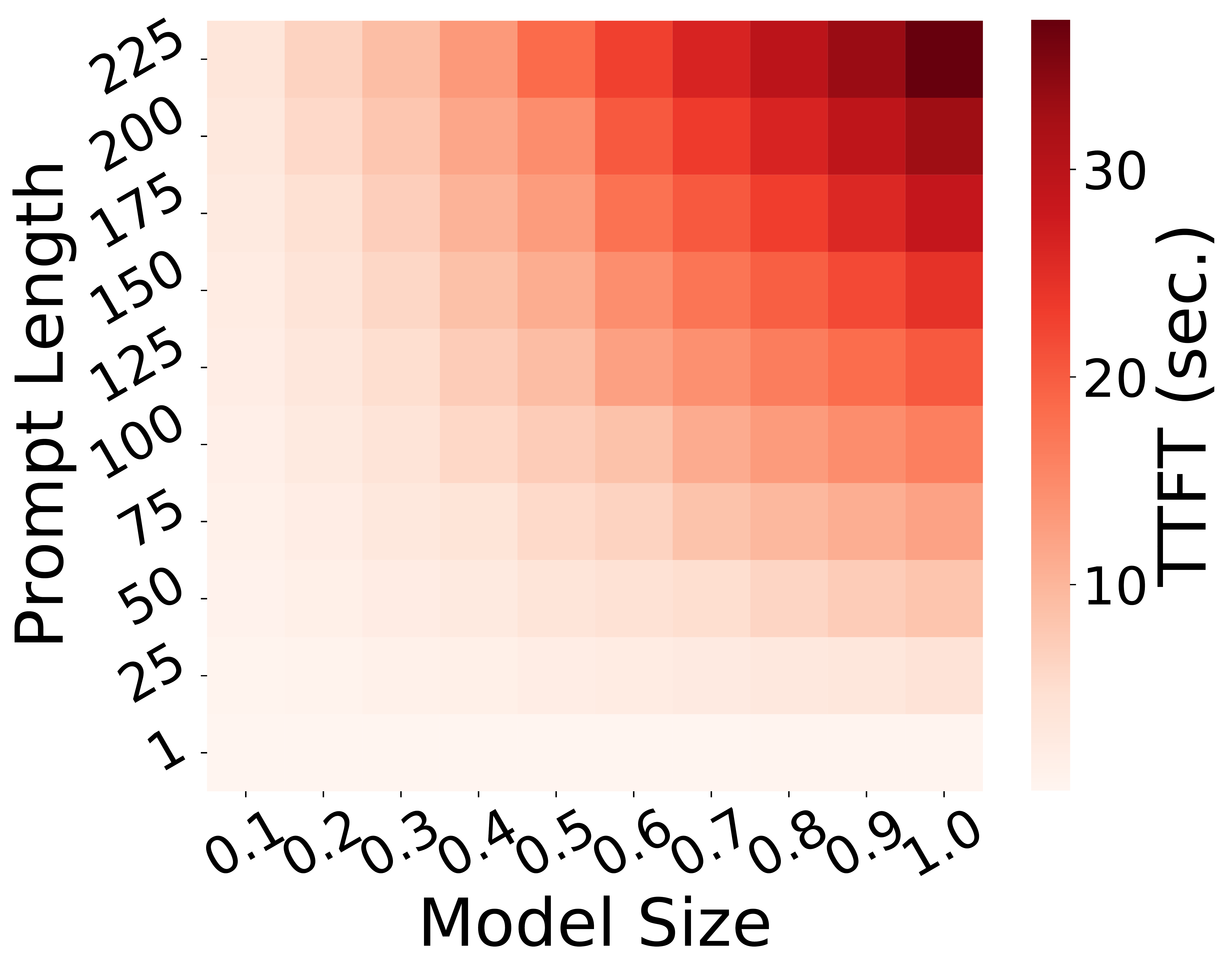}
          \vspace{-12pt}
		\subcaption{Prefill latency.}
		\label{fig:ttft}
	\end{minipage}
	\begin{minipage}[b]{0.23\textwidth}		
            \includegraphics[width=1\textwidth]{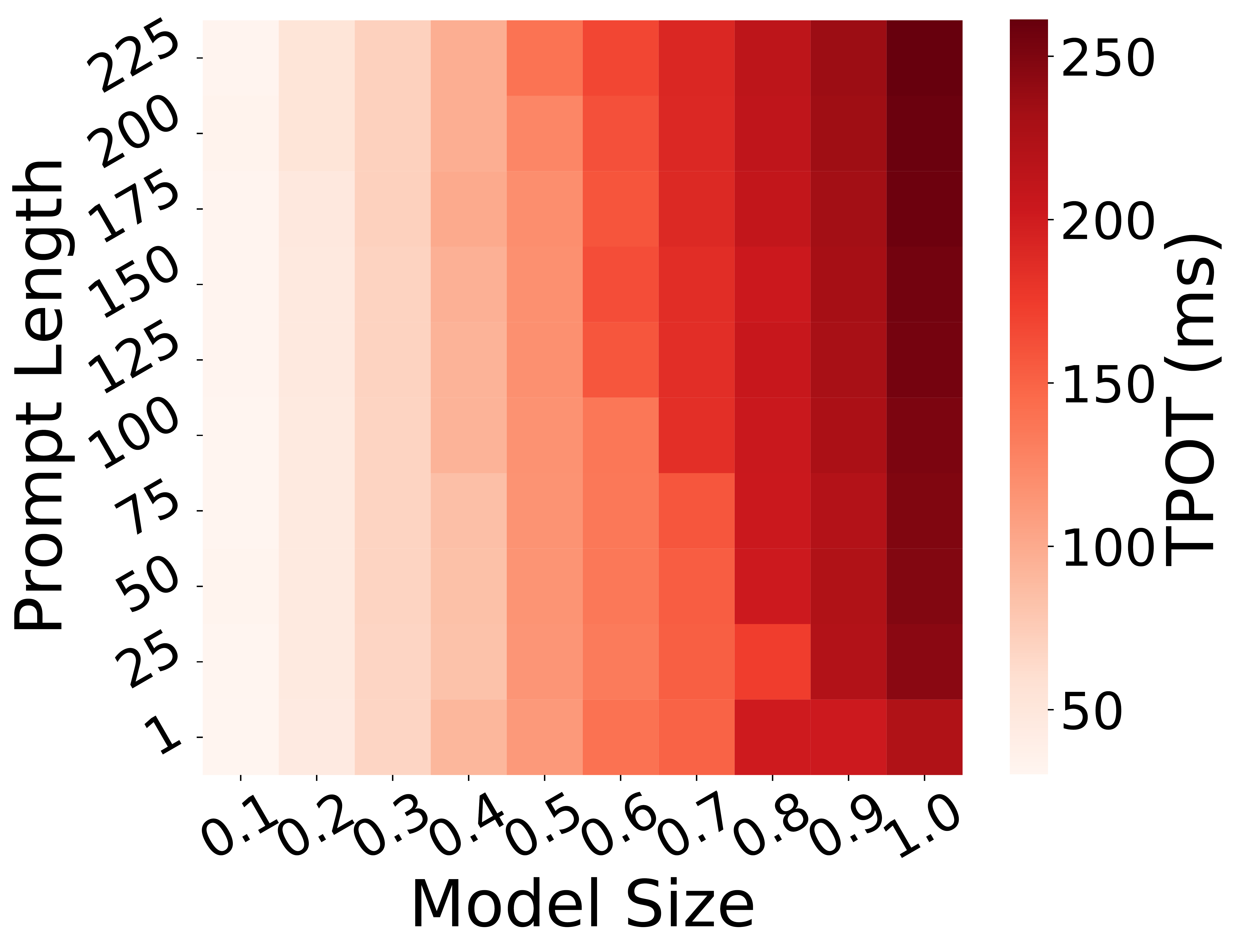}
        \vspace{-12pt}
		\subcaption{Decode latency.}
		\label{fig:tpot}
	\end{minipage}
        \vspace{-12pt}
	\caption{LLM inference latency w.r.t. prompt length and model size. Measured on LLaMA-7B, Redmi K60 Champion (Snapdragon 8Gen2).}
    \label{fig:latency_prefill_decode}
    \vspace{-12pt}
\end{figure}

\begin{figure}[t]
    \centering
    \includegraphics[width=0.46\textwidth]{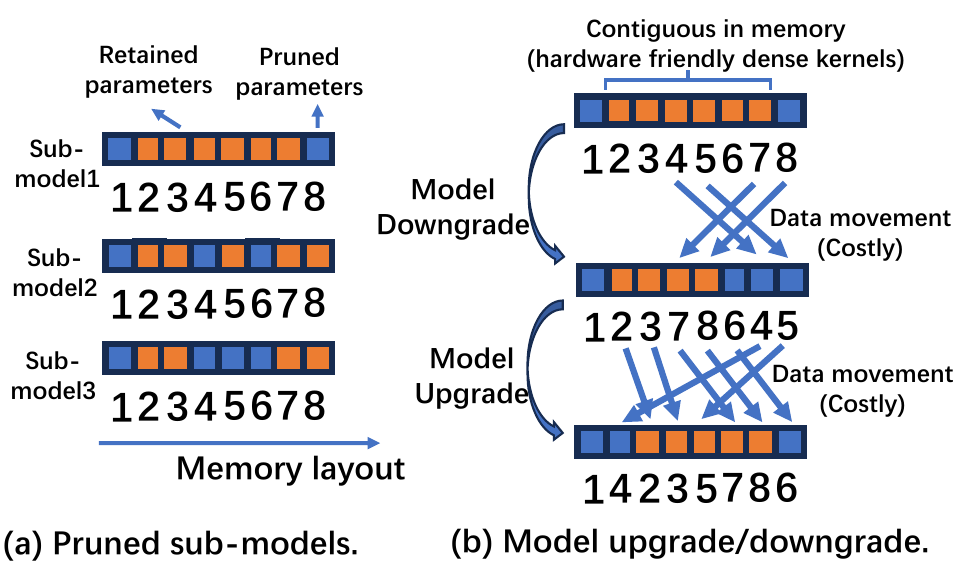}
    \vspace{-12pt}
    \caption{Elastic LLM does not translate to Elastic LLM service. Non-negligible request-level overhead of data movement is still suffered when directly employing model pruning.}
    \label{fig:mem_reorder}
    \vspace{-12pt}
\end{figure}
\begin{figure}[t]
	\centering
	\begin{minipage}[b]{0.22\textwidth}
		\includegraphics[width=1\textwidth]{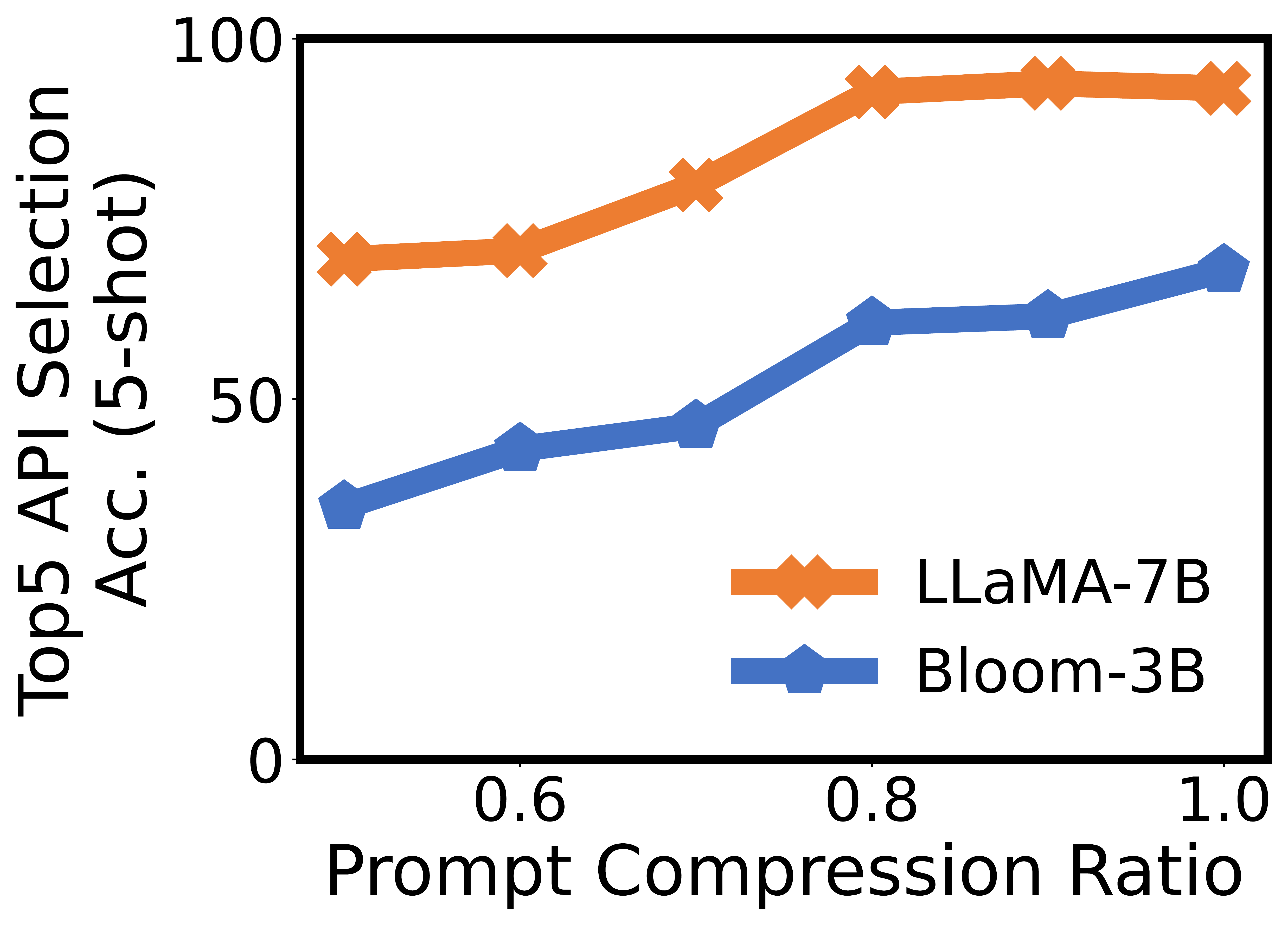}
        \vspace{-17pt}		
  \subcaption{Prompt elastification.}
		\label{fig:prompt_elastification}
	\end{minipage}
	\begin{minipage}[b]{0.22\textwidth}		
            \includegraphics[width=1\textwidth]{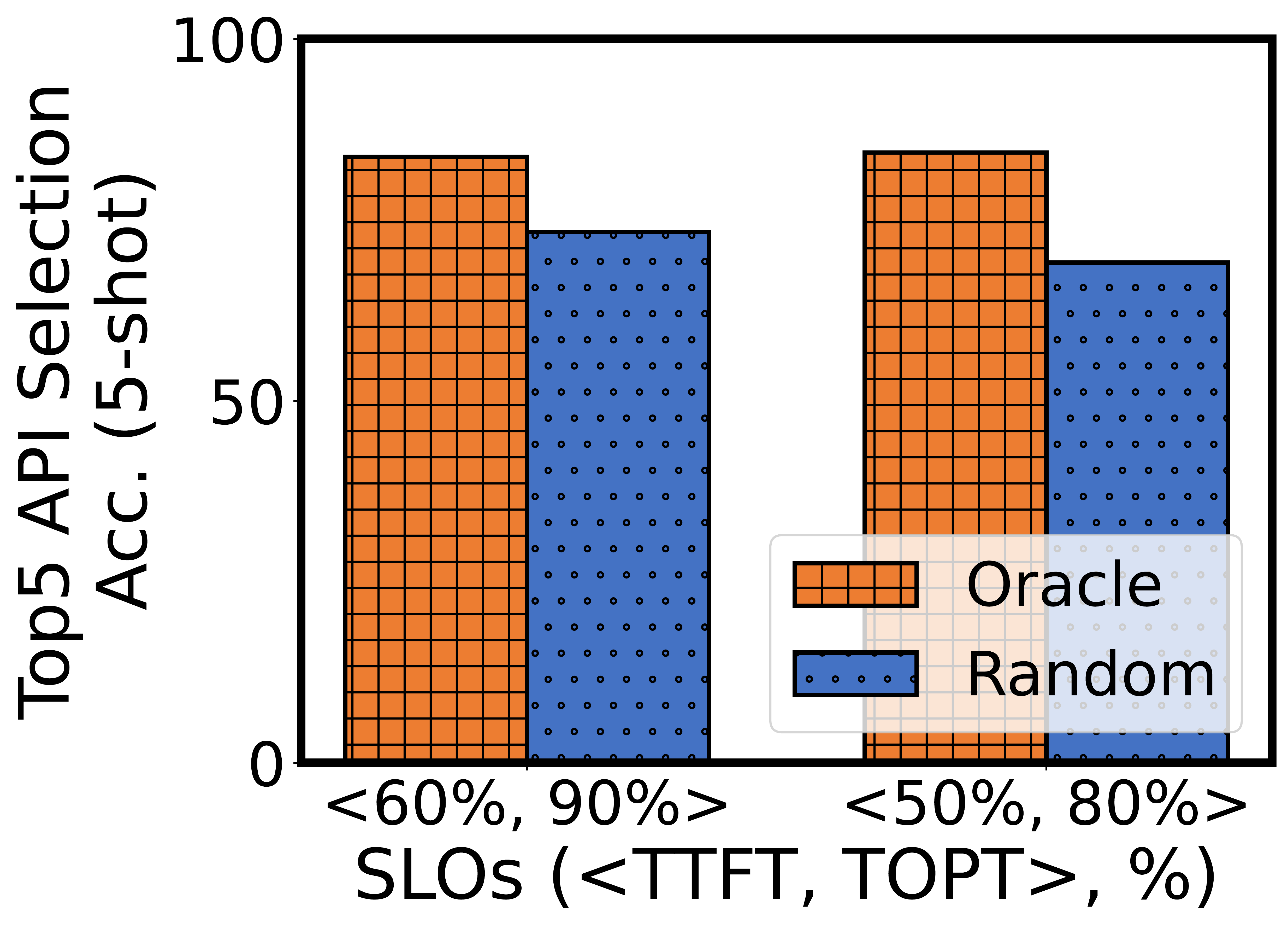}
        \vspace{-17pt}		
  \subcaption{Strategy selection.}
		\label{fig:strategy_selection}
	\end{minipage}
        \vspace{-12pt}
	\caption{Prompts of on-device LLM service are also elasticizable. Yet the ratio needs careful (and content-aware) selection to achieve an optimal orchestration with model elastification.}
    \label{fig:prompt_strategy}
\end{figure}

\begin{figure}[t]
    \centering
    \includegraphics[width=0.48\textwidth]{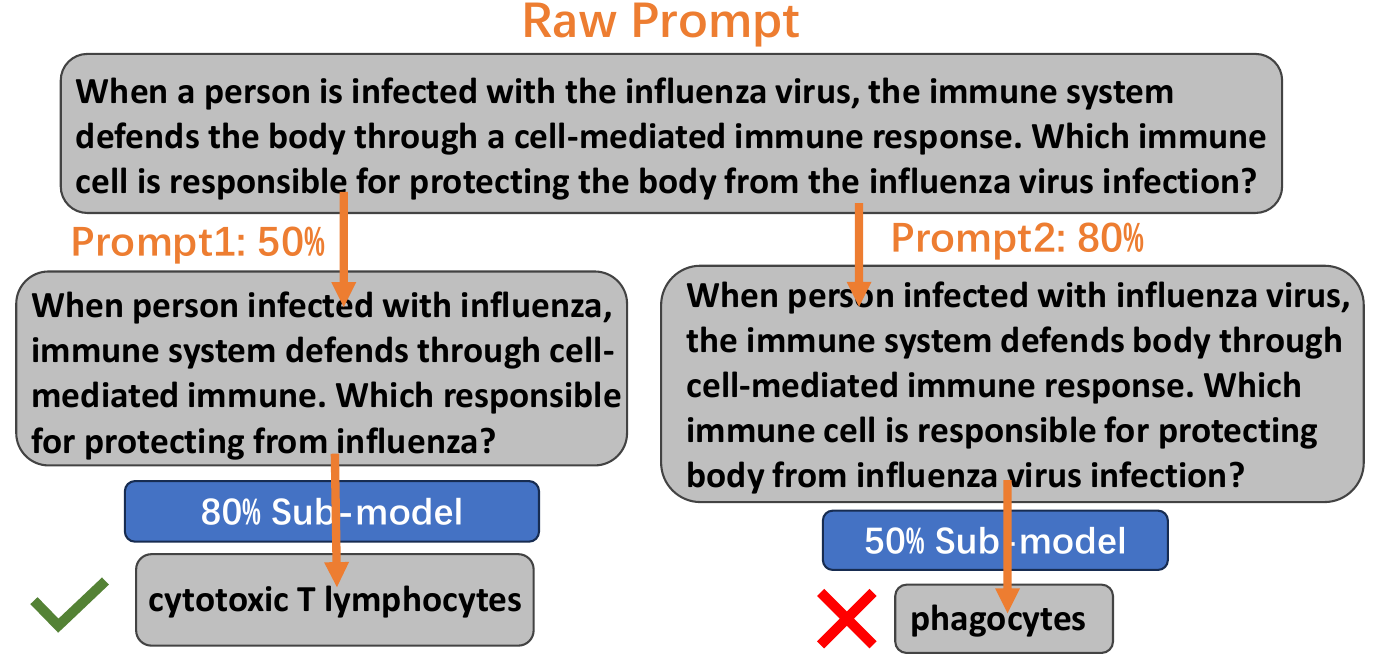}
    \vspace{-17pt}
    \caption{Sensitive prompt-model orchestration. We use a prompt of \texttt{ARC\_E} dataset and the elasticized LLaMA-7B as an exemplification. The SLO is 40\% TTFT of the full LLM.}
    \label{fig:decision_example}
\end{figure}


\noindent \textbf{Observation\#1: LLM inference latency is influenced by two dimensions --- prompt and model.}
An LLM inference workload can be divided into two dimensions: prompt (activations) and model (weights).
We conduct a measurement of LLaMA-7B inference on Redmi K60 Champion smartphone equipped with Snapdragon 8Gen2 SoC.
We use 4 threads (big cores).
The ``model size'' here represents  a sub-model of LLaMA-7B~\cite{touvron2023llamaopenefficientfoundation} (e.g., 0.1 means 10\% parameters). 
In Figure~\ref{fig:latency_prefill_decode}, we observe that both the two dimensions can influence LLM inference latency.
TTFT is influenced by both prompt length and model size; TPOT is mainly determined by model size\footnote{Only when the prompt is extremely long (e.g. over 10K tokens), the \texttt{attention} operator will dominate inference, and TPOT will then be influenced by prompt length.}.
We have the following proportional relationships:
\begin{equation}
\label{eq:prefill_decode}
\begin{split}
        TTFT &\propto PromptLength \times ModelSize, \\
        TPOT &\propto Modelsize,
\end{split}
\end{equation}
Notably, as shown Figure~\ref{fig:latency_prefill_decode}, TTFT is much longer than TPOT (e.g, seconds v.s. milliseconds), which necessitates a more aggressive elastification of TTFT.


\noindent \textbf{Observation\#2: LLMs are elasticizable; yet elastic LLM does not necessarily translate to elastic LLM service.}
DNNs are known to be elasticizable: they can provide various compute-accuracy tradeoffs by a subset of their parameters (known as pruning~\cite{fang2023depgraph, ma2023llmpruner}).
For instance, Sheared-LLaMA~\cite{xia2024shearedllamaacceleratinglanguage} demonstrates that pruning off 60\% parameters of a 7B-size LLM can still retain 74\% of original accuracy on \texttt{ARC\_E} dataset~\cite{allenai:arc}.
LLMPruner~\cite{ma2023llmpruner} further shows that with lightweight Parameter-Efficient Fine-Tuning (PEFT) methods, the accuracy loss incurred by pruning could be recovered.
Since pruning (and PEFT) generates sub-models that share the same superset of parameters, there is no overhead of extra memory or pre-training that mentioned in $\S$\ref{subsec: bkg_llmservice}.

Yet, the challenge is that, since the model upgrades/downgrades itself to adapt to various requests' SLOs, switching between these sub-models is not overhead-free.
As shown in Figure~\ref{fig:mem_reorder}a, although sub-models share the same superset of parameters, they are no longer contiguous in memory.
One may change the deeply optimized dense kernels of on-device NN libraries (e.g., MNN~\cite{alibaba2020mnn} or mllm~\cite{mllm}) to handcrafted sparse kernels.
However, these kernels typically undergo degraded performance without fully exploiting the parallelism of processors (e.g., mobile GPUs/NPUs or multi-core CPUs).
Another compromising method is to perform a movement of parameter data for each model switching, as shown in Figure~\ref{fig:mem_reorder}b.
Although the switching overhead is mitigated from iteration/operator level to request level, it is still non-negligible.
For instance, movement of LLaMA-7B's a $W_Q$ matrix (4096$\times$4096) takes 139 ms on Redmi K60 Champion smartphone in the worst case, and consequently the entire model suffers time overhead at seconds level.


\noindent \textbf{Observation\#3: Prompts of on-device LLM service are also elasticizable.}
Intuitively, as a natural language sequence, a prompt could still preserve essential semantic information when refined to a shorter one.
Especially, the prompts of LLM service callers tend to be verbosely designed in order to maximize the LLM's instruction following~\cite{ouyang2022traininglanguagemodelsfollow, zhou2023instructionfollowingevaluationlargelanguage} and in-context learning~\cite{wies2023learnabilityincontextlearning, dong2024surveyincontextlearning} abilities.
In other words, the prompt dimension can also be elasticized just like the model dimension.
We showcase employing a commonly used prompt compression method \texttt{LLMLingua2}~\cite{wu2024llmlingua2} for \texttt{Octopus}~\cite{chen2024octopus} dataset, which contains traces of an on-device API-calling agent.
\texttt{LLMLingua2} identifies and preserves most semantically significant tokens by a dedicated language model.
We report top5 function match accuracy of \texttt{Octopus}.
As shown in Figure~\ref{fig:prompt_elastification}, the accuracy shows a well-performing tradeoff curve when gradually compressing the prompt.

However, the challenge is the sensitive prompt-model orchestration.
An intuitive example is that, if a request sets its SLO as 40\% TTFT and 80\% TPOT\footnote{See our formal definition of SLO in $\S$\ref{subsec:overview}.}, we cannot know which strategy is golden a priori --- a 50\% prompt with an 80\% model? an 80\% prompt with a 50\% model? or others?\footnote{According to formula~\ref{eq:prefill_decode}, there are multiple combinations of prompt and model that can meet the exemplified SLO.}
Prompts with various content naturally require distinct and customized strategies. 
In Figure~\ref{fig:strategy_selection} and Figure~\ref{fig:decision_example}, we demonstrate that a strategy without careful design (e.g., random) may lead to a significant degradation on accuracy.
We use LLaMA-7B elasticized by our method (elaborated in $\S$\ref{subsec: model_elastification}) on \texttt{Octopus} dataset.
The prompt is elasticized by \texttt{LLMLingua2}.




\section{\sys Design}
\label{sec:design}

\subsection{Overview}
\label{subsec:overview}

\textbf{Design goal.}
\sys aims to provide LLM service that adapts to a specific Service Level Objective (SLO) of resource constraint per request (prompt), while maximizing the service quality (i.e., LLM generation accuracy).

\noindent \textbf{SLO definition.}
In this paper, we define SLO of LLM service as a tuple $<\zeta_{TTFT}, \zeta_{TPOT}>$, where $\zeta$ is the compression ratio to full LLM latency.
The SLOs that an LLM service should serve is pre-defined by the service developers.

\begin{figure}[t]
    \centering
    \includegraphics[width=0.48\textwidth]{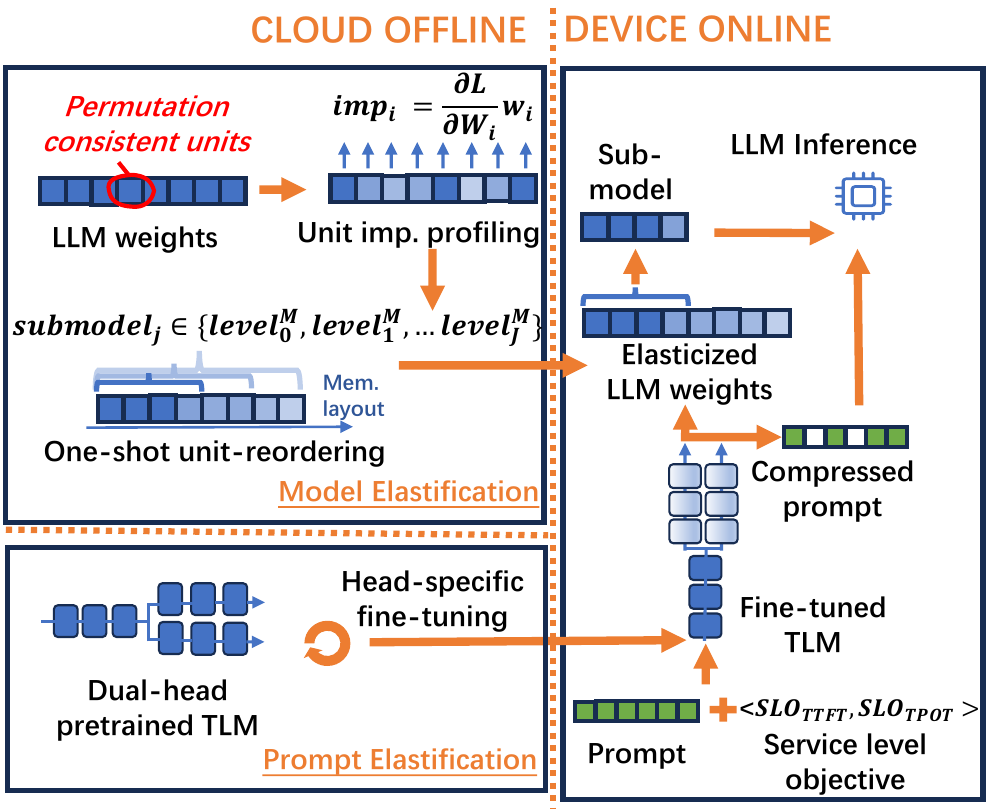}
    \vspace{-17pt}
    \caption{Workflow of \sys.}
    \label{fig:workflow}
    \vspace{-12pt}
\end{figure}

\noindent \textbf{Workflow.}
As shown in Figure~\ref{fig:workflow}, \sys features a two-stage workflow.
At cloud offline stage, on one hand, the model is elasticized to various levels of sub-models that share the memory and can be cost-effectively switched ($\S$\ref{subsec: model_elastification}).
On the other hand, we on-cloud fine-tune a TLM for prompt elastification ($\S$\ref{subsec:prompt_elastification}).
At device online stage, the elasticized LLM and fine-tuned TLM are deployed on mobile devices as a service.
For each LLM service request, the prompt and the corresponding SLO are fed into the fine-tuned TLM.
The TLM then outputs a compressed prompt and selects a sub-model with proper size.
Finally, an LLM inference that satisfies the given SLO is performed atop the sub-model and the compressed prompt.

\subsection{Model elastification}
\label{subsec: model_elastification}

\noindent \textbf{Permutation consistent units of Transformer models.}
We explore a mathematically provable characteristic of Transformer models --- permutation consistency.

\begin{figure}[t]
    \centering
    \includegraphics[width=0.44\textwidth]{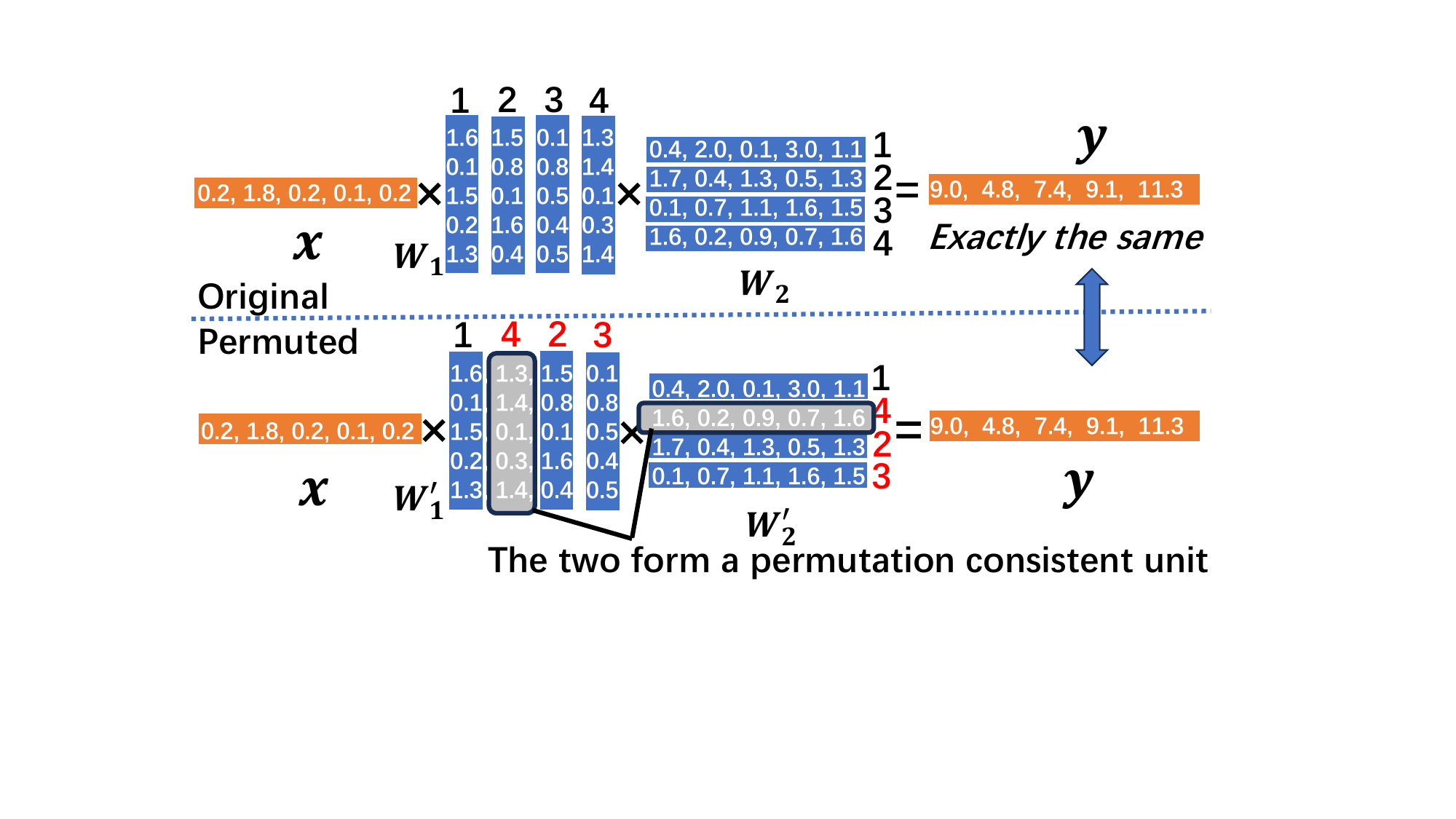}
    \vspace{-10pt}
    \caption{\revision{Illustration of permutation consistency.}}
    \label{fig:permut}
    \vspace{-12pt}
\end{figure}

\vspace{1\baselineskip}
\textit{Property 1. Units of a neuron network are called permutation consistent units if they can be reordered between each other in a block without affecting the block's input/output.}

\vspace{1\baselineskip}
This property indicates that a dense operator kernel can equivalently process these units in arbitrary order without any runtime data re-layout.
The rationale behind it is that the \texttt{Reduce} operator (e.g., sum/min/max) satisfies the commutative and associative laws.
A basic block that contains such units is $y = x W_{1} W_{2}$ in Figure~\ref{fig:permut}.
Its permutation consistent unit is a column of $W_{1}$ together with the corresponding row of $W_{2}$.
If we permute the weights as shown in Figure~\ref{fig:permut}, the intermediate activation $x W'_{1}$ will be permuted in response.
Nevertheless, $W'_{2}$ is also permuted in the same order, so the multiplication of 
\texttt{MatMul} operator can still be performed correctly.
Since the following addition of \texttt{MatMul} operator is a \texttt{Reduce} operator, the calculated $x {W'}_{1} {W'}_{2}$ is exactly the same as $x W_{1} W_{2}$.
Notably, different from prior work~\cite{PIT} that also leverages the \texttt{Reduce} operator to permute DNNs, \sys's key insight is to identify such a joint unit in two-layer blocks that are ubiquitous in Transformers.
Permutation of this unit can be made completely offline, while \cite{PIT} still needs online reordering the input with a single operator level abstraction.


\begin{figure}[t]
    \centering
    \includegraphics[width=0.42\textwidth]{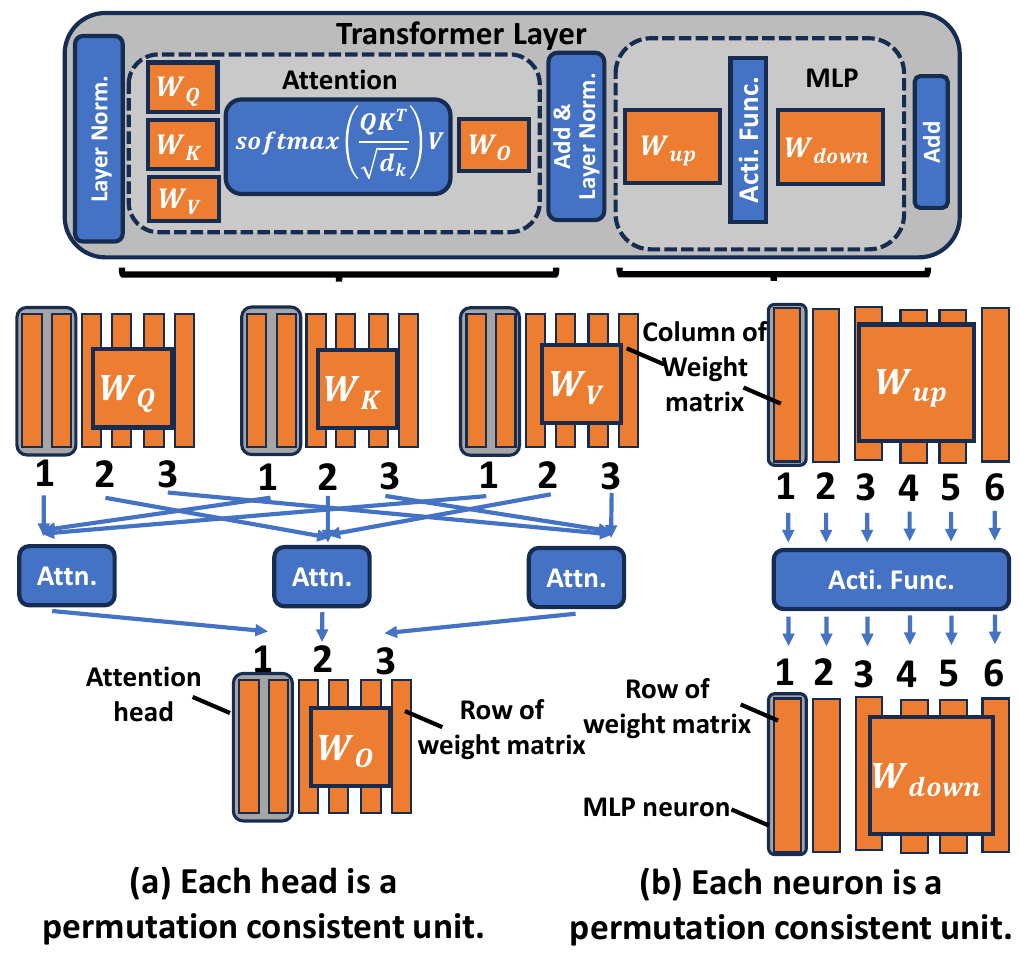}
    \vspace{-10pt}
    \caption{Permutation consistent units in Transformer.}
    \label{fig:transformer_permut}
    \vspace{-12pt}
\end{figure}

\vspace{1\baselineskip}
\textit{Property 2. Attention heads and MLP neurons of Transformer models are permutation consistent units.}
\vspace{1\baselineskip}

As shown in Figure~\ref{fig:transformer_permut}, there are two types of permutation consistent units in the main weights of a Transformer layer, i.e., attention heads and MLP neurons, and they are independent to each other.
Specifically, the contiguous columns/rows with the same indices in $W_{Q}$, $W_{K}$, $W_{V}$ and $W_{O}$ (i.e., an attention head) constitute a permutation consistent unit; a column/row with the same index in $W_{up}$ and $W_{down}$ (i.e., an MLP neuron) also constitute a permutation consistent unit.
The derivation process is similar to the example in Property 1, Figure~\ref{fig:permut} --- the last operator of attention and MLP blocks is \texttt{Reduce}.
Property 2 holds true for mainstream Transformer-based language models with variants like RoPE\footnote{RoPE only introduces position information in sequence axis and intra-head axis.}~\cite{su2023roformerenhancedtransformerrotary}, biased QKV~\cite{qwen}, GQA~\cite{dubey2024llama3herdmodels} or gated MLP~\cite{touvron2023llamaopenefficientfoundation}.

\begin{figure}[t]
    \centering
    \includegraphics[width=0.36\textwidth]{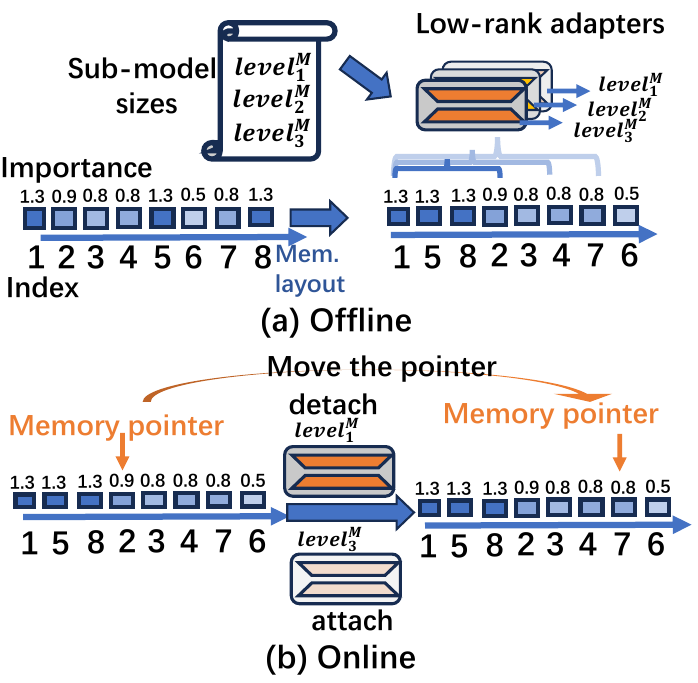}
    \vspace{-10pt}
    \caption{\sys offline profiles and reorders permutation consistent units in one-shot; it online switches (i.e., upgrade/downgrade) sub-models by cost-effectively attaching/detaching the corresponding adapters and moving the memory pointer.}
    \label{fig:unit_reorder}
    \vspace{-13pt}
\end{figure}

\noindent \textbf{Our method: one-shot unit-reordering.}
Based on Property 1 and Property 2, we propose a novel model elastification method as shown in Figure~\ref{fig:unit_reorder}.
Its key idea is to ``atomize'' the Transformer into the units shown in Figure~\ref{fig:transformer_permut}, and then group them to construct a series of sub-models that each is contiguous in memory.
\revision{
To illustrate, basically, in Figure~\ref{fig:permut} each column in $W_1$ and the row with same index in $W_2$ will be assigned with an importance score offline.
Then the columns/rows are reordered in $W_1$/$W_2$ to make a submodel contiguous.
At online, $W'_1$ and $W'_2$ only need to slice a submatrix (e.g., with indices 1,4 out of $W'_1$/$W'_2$) out of the original one by a zero-cost movement of memory pointer. Detailed later.
}

\begin{figure}[t]
	\centering
	\begin{minipage}[b]{0.24\textwidth}
		\includegraphics[width=1\textwidth]{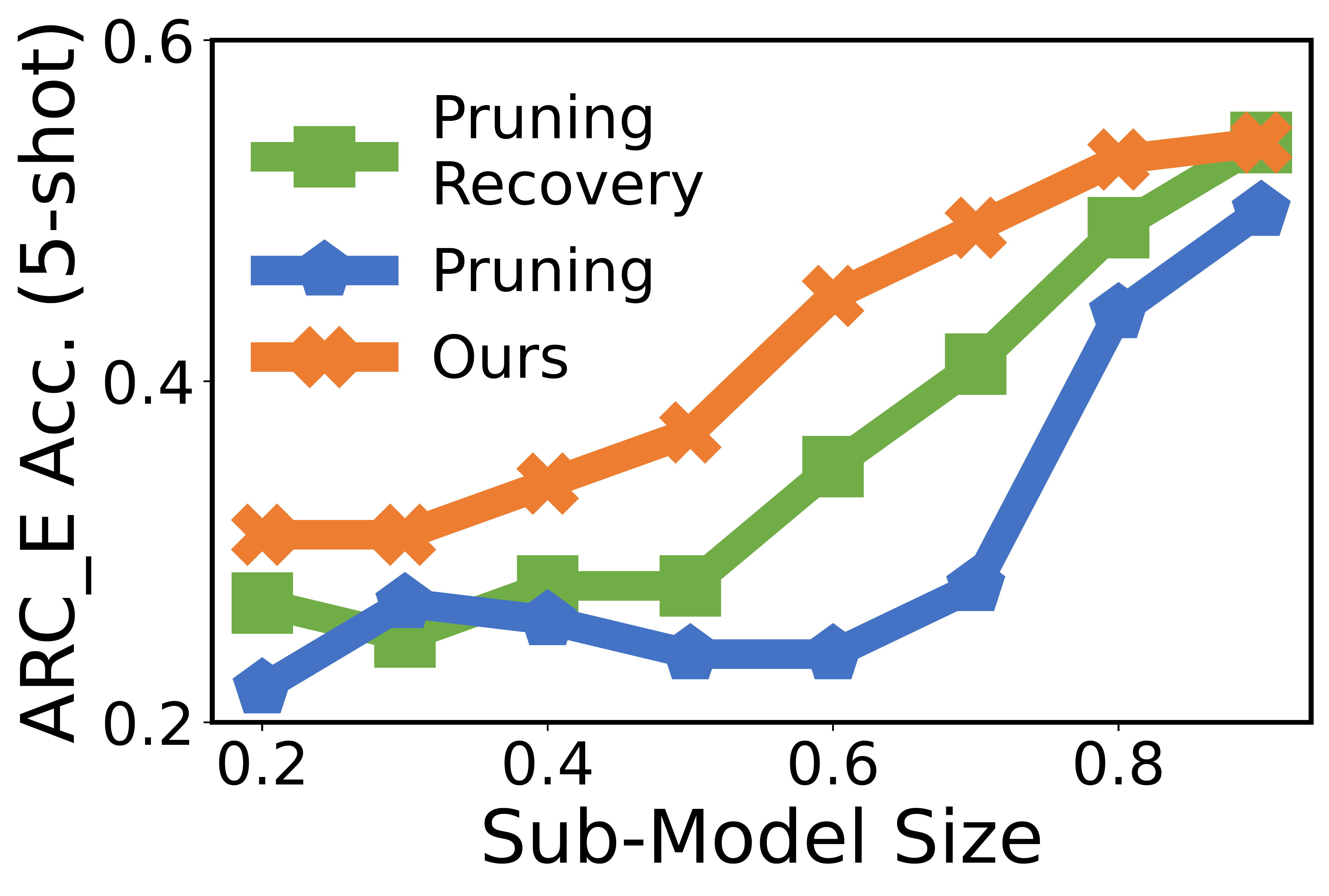}
              \vspace{-17pt}
		\subcaption{ Sub-model quality.}
		\label{fig:sub-model_quality}
	\end{minipage}
	\begin{minipage}[b]{0.19\textwidth}		
            \includegraphics[width=1\textwidth]{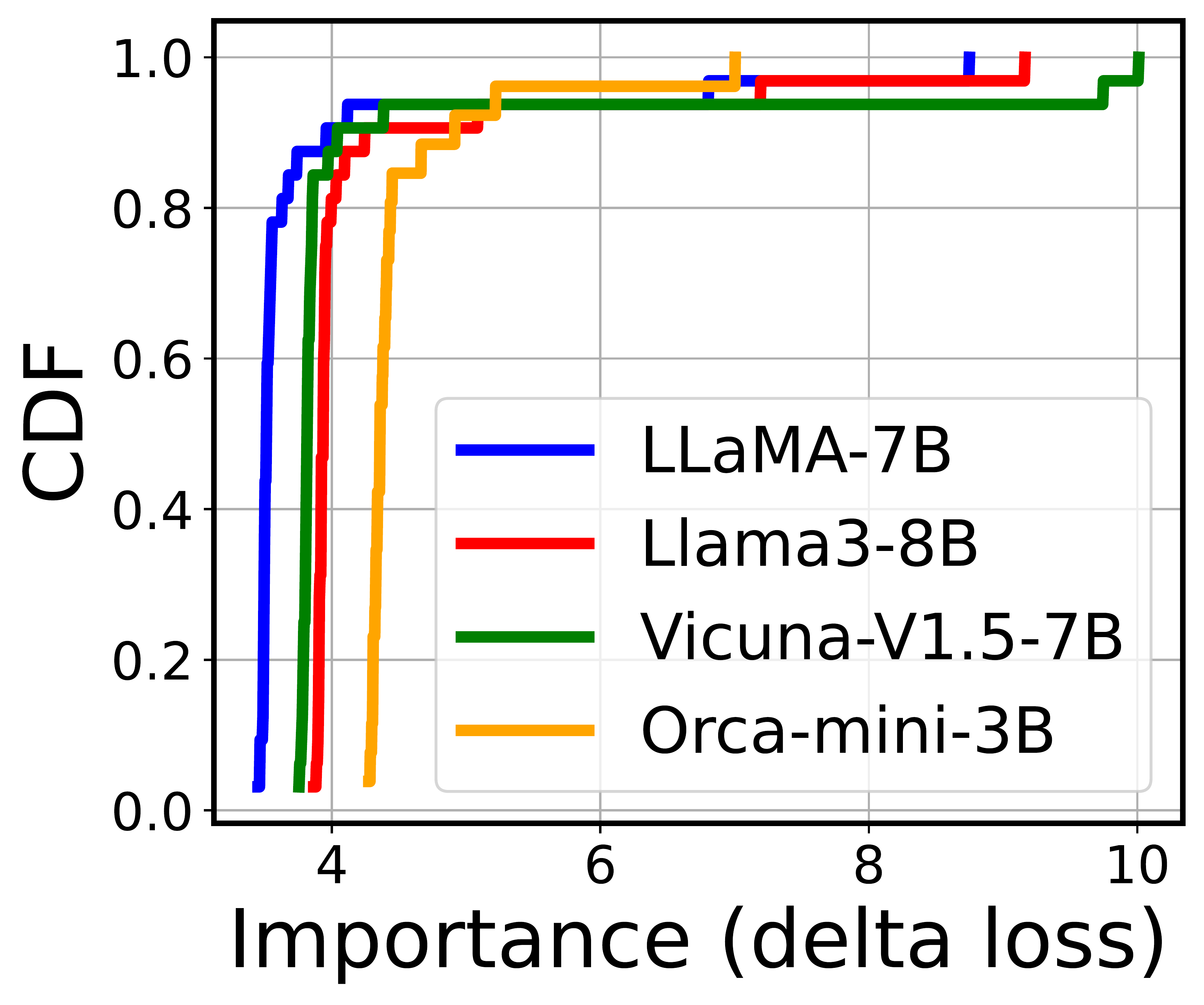}
            \vspace{-17pt}
            \subcaption{Anchor layers.}
            \label{fig:anchor_layers}
	\end{minipage}
        \vspace{-12pt}
	\caption{(a) \sys generates sub-models that with consistently higher quality than pruning and pruning+recovery~\cite{ma2023llmpruner}. (b) A small portion of layers. i.e., ``anchor layers'', is much more important than others.}
    \label{fig:model_elasti_micro_exps}
\end{figure}

\noindent \textit{$\bullet$ Offline.}
Specifically, \sys first profiles importance of each unit offline (detailed later).
Since these units are permutation consistent, \sys freely reorders them by their importance in descending order (if importance is the higher the better), starting from the base address of the weight tensor.
Notably, the reordering is only performed intra-block, e.g., reordering the unit of attention heads in the same attention block.
Then, \sys groups these units into sub-models.
For example, in Figure~\ref{fig:unit_reorder}, sub-model in $level^{M}_{3}$ contains units with indices (not address) ``1 5 8 2'', and sub-model in $level^{M}_{1}$ contains ``1 5 8 2 3 4 7''.
The sub-model sizes and numbers $\{level^{M}_{1}, \cdots, level^{M}_{J}\}$ are pre-defined by the developers.
In practice, we set this to a fine enough granularity (a global ratio of 20\% to 100\% in a step size of 10\%, by default).
Such a ratio is evenly shared by all Transformer layers that need elastification.
After that, low-rank adapters~\cite{hu2021loralowrankadaptationlarge} are attached to each sub-model to recover potential generation quality loss (if there is any) of these sub-models.
We elaborate such a recovery process in the following parts.
So far, the LLM weights have been elasticized into a series of high-quality sub-models that run in various speed.
We demonstrate the quality of generated sub-models in Figure~\ref{fig:sub-model_quality}.

\noindent \textit{$\bullet$ Online.}
In Figure~\ref{fig:unit_reorder}b, the upgrading of model is performed in the following steps:
\sys first detaches the corresponding adapter from sub-model $level^{M}_{1}$, which has served the last request.
Then, it moves the ending memory pointer of the weights tensor from the address of unit with index ``2'' to ``7''.
After that, \sys attaches another adapter to sub-model $level^{M}_{3}$, and the upgrading is finished.
Such a process is very cost-effective on mobile devices --- it does not involve any data movement compared to traditional pruning methods, and can still utilize deeply-optimized dense kernels provided by NN libraries.
For instance, upgrading $W_{Q}$ to  4096$\times$4096 size only takes 2 ms on Redmi K60 Champion smartphone, while a naive pruning method must undergo a 140ms data movement.

\noindent \textbf{Profiling unit importance through explainable-AI.}
Parameters of neuron networks are known to feature diverse importance.
For instance, a weight element with higher magnitude may contribute more to NN capability~\cite{han2015learningweightsconnectionsefficient, PhysRevA.39.6600, li2017pruningfiltersefficientconvnets}.
Inspired by the concept of eXplainable-AI (XAI)~\cite{huang2023elastictrainer, Selvaraju_2019, sundararajan2017axiomaticattributiondeepnetworks}, \sys profiles unit importance with a more accurate method, i.e., by the next-token prediction loss function $L$ on a calibration corpus $C$.
The intuition behind XAI is that, if a unit is more important, it should make $L$ larger on $C$ when been removed.
Specifically, we define importance of unit $i$ as $imp_{i} = |L - L_{W_{i}=0}|$.
By Maclaurin series~\cite{swokowski1979calculus}, we get
\begin{equation}
\label{eq:imp}
\begin{split}
    imp_{i} = |L - L_{W_{i}=0}| = | \frac{\partial L}{\partial W_{i}}W_{i} + o(||W_{i}||^2) |.
\end{split}
\end{equation}
Since the second term is a higher-order infinitesimal, \sys then takes $| \frac{\partial L}{\partial W_{i}}W_{i} |$ as a estimation of unit importance.
By default, $C$ is set to a sub-set of \texttt{bookcorpus}~\cite{Zhu_2015_ICCV}, which is a general-purpose language-modeling dataset.

Besides, interestingly, we find that several layers are much more important than other layers.
We call these layers ``anchor layers''.
We measure the importance of a layer by the increase of loss function when a layer is removed.
As shown in Figure~\ref{fig:anchor_layers}, the importance of layers exhibits a power-law distribution (80/20 rule), which means about 20\% layers are anchor layers.
As a result, we lock these layers from elastification.
For example, if we need a 50\%-size sub-model of a 32 layers LLM, we retain 37.5\% permutation consistent units for each non-anchor layer (26 layers in total).

\noindent \textbf{Task-agnostic low-rank recovery of sub-models.}
We add Low-Rank Adapters (LoRAs)~\cite{hu2021loralowrankadaptationlarge} to the frozen $W_{Q/K/V/O}$ and $W_{up/down}$ of each sub-model to recover them from potential accuracy loss.
A LoRA is two low-rank matrices $A \in \mathcal{R}^{n\times r}$ and $B \in \mathcal{R}^{r\times m}$ that trained as a side branch of main weights.
\sys's default setting of $r$ is 8, an empirically common practice in LoRA-based tunings.
LoRA weights are only 0.1\%--0.5\% of the entire LLM weights, 
and thus such a method only introduces <5\% extra memory overhead even under \sys's fine-grained sub-model settings.
The switching overhead is also minimized since the attaching/detaching operations are all low-rank \texttt{MatMul} and element-wise \texttt{MatAdd}.

Different from traditional language models that need tuning on specific downstream tasks (e.g., Bert~\cite{devlin2019bertpretrainingdeepbidirectional}, T5~\cite{raffel2023exploringlimitstransferlearning}), LLM commonly serves as a generic task solver.
Thereby, \sys's sub-model recovery is \textit{task-agnostic}.
LoRAs are trained with next-token prediction loss that identical to the pre-training process.
Thanks to LoRA's preservation of the LLM backbone's capability, a general, high-quality and moderate-size corpus can handle this recovery well.
By default, each sub-model is recovered on about 50M tokens of \texttt{Alpaca-cleaned}~\cite{alpaca, alpacacleaned} dataset.
We also discuss the impact of recovery data in $\S$\ref{subsec:sensitivity}.

\noindent \textbf{Remarks}
\sys's model elastification generates fine-grained, high-quality sub-models with both acceptable offline overhead and negligible online overhead. 












\subsection{Prompt elastification}
\label{subsec:prompt_elastification}

\begin{figure}[t]
    \centering
    \includegraphics[width=0.45\textwidth]{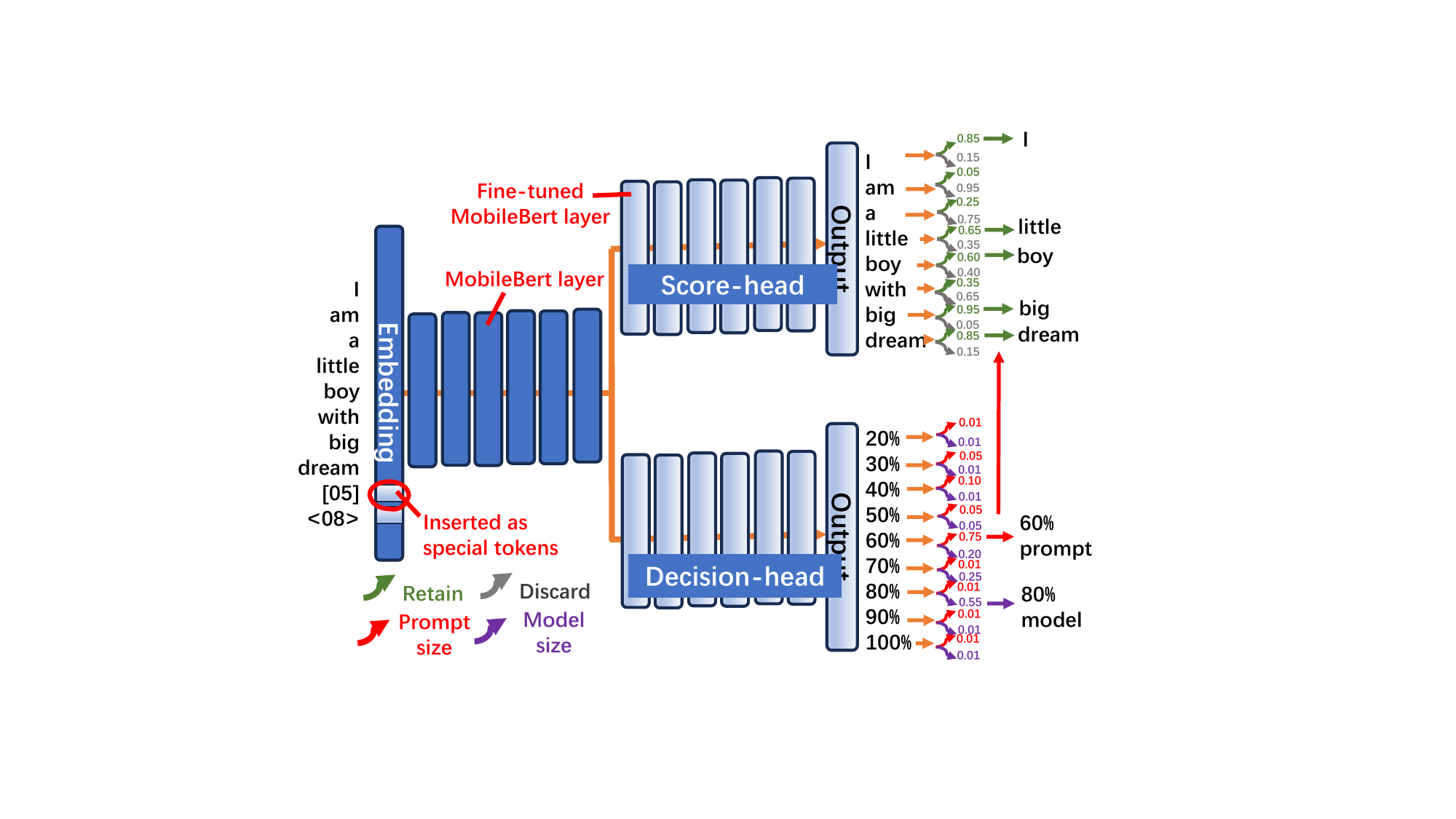}
    \vspace{-10pt}
    \caption{The dual-head TLM.}
    \label{fig:tlm}
\end{figure}

\textbf{Dual-head TLM.}
\sys tackles the challenges mentioned in $\S$\ref{subsec: opportunity} by a dual-head Tiny Language Model (TLM).
As shown in Figure~\ref{fig:tlm}, the TLM is a variant of mobile-friendly pre-trained language model MobileBert~\cite{sun2020mobilebertcompacttaskagnosticbert}, a compact model with only 20\% parameters of BERT\_base~\cite{devlin2019bertpretrainingdeepbidirectional} yet just 0.7\% accuracy loss on GLUE benchmark~\cite{wang2019gluemultitaskbenchmarkanalysis}.
We make the following modifications.
Firstly, the SLO of the current request is marked in natural language and inserted into the embedding layer of MobileBert as special tokens.
For instance, ``[05]'' represents the prefill SLO is 50\% TTFT; ``<08>'' is 80\% TPOT.
These special tokens are initialized to word vectors that orthogonal to each other.
Secondly, the TLM is designed with two separate heads, named score-head and decision-head.
The score-head treats each token of the prompt as a two-class classification problem, where each token can be classified into ``discard'' or ``retain''.
The decision-head treats the entire sequence of prompt with SLO as two multi-class classification problems.
Each possible model elastification level (i.e., the sub-model size, discussed in $\S$\ref{subsec: model_elastification}) is a class of one problem; 
possible prompt elastification levels are classes of the other problem.
Akin to model elastification, the prompt is also pre-defined to multiple fine-grained levels $\{level^{P}_{1}, \cdots, level^{P}_{K}\}$ by the developers.
By default, we set it in alignment with model elastification levels.
Besides, the two heads share the same bottom layers (12 out of 24 layers by default) based on the rationale that bottom layers of DNNs are mainly responsible for capturing basic instead of task-specific information.
In doing so, the overhead of TLM inference/training is further minimized.

\noindent \textbf{TLM inference.}
\revision{
At decision-head, \sys takes the class with the max probability as its decision.
For example, in Figure~\ref{fig:tlm}, the decision-head selects a 60\% prompt and 80\% model.
At score-head, \sys ranks the prompt tokens by the probability of ``retain'', then selects top ones according to decision-head.
For example, in Figure~\ref{fig:tlm}, the scores are ranked by the green probabilities.}
Notably, if the TLM outputs a decision that cannot meet the SLO (which is nearly yet not impossible due to the black-box property of DNNs), \sys will execute a runtime check and the decision will fall back to a random one that stringently meet the SLO.
After inference, we get a compressed prompt and a selected sub-model.

The inference overhead is acceptable.
Its total parameters are about 40M, which means 2 orders of magnitude less memory footprint than the LLM service.
Regarding to latency, the TLM can still perform an on-device inference within 5\% of the original LLM's TTFT even if the LLM's prompt is compressed while TLM's not.

\begin{figure}[t]
    \centering
    \includegraphics[width=0.48\textwidth]{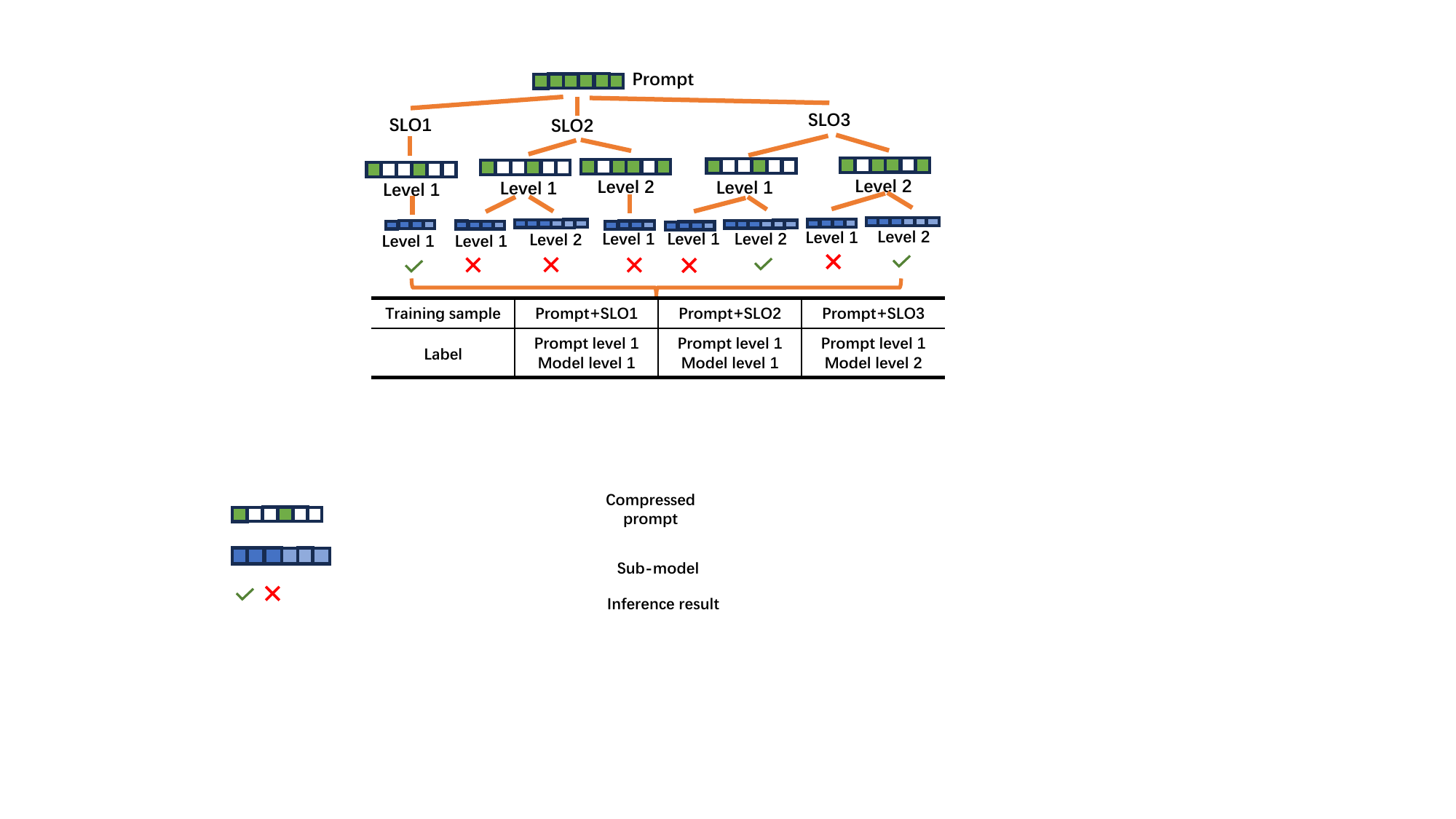}
    \vspace{-22pt}
    \caption{Illustration of the self-induced labelling process of decision-head training data collection.}
    \vspace{-12pt}
    \label{fig:self-induced labelling}
\end{figure}

\noindent \textbf{TLM training.}
The TLM is initialized from the pre-trained weights of MobileBert; each head is further fine-tuned individually.
We keep the pre-trained embedding layer and bottom layers frozen.
When training one head, the other head is also frozen.

\begin{figure}[t]
	\centering
	\begin{minipage}[b]{0.28\textwidth}		
            \includegraphics[width=1\textwidth]{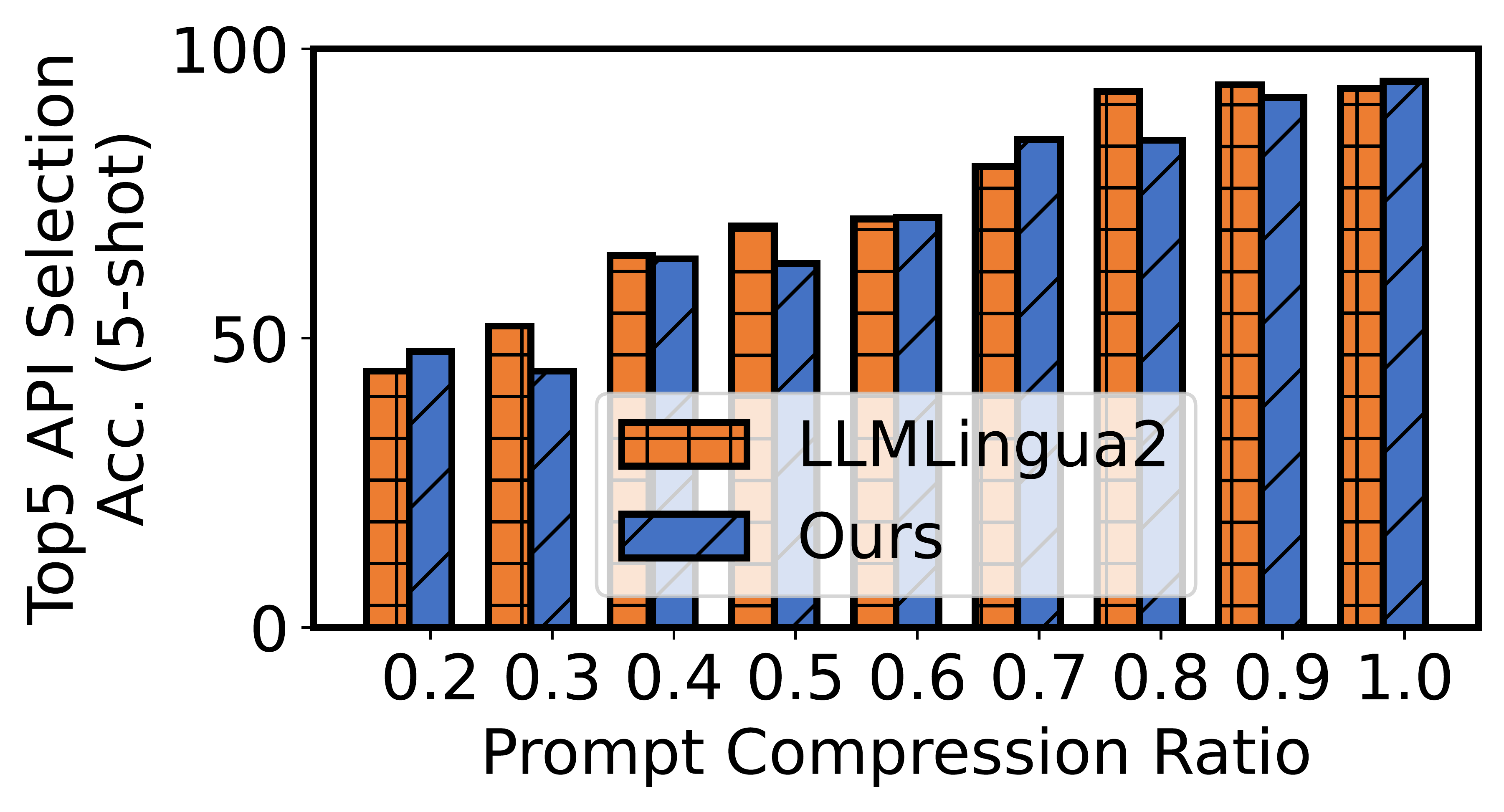}
            \vspace{-17pt}
            \subcaption{Score-head.}
            \label{fig:score-head-effect}
	\end{minipage}
	\begin{minipage}[b]{0.19\textwidth}
		\includegraphics[width=1\textwidth]{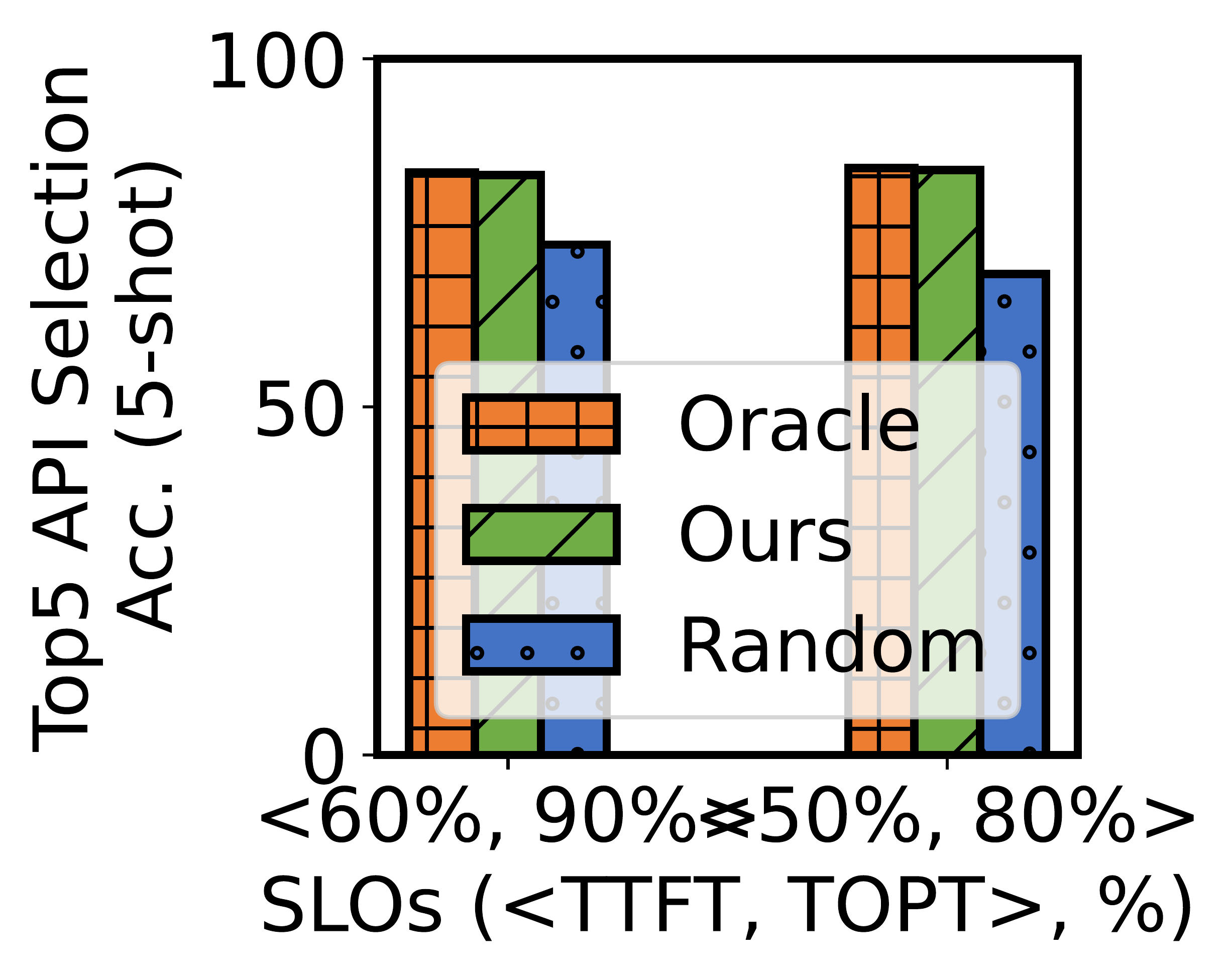}
              \vspace{-17pt}
		\subcaption{Decision-head.}
		\label{fig:decision-head-effect}
	\end{minipage}
        \vspace{-22pt}
	\caption{Effectiveness of dual-head TLM.}
    \label{fig:TLM-effect}
\end{figure}

\noindent $\bullet$ \textit{Score-head. }
We first use \texttt{MeetingBank}~\cite{MeetingBank} dataset to fine-tune the score-head. 
The dataset contains about 50M tokens of corpus that each token is labelled ``discard'' or ``retain'' by GPT-4~\cite{openai2023gpt4}.
Note that the training of score-head is independent to the decision-head and the LLM.
Figure~\ref{fig:score-head-effect} shows the effectiveness of score-head with LLaMA-7B and \texttt{Octopus} dataset.
Our method achieves on-par prompt refining ability compared to \texttt{LLMLingua2}.

\noindent $\bullet$ \textit{Decision-head. }
After training score-head, we train decision-head with the assistance of both the elasticized LLM and the score-head.
The intuition is that we can traverse the decision space and learn the optimal one offline.
Specifically, the training data is collected by a self-induced labelling process.
In Figure~\ref{fig:self-induced labelling}, we provide an illustrative example.
For each prompt and SLO (pre-defined by developer, see $\S$\ref{subsec:overview}), we enumerate all its possible\footnote{The developers pre-execute a one-shot profiling on a testing device within 1 hour under the guidance of Formula~\ref{eq:prefill_decode} and leverage the profiled data to judge whether a decision can meet a SLO.} decisions, and label it with \textit{the most lightweight one under which the LLM can output correct answer}.
If all possible decisions fail, the decision is assigned to a default one (e.g., random).
By default, \sys collects training samples from a comprehensive benchmark \texttt{MMLU-Pro}~\cite{wang2024mmluprorobustchallengingmultitask}, which contains 17 domains of question-answer pairs.
In Figure~\ref{fig:decision-head-effect}, the decision head demonstrates significantly higher  quality than random decision, and approaches oracle.
We also discuss the impact of TLM training data in $\S$\ref{subsec:sensitivity}.

\noindent \textbf{Remarks}
\sys's prompt elastification efficiently and effectively elasticizes prompt and orchestrates model- and prompt-dimension of elastic LLM service.

\section{Implementation}

\textbf{Offline stage.}
We build the elasticized LLM and the TLM on top of Pytorch~\cite{paszke2019pytorchimperativestylehighperformance}.
We modify the \texttt{modeling.py} of \texttt{Huggingface Transformers}\footnote{Commit hash c31473ed4492fdf26aec4173451f31590021862f}~\cite{wolf-etal-2020-transformers} library to identify the permutation consistent units and profile their importance.
In doing so, \sys can easily be made forward-compatible with LLMs released in the future.
The offline elastification is performed on a cloud server equipped with 8 NVIDIA A40 GPUs.

\noindent \textbf{On-device stage.}
We build an LLM service on top of \texttt{mllm}\footnote{Commit hash ed766fae54d9f1bf2b6b25018e6cc434ce223303}~\cite{mllm}, which is a lightweight yet powerful on-device LLM inference library written in pure C++ and assembly language.
We pack the LLM inference program as a standalone binary file and run it as an independent process.
The apps interact with it through interfaces like \texttt{bindLLMService()} and \texttt{callLLM()}.
To facilitate the upgrade/downgrade of elasticized sub-models without incurring inference-time performance degradation, the \texttt{Linear} is replaced by \texttt{ElasticLinear}.
Specifically, we wrap the original dense kernel with an additional memory pointer that specifies the addresses of sub-model weights.
We optimize the low-rank \texttt{MatMul} and \texttt{MatAdd} of LoRA with ARM \texttt{NEON}~\cite{armneon}.
The LLM service runs on Commercial Off-The-Shelf (COTS) smartphones.
\section{Evaluation}

\subsection{Experimental settings}
\label{susec: exp_settings}

\textbf{Testbed.}
We conduct experiments on the following testbeds.
On cloud, we use a server with a 64-core CPU (750GB RAM) and 8 A40 GPUs (45GB HBM each).
On device, we test \sys across COTS smartphones listed in Table~\ref{tab:devices}.

\begin{table}[]
\footnotesize
\begin{tabular}{c|c|c}
\hline
\textbf{Name}          & \textbf{SoC}     & \textbf{RAM} \\ \hline
Redmi K60 Champion Edition~\cite{redmik60champion} & Snapdragon 8gen2 & 19GB         \\ \hline
Mi 14~\cite{mi14}                  & Snapdragon 8gen3 & 22GB         \\ \hline
Redmi K70 Pro~\cite{redmik70pro}          & Snapdragon 8gen3 & 24GB         \\ \hline
\end{tabular}
\caption{Devices we use in our experiments.}
\vspace{-17pt}
\label{tab:devices}
\end{table}

\noindent \textbf{Models.}
We test the following LLMs.
(1) \textit{Two base LLMs:} LLaMA-7B~\cite{touvron2023llamaopenefficientfoundation} and Llama3-8B~\cite{dubey2024llama3herdmodels}.
(2) \textit{Two instruction-tuned LLMs:} Vicuna-V1.5-7B~\cite{zheng2023judging} and Llama3-instruct-8B~\cite{dubey2024llama3herdmodels}.
(3) \textit{One sub-7b LLM:} Orca-mini-3B~\cite{mukherjee2023orca}.

\noindent \textbf{SLOs.}
We randomly set 6 SLOs based on a stepwise sensitivity hierarchy as shown in Table~\ref{tab:trace}.
We also enumerate more varying SLOs in Figure~\ref{fig:moreslo} for a real-world discussion.
\yws{Rewind is a content comprehension app, which requires relaxed SLO; GMail is a natural language summary and reasoning task, which requires relatively relaxed SLO; Octopus and Shortcuts are on-device agent tasks, which need moderate SLO as they involve interactions with the environment; Gboard and XiaoAi are AI assistants, which need real-time response, i.e., a relatively tight SLO.}

\noindent \textbf{Workload.} We evaluate \sys on both standalone datasets and end-to-end synthesized traces.

\revision{
\noindent $\bullet$\textit{Datasets.} We select 6 representative datasets/benchmarks:
\texttt{ARC\_E}~\cite{allenai:arc}, \texttt{OBQA}~\cite{OpenBookQA2018}, 
\texttt{Octopus}~\cite{chen2024octopus},
\texttt{PIQA}~\cite{PIQA},
\texttt{SCIQ}~\cite{SciQ} and
\texttt{LlamaTouch}~\cite{zhang2024llamatouch}.
We report option select accuracy of \texttt{ARC\_E}, \texttt{PIQA}, \texttt{SCIQ} and \texttt{OBQA}, top-5 function (without parameters) selection accuracy of \texttt{Octopus}, and app invocation accuracy of \texttt{LlamaTouch}.
Each entry is augmented by in-context learning in 5-shots.
}
\yws{\texttt{ARC\_E}, \texttt{PIQA}, \texttt{SCIQ} and \texttt{OBQA} are natural language comprehension and common-sense reasoning tasks that are ubiquitous on mobile devices; \texttt{Octopus} is an on-device API-calling benchmark that follows the natural language instruction of users and selects the most suitable functions; \texttt{LlamaTouch} is a realistic and complicated on-device UI-automation agent benchmark that manipulates mobile apps following user instructions.}

\noindent $\bullet$\textit{End-to-end traces.}
We further synthesize end-to-end traces on top of the above datasets.
In Table~\ref{tab:trace}, we list 6 conceived apps in response to the pre-defined SLOs.
The requests and groundtruths of an app are synthesized from a dataset in similar domain, since there is no public available user data in the wild.
Specifically, we collect 600 entries of requests in total for a trace.
To comprehensively evaluate \sys, we emulate the distribution skewness of requests by $Num(i) =  \frac{600\times e^{\alpha i}}{\sum^{k=6}_{k=1} e^{\alpha k}}$, where $Num(i)$ is an app's \# of request, $i$ is the SLO level (the lower, the tighter), and $\alpha$ is a controlling factor.
The larger the value of $\alpha$, the greater the proportion of more relaxed SLOs in the trace.
When $\alpha=0$, all SLOs are evenly distributed in the trace.
We synthesize multiple traces with various distributions.
For each trace, we randomly shuffle the requests and set the arrival timestamp by a Poisson distribution.

\begin{table}[]
\footnotesize
\begin{tabular}{@{}ccc|ccc@{}}
\toprule
\textbf{\footnotesize Apps} & \textbf{\footnotesize SLO} & \textbf{ \footnotesize Dataset} & \textbf{\footnotesize Apps} & \textbf{\footnotesize SLO} & \textbf{\footnotesize Dataset} \\ \midrule
Rewind  & {\scriptsize <100\%, 100\%>} & \texttt{OBQA}    & Shortcuts & <40\%, 70\%> & \texttt{\scriptsize LTouch} \\
GMail   & <80\%, 90\%>   & \texttt{ARC\_E}  & Gboard    & <20\%, 60\%> & \texttt{PIQA}       \\
Octopus & <60\%, 80\%>   & \texttt{\scriptsize Octopus} & XiaoAi    & <20\%, 50\%> & \texttt{SCIQ}     \\ \bottomrule
\end{tabular}
\caption{\revision{Apps, SLOs and datasets for trace synthesis.}}
\label{tab:trace}
\vspace{-17pt}
\end{table}


\noindent \textbf{\sys configurations.}
The training/fine-tuning/calibration data, elastification levels and TLM configuration are the same as those described in $\S$\ref{sec:design}.

\noindent \textbf{Baselines.}
We compare \sys to the following alternatives.
(1) Directly employing pre-trained from scratch LLMs (\texttt{PFS}) for diverse SLOs is a strong yet plausible baseline.
Due to the unaffordable GPU resource consumption, we select the off-the-shelf OPT~\cite{zhang2022optopenpretrainedtransformer} family. 
As far as we know, it provides the richest variants with fewer than 7B parameters (5 models from 125M to 6.7B).
(2) LLMPruner~\cite{ma2023llmpruner} (\texttt{LPruner}) is a State-of-The-Art (SoTA) parameter pruning method for elastification.
(3) Layer-wise elastification~\cite{layerelastification} (\texttt{LE}) prunes parameters at layer level.
(4) LLMLingua2~\cite{wu2024llmlingua2} + Contextual sparsity~\cite{pmlr-v202-liu23am} (\texttt{LG2+CS}) compresses prompts at prefill stage, and dynamically activates MLP neurons for each token at decode stage.
\revision{
(5) Other strong pruning baselines: \texttt{LaCo}~\cite{yang2024laco}, \texttt{ShortGPT}~\cite{men2024shortgptlayerslargelanguage} and \texttt{AttnDrop}~\cite{he2024matterstransformersattentionneeded}.
}

\subsection{End-to-end performance}






We first evaluate end-to-end performance on traces in $\S$\ref{susec: exp_settings}.

\begin{figure*}[t]
    \centering
    \includegraphics[width=0.99\textwidth]{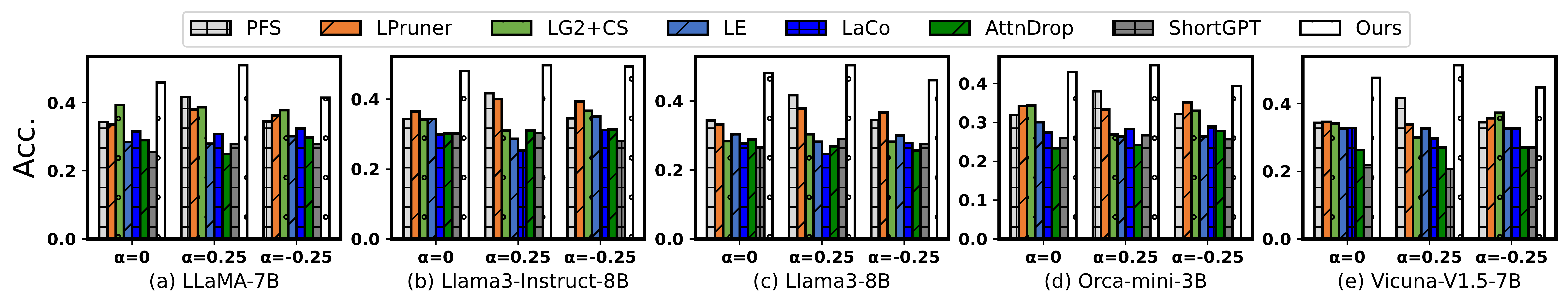}
    \vspace{-10pt}
    \caption{\revision{End-to-end request-response accuracy on the traces. $\alpha$ controls the SLO distribution.}}
    \label{fig:acc_on_trace}
    \vspace{-12pt}
\end{figure*}
\begin{figure*}[t]
    \centering
    \includegraphics[width=0.99\textwidth]{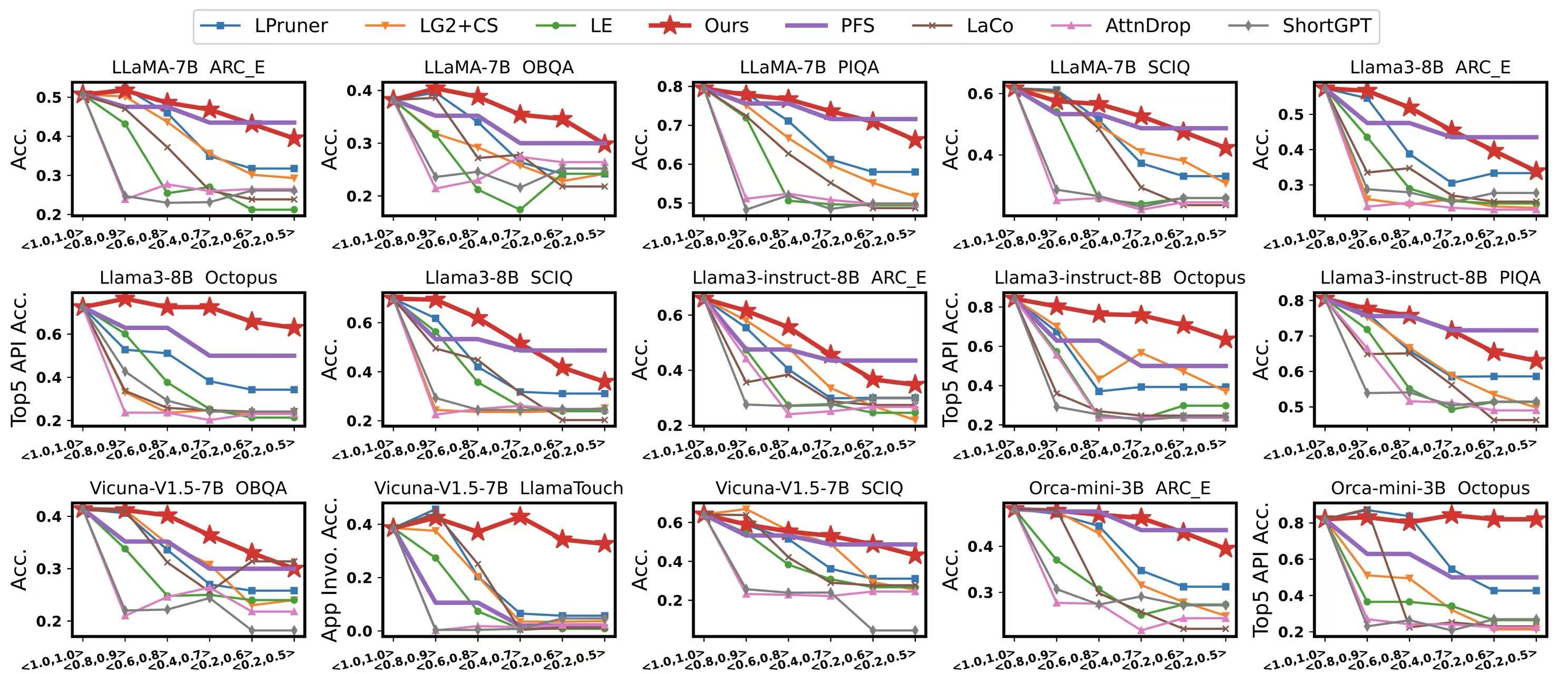}
    \vspace{-10pt}
    \caption{\revision{Performance under one SLO on an entire standalone dataset without considering switching overhead.}}
    \label{fig:dataset_standalone}
\end{figure*}
\noindent \textbf{Accuracy.}
We report the request-response accuracy when all requests' SLOs are met in a trace.
We compare the correctness of the LLM's answer to groundtruth.
We set three levels of trace skewness: $\alpha=0$ (even), $\alpha=0.25$ (towards relaxed) and $\alpha=-0.25$ (towards tight).
Each trace is executed on three diverse COTS devices listed in Table~\ref{tab:devices}.
The results are averaged across these three devices and shown in Figure~\ref{fig:acc_on_trace}.

\sys significantly outperforms the baselines by 6.60\%--14.83\% (10.45\% on average) in absolute accuracy.
Compared to \texttt{PFS}, \sys further involves prompt elastification and does not introduce costly switching overhead.
Besides, \texttt{PFS} does not fully make use of the room below a given SLO.
Another potential reason is that \sys derives sub-models from a bulky one, which may feature stronger emergent ability than the SLMs pretrained from scratch.
Compared to \texttt{LPruner}, \sys involves prompt elastification, and minimizes switching overhead.
\revision{
\textit{Uniqueness to \texttt{LG2+CS}.}
Firstly, the heavy SLM of \texttt{LLMLingua2} must compress the prompt very aggresively to meet a tight TTFT SLO since contextual sparsity shows limited acceleration on prefill stage due to the low locality.
Secondly, the sparsity ratio is also compromised due to non-relu~\cite{agarap2019deeplearningusingrectified,touvron2023llamaopenefficientfoundation} activation functions, non-elastictizable attention~\cite{pmlr-v202-liu23am} and degraded performance of sparse kernels.
In contrast, \sys's model elastification identifies and leverages the permutation consistency in Transformer models and works for both prefill and decoding, and its prompt elastification considers and tackles the unique challenge of prompt-model orchestration.
Compared to \texttt{LE}/\texttt{LaCo}/\texttt{ShortGPT}/\texttt{AttnDrop}, \sys shows steady performance gain.
This is due to that these methods only elasticizes the model in layer level.
In contrast, \sys's unit-level fine-grained pruning traverses a much larger space.
Also, \texttt{LaCo} will generate sub-models that cannot fully share the weights between each other, resulting in extra switching overhead.
}

\begin{figure}[t]
	\centering
	\begin{minipage}[b]{0.25\textwidth}		
            \includegraphics[width=1\textwidth]{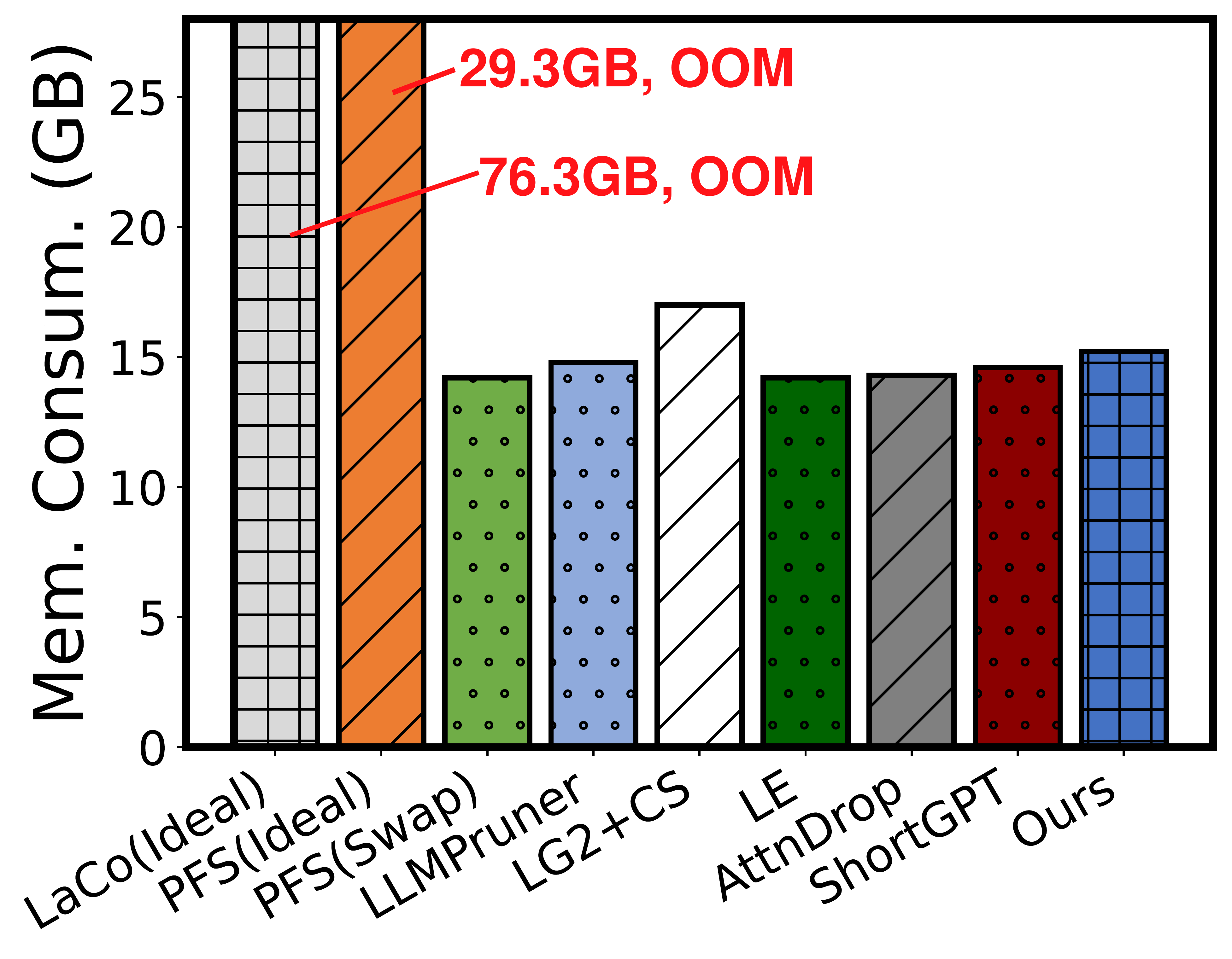}
            \vspace{-17pt}
            \subcaption{Memory consumption.}
            \label{fig:mem consum.}
	\end{minipage}
	\begin{minipage}[b]{0.195\textwidth}
		\includegraphics[width=1\textwidth]{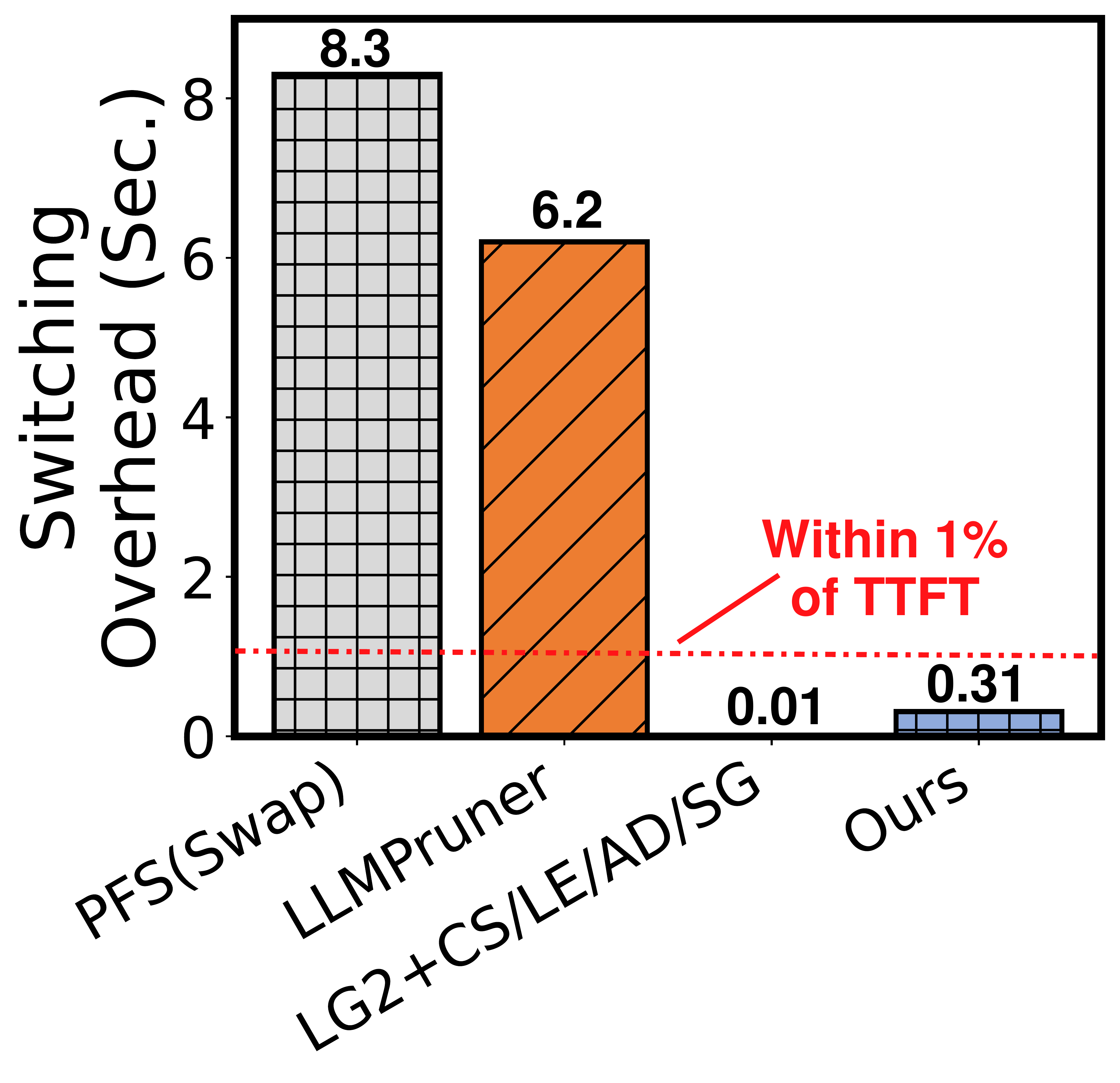}
              \vspace{-17pt}
		\subcaption{Switching time.}
		\label{fig:switching time}
	\end{minipage}
        \vspace{-12pt}
	\caption{Online overhead.}
    \label{fig:online-overhead}
    \vspace{-17pt}
\end{figure}

\noindent \textbf{Memory consumption.}
We discuss peak memory consumption in Figure~\ref{fig:mem consum.}.
Without loss of representativeness. we report LLaMA-7B in the trace with $\alpha=0$ on Redmi K60 Champion.
\sys consumes on-par memory compared to the baselines (15--17GB).
Notably, directly deploying all the dedicated size LLMs in memory is impractical.
As marked as \texttt{PFS(Ideal)}/\texttt{LaCo(Ideal)} in Figure~\ref{fig:mem consum.}, it consumes 29.3GB/76.3GB memory in total, which is OOM on all the COTS devices in Table~\ref{tab:devices}.

\noindent \textbf{Switching overhead.}
With the same setting, we report the breakdown latency of switching between different model elastification levels (i.e., sub-models) in Figure~\ref{fig:switching time}.
Switching time is actually part of TTFT, and a too long switching time will preempt the room of model/prompt capacity, leading to a lower accuracy.
\texttt{PFS (Swap)} and \texttt{LPruner} incur unacceptable time overhead that up to 8.3/6.2 seconds per request.
This is mainly attributed to the costly swapping and in-memory data movement.
\sys only takes 0.31 second to switch to a new submodel.
\revision{
Such a number is lower than 1\% of average TTFT of LLM service, thus being completely acceptable.
The end-to-end experiments with switching overhead considered in Figure~\ref{fig:acc_on_trace} also show that the switching time of \sys is an acceptable tradeoff for LLM service.
}

\noindent \textbf{Remarks }
\sys is the most high-quality and feasible solution for end-to-end on-device elastic LLM service.

\subsection{Performance on standalone datasets}

With a specific SLO, we further report the performance on an entire standalone dataset to show \sys's superiority.
The switching overhead is dismissed as there is no upgrade/downgrade.
The results are obtained on cloud server with SLO statistics on Mi 14 smartphone, and are shown in Figure~\ref{fig:dataset_standalone}.
We have the following observations.
\sys significantly outperforms all its baselines by up to 40\% on accuracy.
Specifically, on all the SLOs, \sys \textit{always} provides a much higher accuracy than the baselines (\texttt{LE/LG2+CS/LPruner}) that derive sub-models from the original one.
Compared to \texttt{PFS}, \sys provides a higher accuracy on 77.8\% of all SLOs.
The reasons are as discussed before.
Please also note that \texttt{PFS} and \texttt{LaCo} are costly in terms of switching and memory at runtime.
\revision{
Performance gains are minimal in some high-SLO scenarios, this is due to a slight pruning (i.e., high SLO) will not significantly differentiate \sys from the pruning baselines.
However, only \sys can work for all the varying SLOs for apps, since there will not always be high-SLO scenarios, and switching overhead (Figure~\ref{fig:switching time}) will also be considered when serving multiple apps.
}

\subsection{Offline stage overhead}
\begin{figure}[t]
	\centering
	\begin{minipage}[b]{0.24\textwidth}		
            \includegraphics[width=1\textwidth]{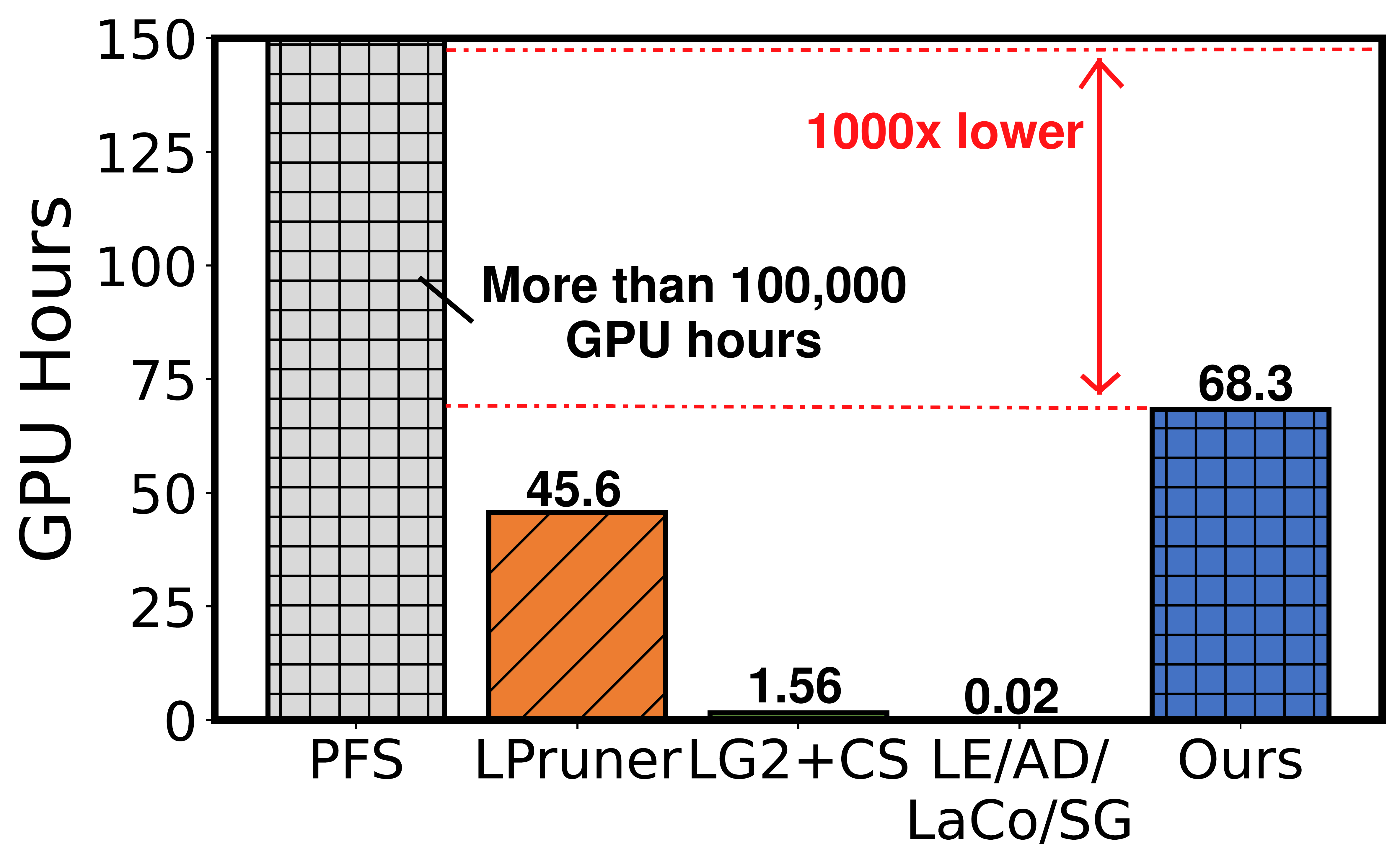}
            \vspace{-17pt}
            \subcaption{GPU hours.}
            \label{fig:gpuhours}
	\end{minipage}
	\begin{minipage}[b]{0.20\textwidth}
		\includegraphics[width=1\textwidth]{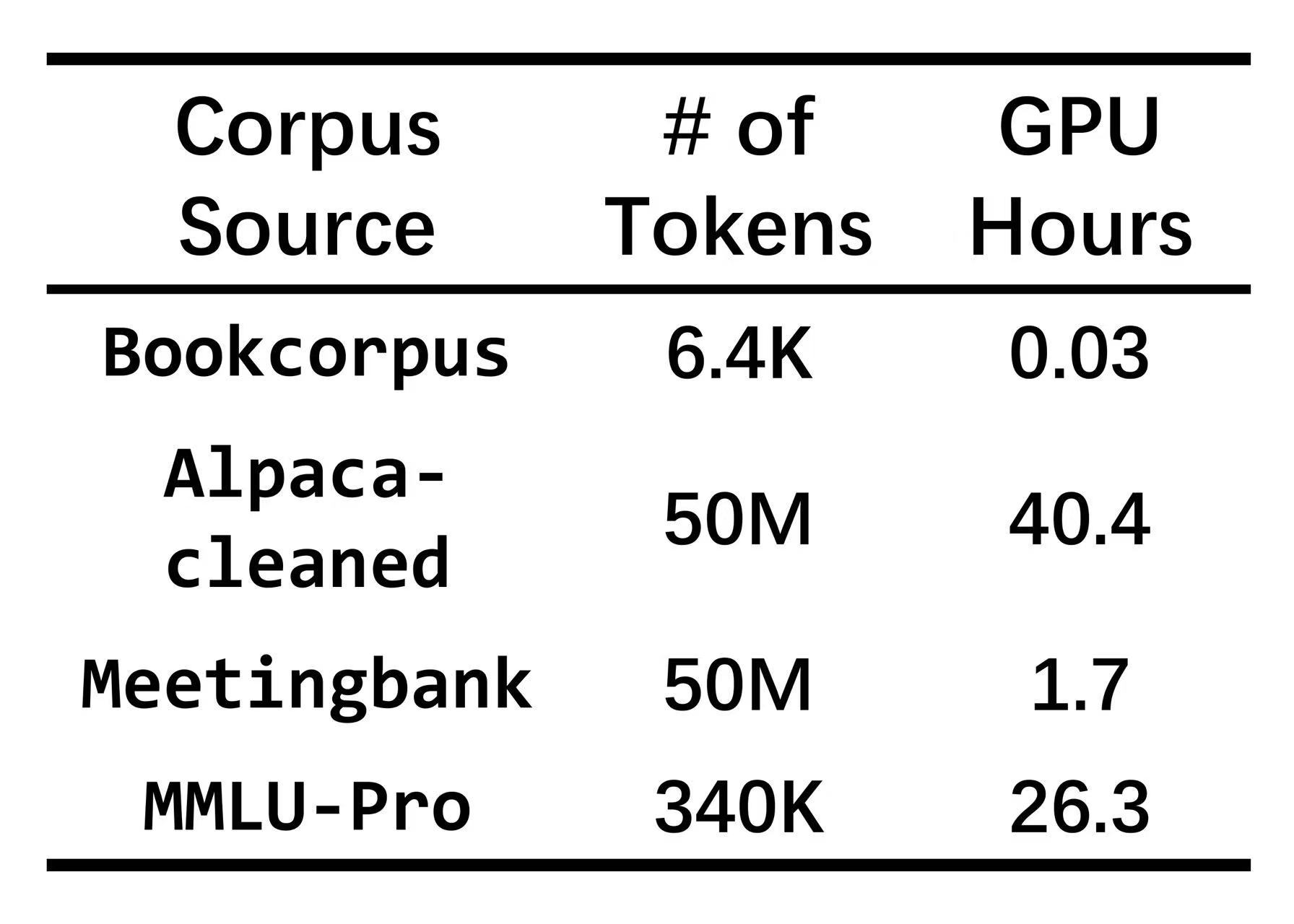}
              \vspace{-17pt}
		\subcaption{Breakdown.}
		\label{fig:corpusbreakdown}
	\end{minipage}
        \vspace{-12pt}
	\caption{Offline overhead of \sys.}
        \vspace{-12pt}
    \label{fig:offline-overhead}
\end{figure}

\begin{figure*}[t]
    \centering
	\begin{minipage}[b]{0.34\textwidth}		
            \includegraphics[width=1\textwidth]{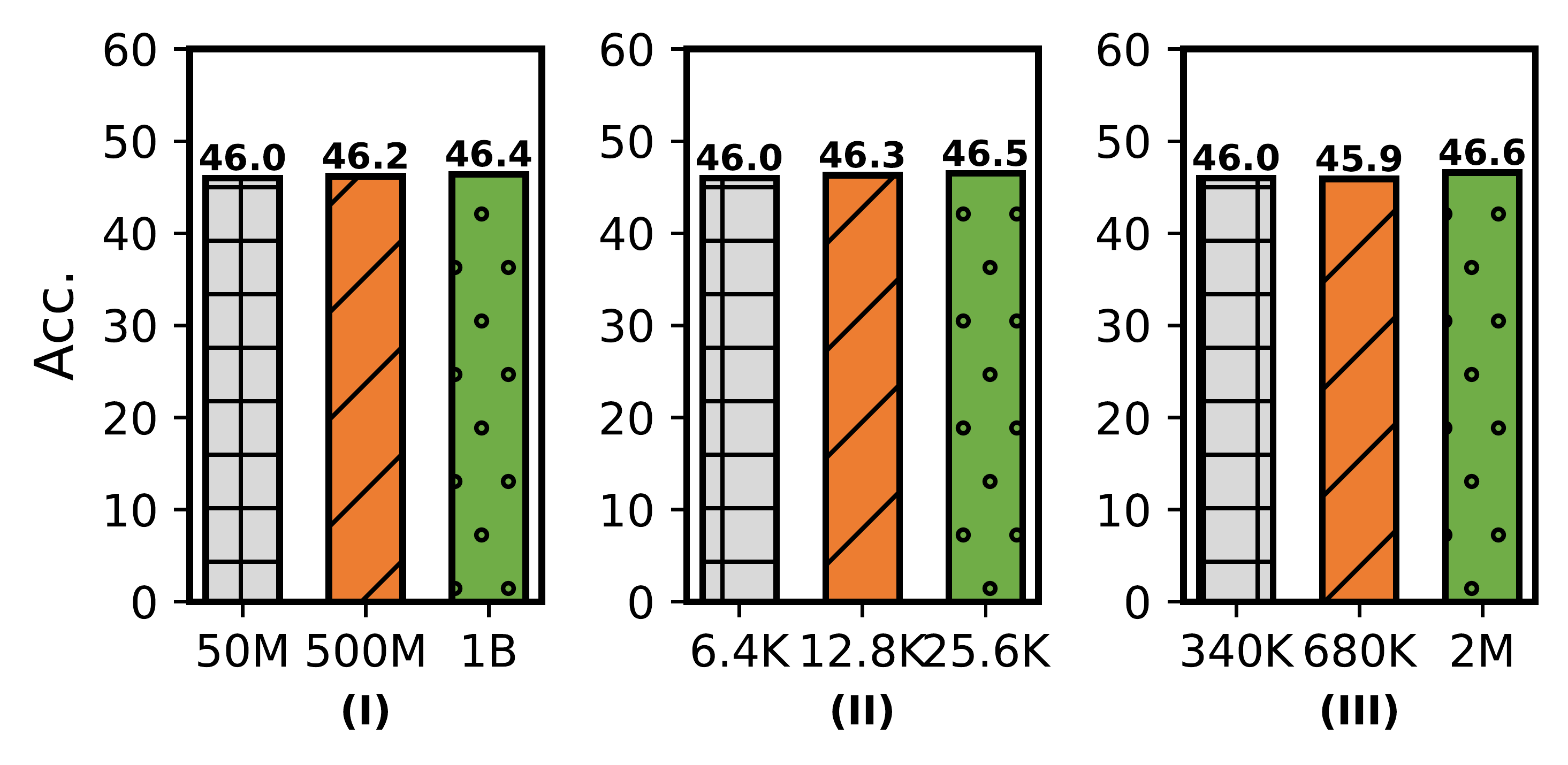}
            \vspace{-17pt}
            \subcaption{\textit{I}. Model Recovery. \textit{II.} Importance profiling. \textit{III}. Decision-head training.}
            \label{fig:data_sensitivity}
	\end{minipage}
	\begin{minipage}[b]{0.23\textwidth}
		\includegraphics[width=1\textwidth]{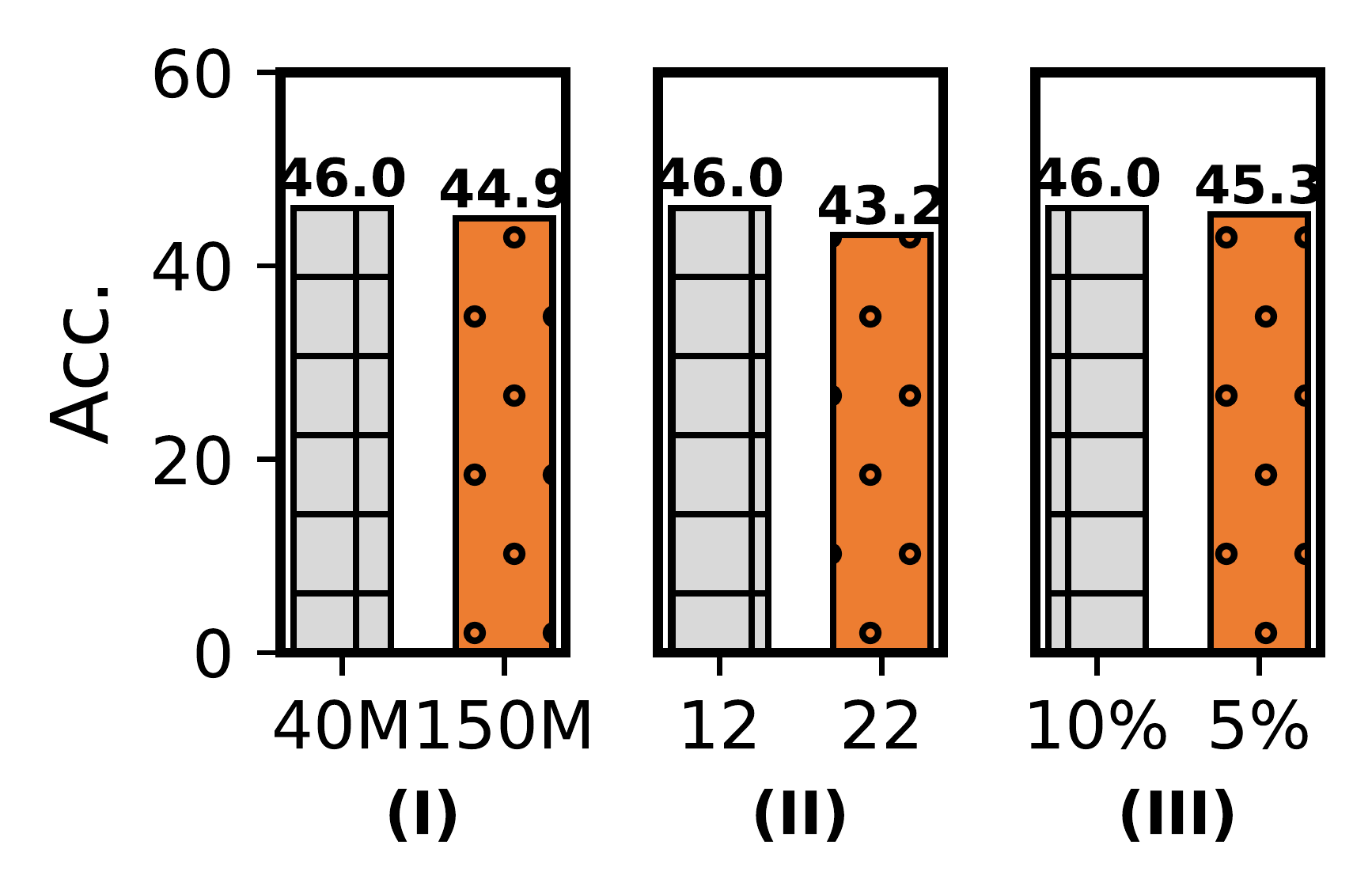}
              \vspace{-17pt}
		\subcaption{\textit{I.} TLM size. \textit{II.} \# of shared layers. \textit{III.} Elasticity stepsize.}
		\label{fig:configuration_sensitivity}
	\end{minipage}
	\begin{minipage}[b]{0.17\textwidth}
		\includegraphics[width=1\textwidth]{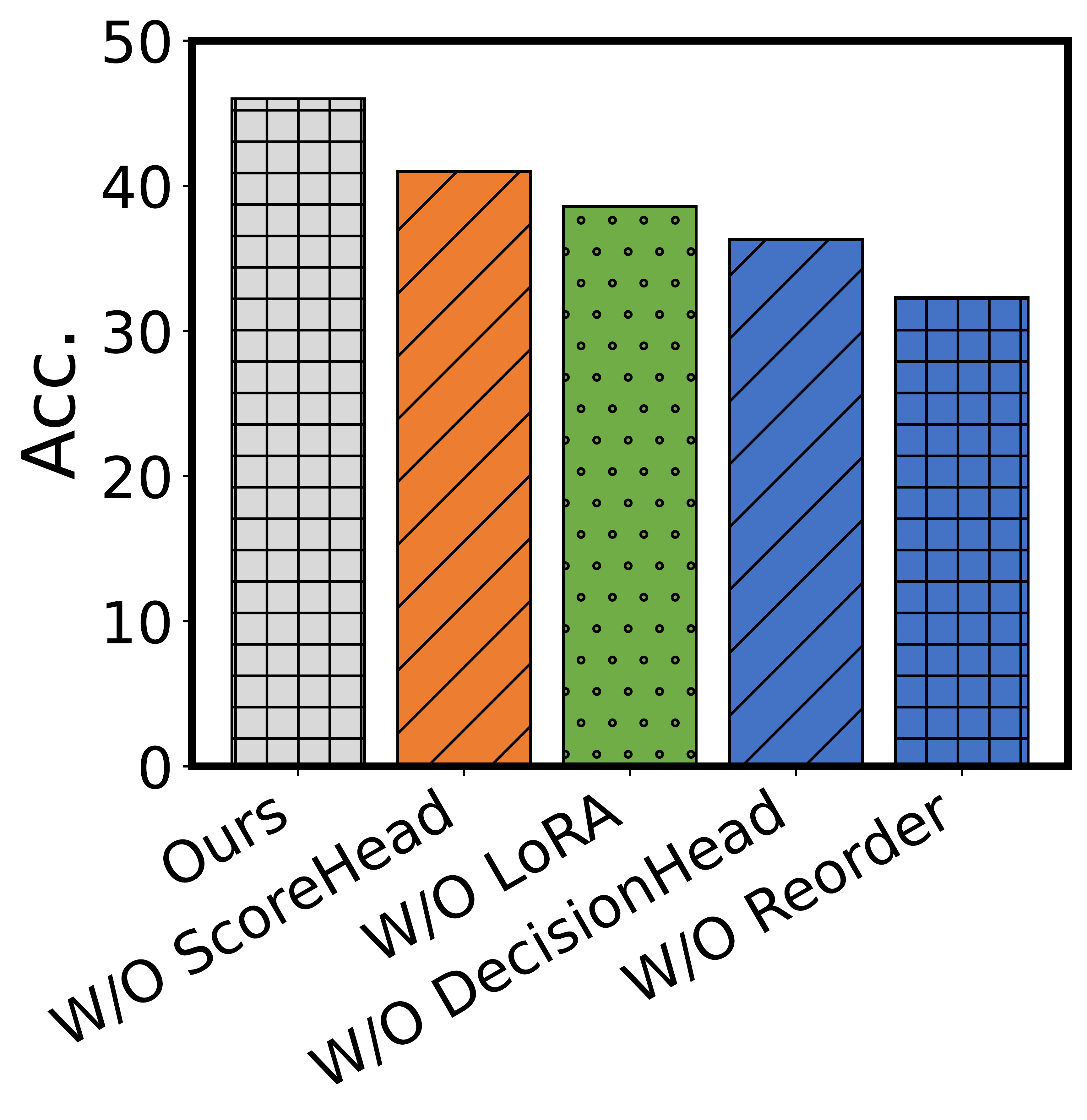}
		\subcaption{Ablation study.}
		\label{fig:ablation}
	\end{minipage}
	\begin{minipage}[b]{0.24\textwidth}
		\includegraphics[width=1\textwidth]{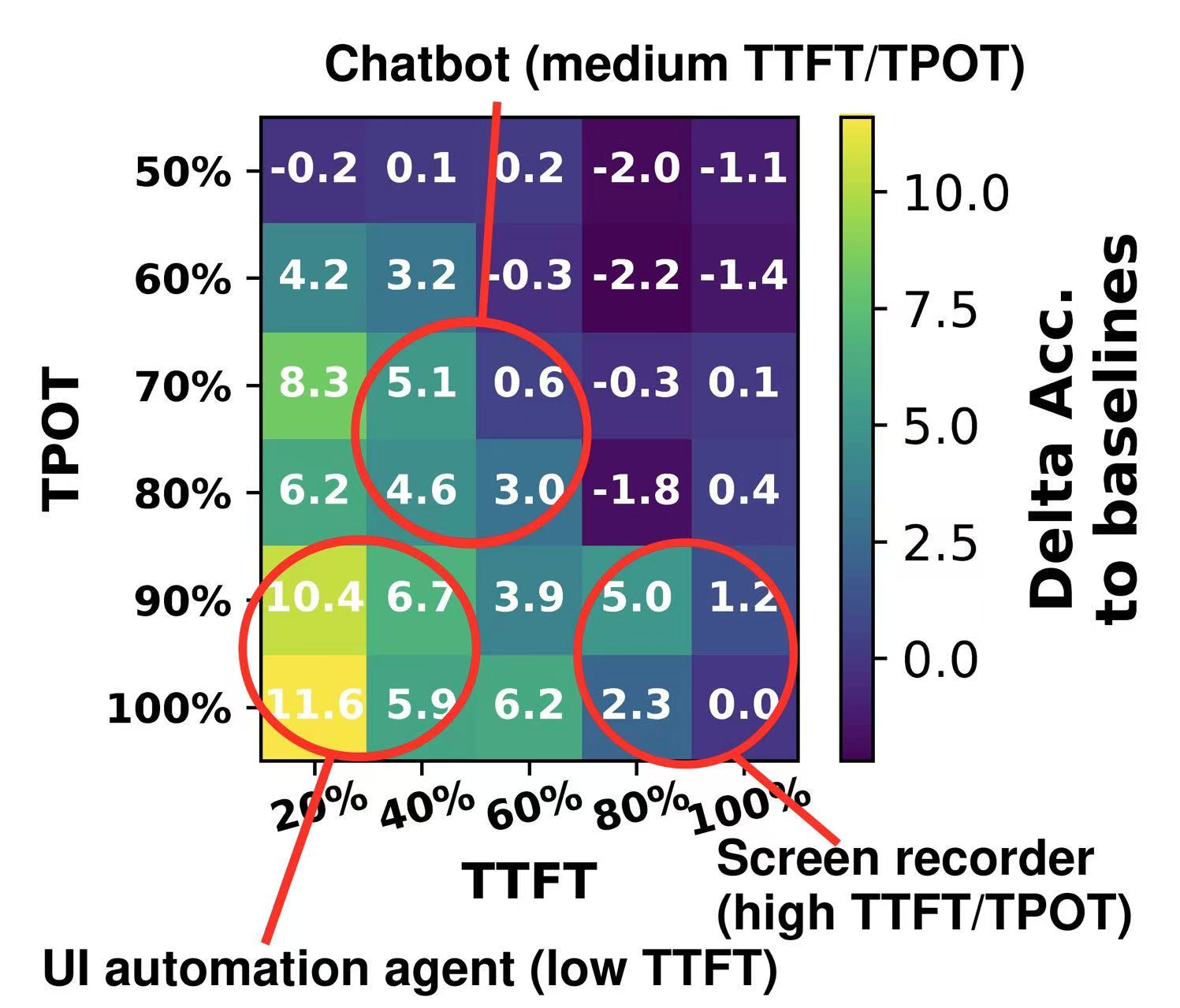}
		\subcaption{Varying real-world SLOs.}
		\label{fig:moreslo}
	\end{minipage}
	\caption{\revision{Sensitivity analysis and ablation study. (a)/(b)/(c): LLaMA-7B, on trace with $\alpha$=0; (d): LLaMA-7B, ARC\_E.}}
    \label{fig:sensitivity}
\end{figure*}

We further analysis the offline overhead in Figure~\ref{fig:offline-overhead}.
We measure the elastification of LLaMA-7B for MI14 on our A40 cloud server.
Compared to \texttt{PFS} that trains a dedicated LLM for each SLO, \sys derives elasticized LLMs from the original LLaMA-7B.
Thereby, as shown in Figure~\ref{fig:gpuhours}, the entire offline stage of \sys only consumes 68.3 GPU hours (translates to about \$100 GPU renting price~\cite{hfgpurenting}), making it affordable for most developers.
Compared to other baselines that also derive elasticized LLMs from the original one, \sys takes 21.7--68.2 more GPU hours, which is acceptable since the offline stage is performed once for all.
Note that in the above sections, we have demonstrated that \sys delivers an elastic LLM service with much higher quality than these baselines.

In Figure~\ref{fig:corpusbreakdown}, we provide a detailed breakdown of \sys's offline stage.
The model recovery ($\S$\ref{subsec: model_elastification}) and self-induced labelling ($\S$\ref{subsec:prompt_elastification}) dominate the offline stage, taking 40.4/26.3 hours.
The reason the latter requires tremendous time is due to the lower GPU utilization caused by interactions with the score-head and sub-models.

\subsection{Sensitivity analysis}
\label{subsec:sensitivity}

\noindent \textbf{Data sensitivity.}
As shown in Figure~\ref{fig:data_sensitivity}, the data scale of elastification exhibits a \textit{marginal effect}.
Regarding to model recovery, we further collect training data from \texttt{LaMini}~\cite{lamini-lm} dataset that akin to \texttt{Alpaca-cleaned}.
The final accuracy only increases 0.2\%/0.4\% when the recovery corpus is 10$\times$/20$\times$ larger.
Regarding to unit importance profiling and decision training data, we expand them from \texttt{Bookcorpus} and \texttt{MMLU}~\cite{hendryckstest2021}, respectively.
We also observe the similar marginal effect.
Since \texttt{Meetingbank} is currently the largest corpus for token importance scoring to the best of our knowledge, we leave expanding it as a further work.
In a nutshell, \sys's data scale achieves a strong and sweet spot for high-quality elastification.








\noindent \textbf{Configuration sensitivity.}
We have the following conclusions in Figure~\ref{fig:configuration_sensitivity}.
(1) 40M parameters are already a sweet configuration of TLM scale.
A larger TLM's gain on scoring and decision-making accuracy quickly gets eliminated by the overhead.
(2) 12 shared bottom layers are reasonable for TLM, since more will lead to a smaller capacity of the heads, and less will incur a higher inference/training overhead.
(3) A step size of 10\% for model and prompt elastification is fine-grained enough for serving the diversified SLOs.
Shrinking it to 5\% makes almost no difference.

\revision{
\noindent \textbf{Ablation study.}
We show the effectiveness of \sys's key designs in Figure~\ref{fig:ablation}.
We report the accuracy under a given resource SLO.
}
\yws{Specifically, we respectively remove the key designs from our full system. Each removal of key design shows a clear performance degradation.
The most significant loss comes from removing the reordering technique, showing over 10\% accuracy loss.
The reason is mainly due to that the reordering preserves the most important permutation consistent units.
The score-head shows the least degradation, yet it is still obvious (about 5\%).
In a nutshell, the design of neuron-reordering, LoRA recovery and each head of the TLM are all non-trivial.
}

\revision{
\noindent \textbf{Varying real-world SLOs.}
In Figure~\ref{fig:moreslo} we report the delta accuracy of \sys compared to its strongest baseline \texttt{PFS} (without considering switching overhead) on \texttt{ARC\_E} LLaMA-7B.
\sys is close to or much better than \texttt{PFS} on all SLOs, especially those are more common on mobile LLM tasks (e.g., chatbot, UI automation agent or screen recorder).
}


\begin{figure}[t]
	\centering
	\begin{minipage}[b]{0.21\textwidth}		
            \includegraphics[width=1\textwidth]{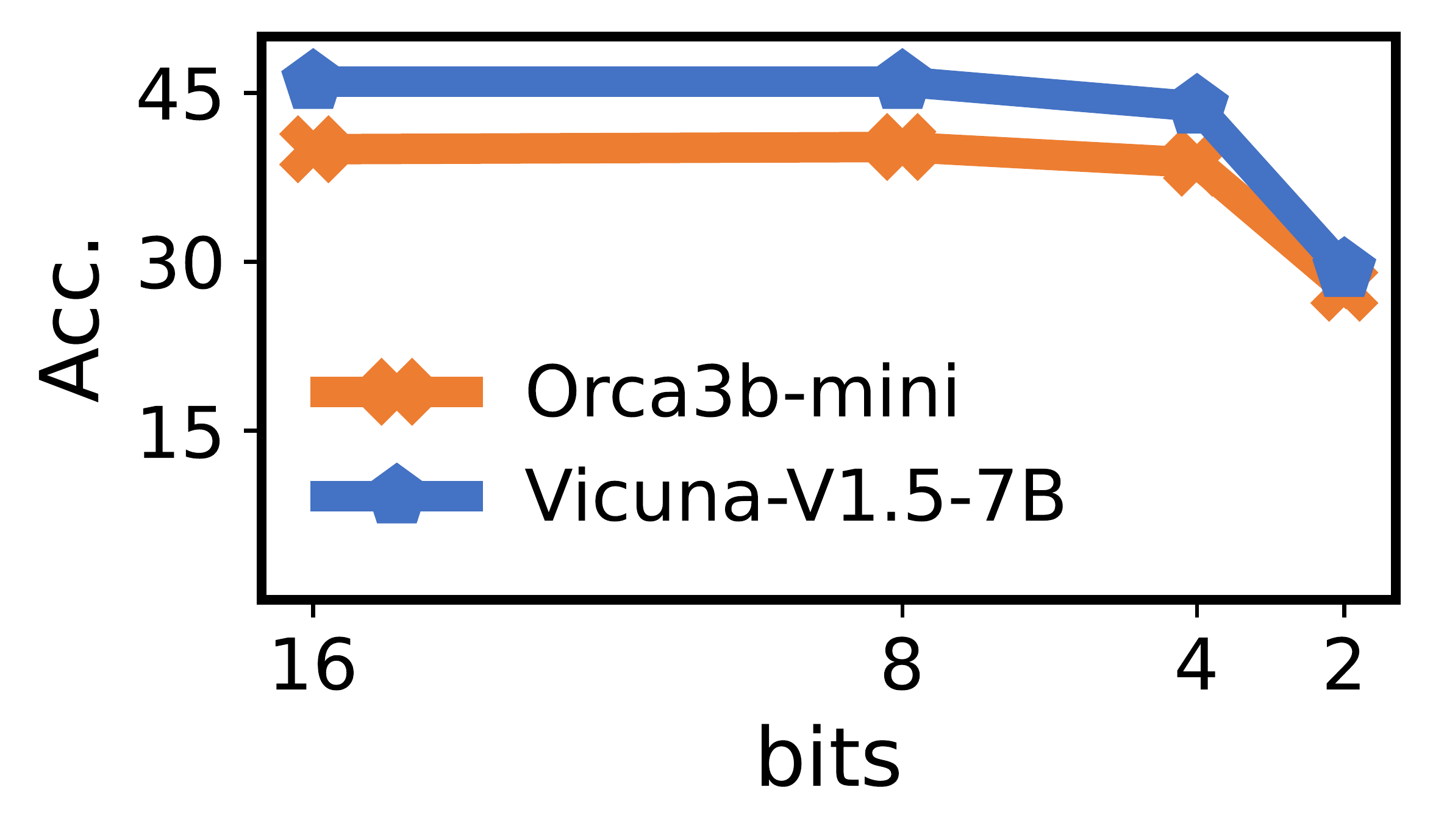}
            \vspace{-17pt}
            \subcaption{Quantization.}
            \label{fig:quant}
	\end{minipage}
	\begin{minipage}[b]{0.26\textwidth}
		\includegraphics[width=1\textwidth]{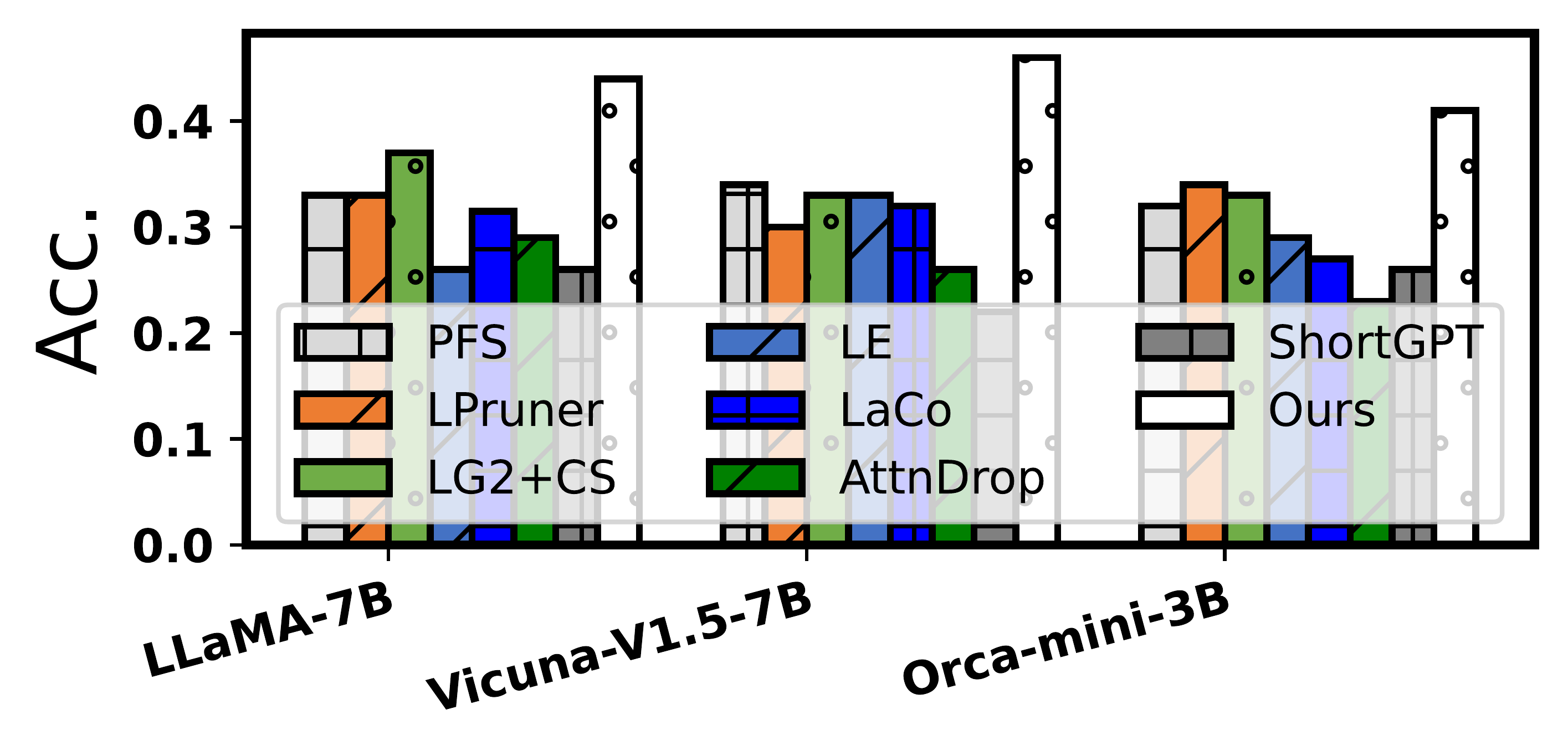}
              \vspace{-17pt}
		\subcaption{Concise prompts.}
		\label{fig:concise}
	\end{minipage}
        \vspace{-12pt}
	\caption{Model quantization and prompt conciseness.}
        \vspace{-12pt}
    \label{fig:quant_and_concise}
\end{figure}

\noindent \textbf{Quantization.}
Quantization is a widely-used technique for deploying LLMs on mobile devices.
Here we employ a linear and weights only method on the $\alpha=0$ trace in Figure~\ref{fig:quant}.
\sys can deliver LLM service with almost lossless accuracy under 8bits integers, and acceptable (3\% lower absolute accuracy) under 4bits.

\noindent \textbf{Concise prompts}.
We further emulate the scenario where the prompts have been consciously streamlined by the LLMaaS callers.
We use \texttt{LLMLingua2} to filter out about 15\% verbose tokens in the prompt.
The results (on trace with $\alpha$=0) in Figure~\ref{fig:concise} show that \sys can still significantly outperform its baselines.

\subsection{Discussion}

\begin{figure}[t]
	\centering
	\begin{minipage}[b]{0.22\textwidth}		
            \includegraphics[width=1\textwidth]{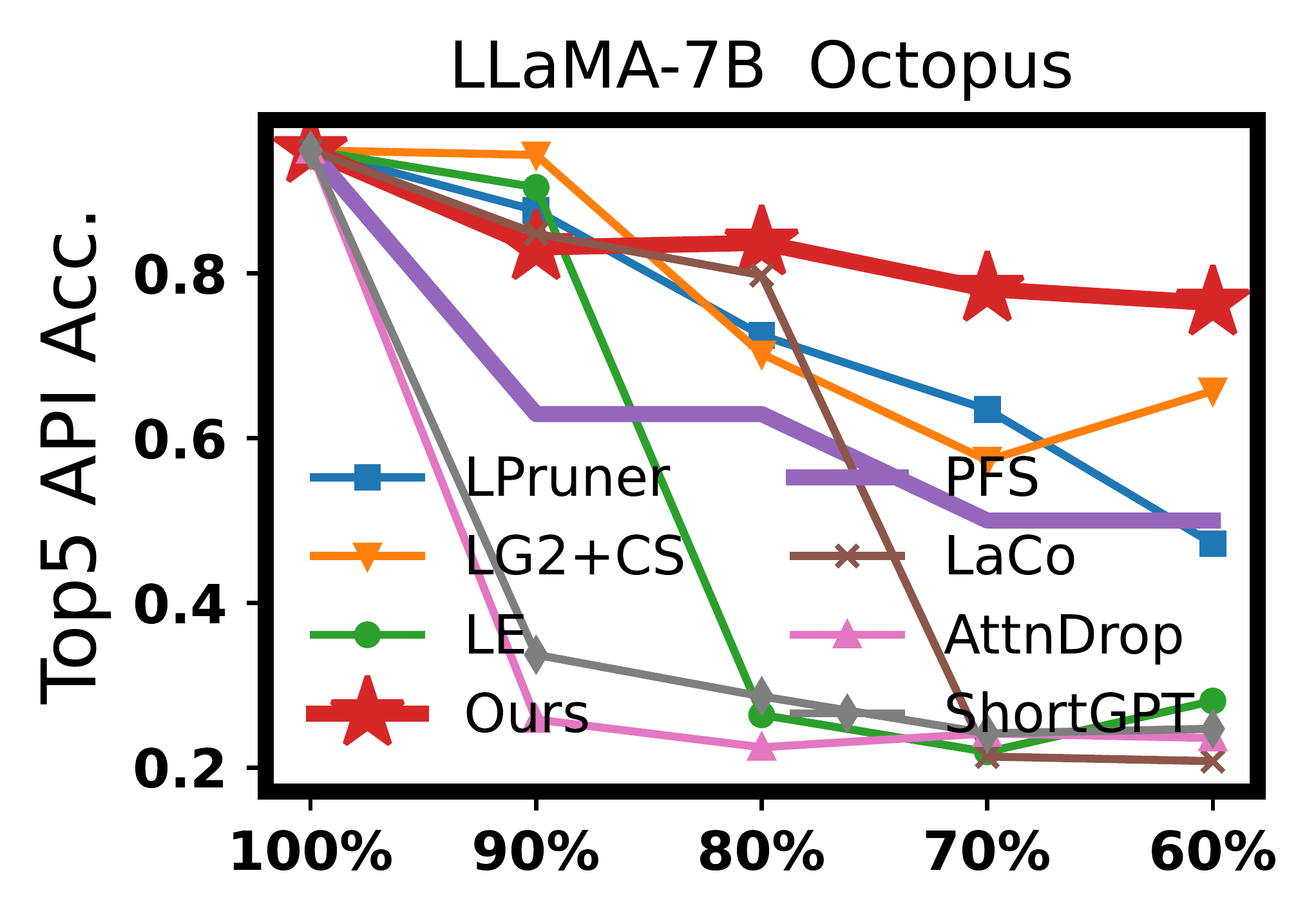}
            \vspace{-17pt}
            \subcaption{E2e inference time.}
            \label{fig:slo_generalization_time}
	\end{minipage}
	\begin{minipage}[b]{0.22\textwidth}
		\includegraphics[width=1\textwidth]{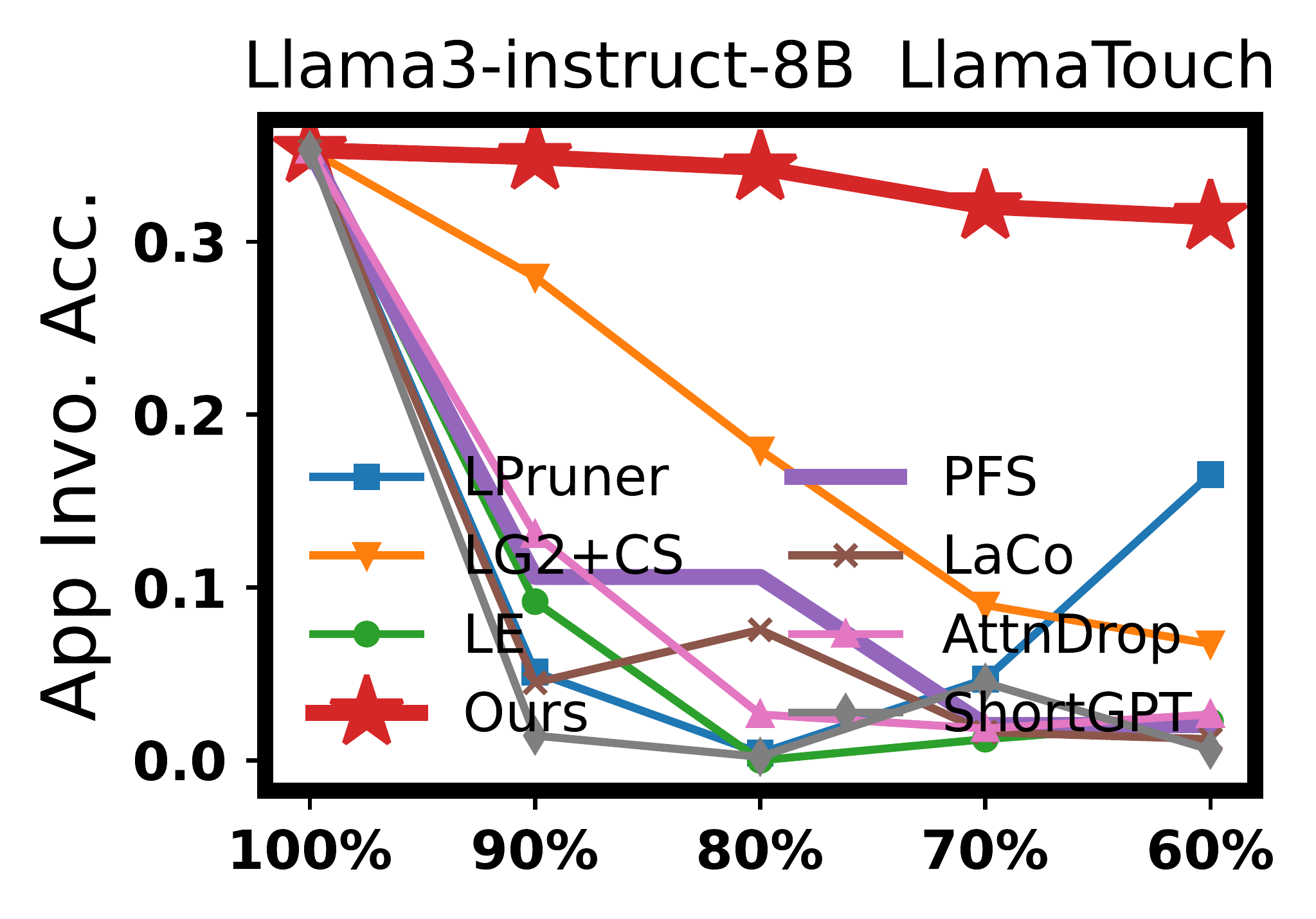}
              \vspace{-17pt}
		\subcaption{Energy consumption.}
		\label{fig:SLO_generalization_energy}
	\end{minipage}
        \vspace{-12pt}
	\caption{Generalization to other kinds of SLOs.}
         \vspace{-12pt}
    \label{fig:SLO_generalization}
\end{figure}

\textbf{Generalization to other SLOs.}
Although in this paper we identify LLM inference latency as the SLO of requests, the SLO can also been easily generalized to other metrics with our proposed method.
Here we define another two SLOs: end-to-end request inference time $SLO_{time}$ (i.e., prefill + decode stage) and inference energy consumption $SLO_{energy}$, each of which is with a step size of 10\%.
The results are obtained on Mi 14 smartphone.
As shown in Figure~\ref{fig:SLO_generalization}, \sys consistently outperforms its baselines on these new SLOs.
the rationale is that these SLOs are all a specific kind of resource requirement and can be break down into prompt- and model- level elastifications.
We believe that \sys can serve various metrics of SLOs for diverse apps' requirements.

\revision{
\noindent \textbf{Apps with competing demands.}
In regular on-device LLM service model, there is no concurrent requests since the device only has one user.
However, in some extreme scenarios, other apps might have competing demands.
\sys will make it transparent to developers/apps.
For instance, the LLM service may introduce upper-layer scheduling mechanisms like batching~\cite{Orca_osdi} to handle concurrency.
}

\begin{figure}[t]
	\centering
	\begin{minipage}[b]{0.23\textwidth}		
            \includegraphics[width=1\textwidth]{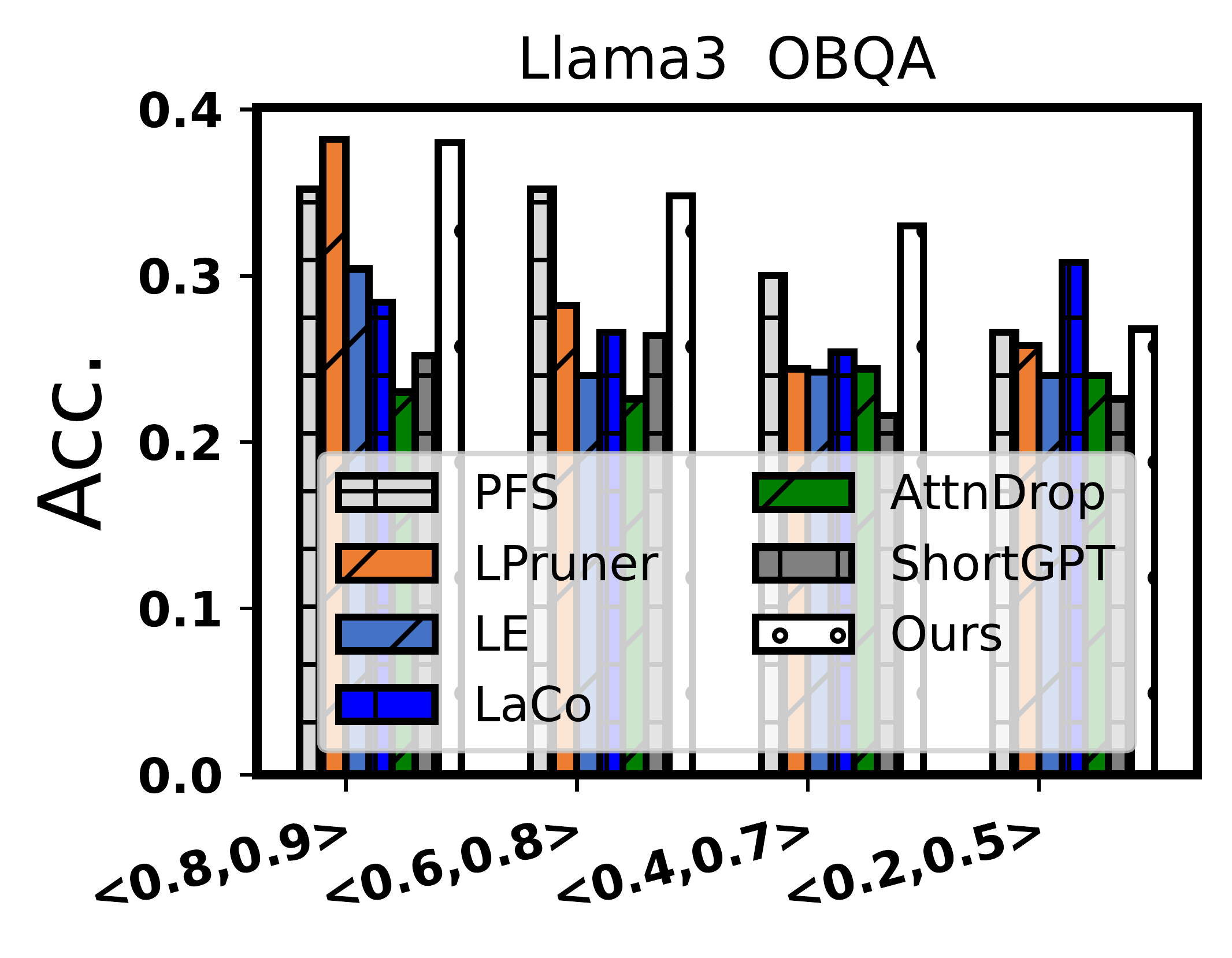}
            \vspace{-17pt}
            \subcaption{On mobile GPU.}
            \label{fig:on_gpu}
	\end{minipage}
	\begin{minipage}[b]{0.23\textwidth}
		\includegraphics[width=1\textwidth]{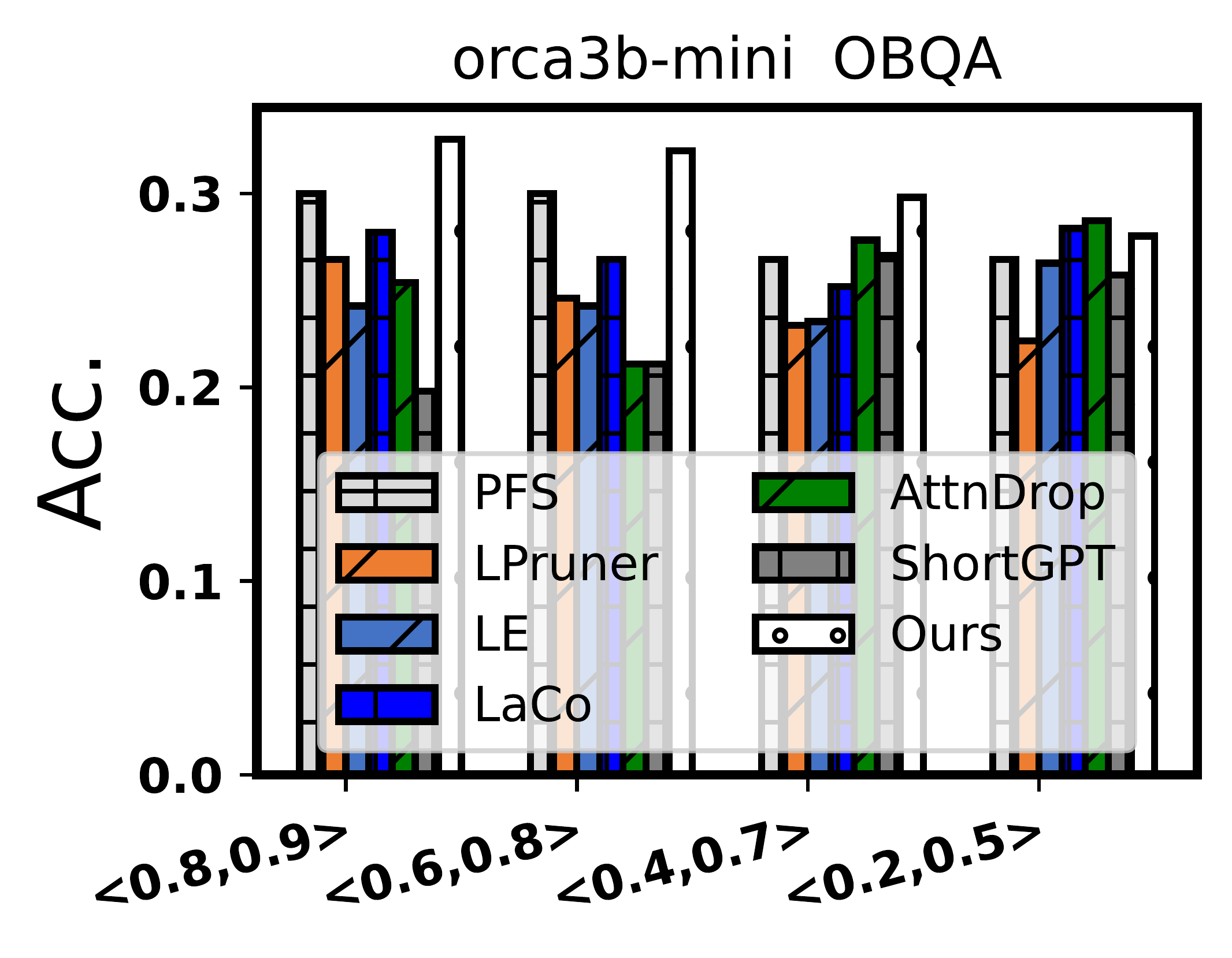}
              \vspace{-17pt}
		\subcaption{On mobile NPU.}
		\label{fig:on_npu}
	\end{minipage}
        \vspace{-12pt}
	\caption{``What if'' experiment on mobile accelerators that cannot seamlessly run \sys and its baselines currently. \texttt{LG2+CS} is omitted due to its sparse compute.}
         \vspace{-12pt}
    \label{fig:accelerators}
\end{figure}
\noindent \textbf{On mobile SoC accelerators. }
To the best of our knowledge currently (Aug., 2024) none of mobile DSAs can seamlessly run prefill and decode of \sys and all its baselines due to legacy SDK/hardware issues like dynamic shape, graph building or non-sparse kernels.
Thus, we conduct a ``what if'' experiment by mapping the theoretical compute/memory load to the profiled latency on these processors.
We use \texttt{MNN}~\cite{alibaba2020mnn} and \texttt{mllm-NPU}~\cite{xu2024empowering1000tokenssecondondevice} for GPU/NPU profiling on Mi 14, respectively.
The results are shown in Figure~\ref{fig:accelerators}.
\sys can still significantly outperform the baselines.
We believe that \sys will become more friendly and practical to practitioners with the maturity of accelerators.
\section{Related Work}

\textbf{Elastic neuron networks.}
Elastic neuron networks can change their capacity at runtime to dynamically adapt to various resource/accuracy constraints.
Early exit networks~\cite{teerapittayanon2017branchynetfastinferenceearly, tambe2021edgebertsentencelevelenergyoptimizations} only perform inference on bottom layers since they are empirically more critical.
Yet, early exit is not suitable for elastic LLM service.
On one hand, due to LLMs' autoregressive inference nature, a skipped layer's KV cache may be accessed later.
On the other hand, using layers as the granularity for trade-offs is not fine-grained enough.
Parameter sharing networks~\cite{Fang_2018, Han_2021, wen2023adaptivenetpostdeploymentneuralarchitecture} generate memory-efficient sub-models that shares parameters with each other.
For instance, NestDNN~\cite{Fang_2018} and LegoDNN~\cite{Han_2021} create sub-models of CNNs via offline re-pretraining, which is costly for foundation models.
Adaptivenet~\cite{wen2023adaptivenetpostdeploymentneuralarchitecture} employs a CNN-oriented supernet, which provides diverse accuracy-latency trade-offs yet needs extensive pre-training and memory resources.
Activation sparsity~\cite{pmlr-v202-liu23am, song2023powerinfer, seernet, convreluplusplus} elasticitizes DNNs via sparsifying weights according to the NN inputs.
An accuracy-latency trade-off can be achieved by setting the proportion of activated weights.
However, the prefill stage cannot be elastictized due to the low locality.

\noindent \textbf{Efficient on-device LLM inference.}
Tremendous work~\cite{mlc-llm, alibaba2020mnn, mllm, xue2024powerinfer2fastlargelanguage} shed light on resource-efficiently deploying LLMs on mobile devices.
For instance, MLC-LLM~\cite{mlc-llm} is an NN-compiler with operator- and kernel- level optimizations for mobile devices.
MNN~\cite{alibaba2020mnn} and mllm~\cite{mllm} are on-device inference libraries for LLMs.
PowerinferV2~\cite{xue2024powerinfer2fastlargelanguage} addresses the memory issue of mobile devices by introducing swapping and activation sparsity to LLM inference.
Targeting at elastic LLM service, \sys is orthogonal to these work. 

\noindent \textbf{Foundation models as a service.}
As a general task solver, or so-called ``AGI'', a single foundation model is deployed as a service to process heterogeneous tasks~\cite{yin2024llmservicemobiledevices, yuanrethinking, ge2023llmosagentsapps}.
For instance, AIOS~\cite{ge2023llmosagentsapps} is a system where a single LLM runs as the ``brain'' to serve all apps on the device.
\sys makes it more practical on resource-constrained mobile devices.

\noindent \textbf{Model collaboration.}
\sys employs a dual-head tiny language model to elasticize the prompt of the LLM.
Using a small model to collaborate with the big model is common in ML systems.
For instance, speculative decoding~\cite{Miao_2024, xu2023llmcadfastscalableondevice} accelerates the decode stage with a draft SLM.
However, the prefill stage cannot be accelerated since it is typically compute-bound.
LLMlingua~\cite{jiang2023llmlinguacompressingpromptsaccelerated} uses an SLM to refine the LLM prompt, which has several untackled issues for elastic LLM service as discussed in this paper.
\section{Conclusion}

This work has proposed \sys, an elastic on-device LLM service that serves apps with diverse SLOs.
\sys incorporates two novel designs, i.e., one-shot reordering of permutation consistent units and dual-head tiny language model to fully unleash the potential of model- and prompt- elastification.
\sys significantly outperforms competitive baselines by up to 14.83\% and 10.45\% on average in accuracy.

\section{Acknowledgment}
This work was supported by the National Natural Science Foundation of China under Grant 62325201.
Mengwei Xu was supported in part by the Shenzhen Science and Technology Program with Grant No. JCYJ20241202124021028.

\bibliographystyle{ACM-Reference-Format}
\bibliography{ref}


\begin{thebibliography}{93}


\ifx \showCODEN    \undefined \def \showCODEN     #1{\unskip}     \fi
\ifx \showDOI      \undefined \def \showDOI       #1{#1}\fi
\ifx \showISBNx    \undefined \def \showISBNx     #1{\unskip}     \fi
\ifx \showISBNxiii \undefined \def \showISBNxiii  #1{\unskip}     \fi
\ifx \showISSN     \undefined \def \showISSN      #1{\unskip}     \fi
\ifx \showLCCN     \undefined \def \showLCCN      #1{\unskip}     \fi
\ifx \shownote     \undefined \def \shownote      #1{#1}          \fi
\ifx \showarticletitle \undefined \def \showarticletitle #1{#1}   \fi
\ifx \showURL      \undefined \def \showURL       {\relax}        \fi
\providecommand\bibfield[2]{#2}
\providecommand\bibinfo[2]{#2}
\providecommand\natexlab[1]{#1}
\providecommand\showeprint[2][]{arXiv:#2}

\bibitem[AIC(2024)]%
        {AICore}
 \bibinfo{year}{2024}\natexlab{}.
\newblock \bibinfo{title}{AICore}.
\newblock \bibinfo{howpublished}{\url{https://developer.android.com/ml/aicore}}.
\newblock


\bibitem[alp(2024)]%
        {alpacacleaned}
 \bibinfo{year}{2024}\natexlab{}.
\newblock \bibinfo{title}{alpaca cleaned}.
\newblock \bibinfo{howpublished}{\url{https://huggingface.co/datasets/yahma/alpaca-cleaned}}.
\newblock


\bibitem[App(2024)]%
        {Apple_Intelligence}
 \bibinfo{year}{2024}\natexlab{}.
\newblock \bibinfo{title}{Apple Intelligence}.
\newblock \bibinfo{howpublished}{\url{https://www.apple.com/apple-intelligence/}}.
\newblock


\bibitem[arm(2024)]%
        {armneon}
 \bibinfo{year}{2024}\natexlab{}.
\newblock \bibinfo{title}{ARM NEON}.
\newblock \bibinfo{howpublished}{\url{https://developer.arm.com/Architectures/Neon}}.
\newblock


\bibitem[gbo(2024)]%
        {gboard}
 \bibinfo{year}{2024}\natexlab{}.
\newblock \bibinfo{title}{Gboard Smart Reply.}
\newblock \bibinfo{howpublished}{\url{https://developers.google.com/ml-kit/language/smart-reply}}.
\newblock


\bibitem[sir(2024)]%
        {siri}
 \bibinfo{year}{2024}\natexlab{}.
\newblock \bibinfo{title}{Hey Siri: An On-device DNN-powered Voice Trigger for Apple’s Personal Assistant}.
\newblock \bibinfo{howpublished}{\url{https://machinelearning.apple.com/research/hey-siri}}.
\newblock


\bibitem[hfg(2024)]%
        {hfgpurenting}
 \bibinfo{year}{2024}\natexlab{}.
\newblock \bibinfo{title}{Huggingface GPU pricing}.
\newblock \bibinfo{howpublished}{\url{https://huggingface.co/pricing}}.
\newblock


\bibitem[lla(2024)]%
        {llama.cpp}
 \bibinfo{year}{2024}\natexlab{}.
\newblock \bibinfo{title}{Llama.cpp.}
\newblock \bibinfo{howpublished}{\url{https://github.com/ggerganov/llama.cpp}}.
\newblock


\bibitem[lay(2024)]%
        {layerelastification}
 \bibinfo{year}{2024}\natexlab{}.
\newblock \bibinfo{title}{LLM layer pruning}.
\newblock \bibinfo{howpublished}{\url{https://github.com/horseee/LLM-Pruner/blob/cbe488944ed772f342e99d3d0efbab9df6520c21/hf_prune.py\#L219}}.
\newblock


\bibitem[Mee(2024)]%
        {MeetingBank}
 \bibinfo{year}{2024}\natexlab{}.
\newblock \bibinfo{title}{MeetingBank compressed}.
\newblock \bibinfo{howpublished}{\url{https://huggingface.co/datasets/microsoft/MeetingBank-LLMCompressed}}.
\newblock


\bibitem[mll(2024)]%
        {mllm}
 \bibinfo{year}{2024}\natexlab{}.
\newblock \bibinfo{title}{mllm}.
\newblock \bibinfo{howpublished}{\url{https://github.com/UbiquitousLearning/mllm}}.
\newblock


\bibitem[red(2024a)]%
        {redmik60champion}
 \bibinfo{year}{2024}\natexlab{a}.
\newblock \bibinfo{title}{redmi-k60-champion-edition}.
\newblock \bibinfo{howpublished}{\url{https://www.giztop.com/redmi-k60-champion-edition.html}}.
\newblock


\bibitem[red(2024b)]%
        {redmik70pro}
 \bibinfo{year}{2024}\natexlab{b}.
\newblock \bibinfo{title}{Redmi K70 Pro}.
\newblock \bibinfo{howpublished}{\url{https://www.mi.com/redmi-k70-pro}}.
\newblock


\bibitem[rew(2024)]%
        {rewind}
 \bibinfo{year}{2024}\natexlab{}.
\newblock \bibinfo{title}{rewind}.
\newblock \bibinfo{howpublished}{\url{https://www.rewind.ai/}}.
\newblock


\bibitem[xia(2024)]%
        {xiaoai}
 \bibinfo{year}{2024}\natexlab{}.
\newblock \bibinfo{title}{XiaoAi smart assistant}.
\newblock \bibinfo{howpublished}{\url{https://xiaoai.mi.com/}}.
\newblock


\bibitem[mi1(2024)]%
        {mi14}
 \bibinfo{year}{2024}\natexlab{}.
\newblock \bibinfo{title}{xiaomi-14}.
\newblock \bibinfo{howpublished}{\url{https://www.mi.com/global/product/xiaomi-14/}}.
\newblock


\bibitem[Achiam et~al\mbox{.}(2023)]%
        {openai2023gpt4}
\bibfield{author}{\bibinfo{person}{OpenAI:~Josh Achiam}, \bibinfo{person}{Steven Adler}, \bibinfo{person}{Sandhini Agarwal}, \bibinfo{person}{Lama Ahmad}, \bibinfo{person}{Ilge Akkaya}, {et~al\mbox{.}}} \bibinfo{year}{2023}\natexlab{}.
\newblock \bibinfo{title}{GPT-4 Technical Report}.
\newblock
\newblock
\showeprint[arxiv]{2303.08774}~[cs.CL]


\bibitem[Agarap(2019)]%
        {agarap2019deeplearningusingrectified}
\bibfield{author}{\bibinfo{person}{Abien~Fred Agarap}.} \bibinfo{year}{2019}\natexlab{}.
\newblock \bibinfo{title}{Deep Learning using Rectified Linear Units (ReLU)}.
\newblock
\newblock
\showeprint[arxiv]{1803.08375}~[cs.NE]
\urldef\tempurl%
\url{https://arxiv.org/abs/1803.08375}
\showURL{%
\tempurl}


\bibitem[Bai et~al\mbox{.}(2023)]%
        {qwen}
\bibfield{author}{\bibinfo{person}{Jinze Bai}, \bibinfo{person}{Shuai Bai}, \bibinfo{person}{Yunfei Chu}, \bibinfo{person}{Zeyu Cui}, \bibinfo{person}{Kai Dang}, \bibinfo{person}{Xiaodong Deng}, \bibinfo{person}{Yang Fan}, {et~al\mbox{.}}} \bibinfo{year}{2023}\natexlab{}.
\newblock \showarticletitle{Qwen Technical Report}.
\newblock \bibinfo{journal}{\emph{arXiv preprint arXiv:2309.16609}} (\bibinfo{year}{2023}).
\newblock


\bibitem[Cai et~al\mbox{.}(2020)]%
        {cai2020onceforalltrainnetworkspecialize}
\bibfield{author}{\bibinfo{person}{Han Cai}, \bibinfo{person}{Chuang Gan}, \bibinfo{person}{Tianzhe Wang}, \bibinfo{person}{Zhekai Zhang}, {and} \bibinfo{person}{Song Han}.} \bibinfo{year}{2020}\natexlab{}.
\newblock \bibinfo{title}{Once-for-All: Train One Network and Specialize it for Efficient Deployment}.
\newblock
\newblock
\showeprint[arxiv]{1908.09791}~[cs.LG]
\urldef\tempurl%
\url{https://arxiv.org/abs/1908.09791}
\showURL{%
\tempurl}


\bibitem[Cao et~al\mbox{.}(2019)]%
        {seernet}
\bibfield{author}{\bibinfo{person}{Shijie Cao}, \bibinfo{person}{Lingxiao Ma}, \bibinfo{person}{Wencong Xiao}, \bibinfo{person}{Chen Zhang}, \bibinfo{person}{Yunxin Liu}, \bibinfo{person}{Lintao Zhang}, \bibinfo{person}{Lanshun Nie}, {and} \bibinfo{person}{Zhi Yang}.} \bibinfo{year}{2019}\natexlab{}.
\newblock \showarticletitle{SeerNet: Predicting Convolutional Neural Network Feature-Map Sparsity Through Low-Bit Quantization}. In \bibinfo{booktitle}{\emph{2019 IEEE/CVF Conference on Computer Vision and Pattern Recognition (CVPR)}}. \bibinfo{pages}{11208--11217}.
\newblock
\urldef\tempurl%
\url{https://doi.org/10.1109/CVPR.2019.01147}
\showDOI{\tempurl}


\bibitem[Chen and Li(2024)]%
        {chen2024octopus}
\bibfield{author}{\bibinfo{person}{Wei Chen} {and} \bibinfo{person}{Zhiyuan Li}.} \bibinfo{year}{2024}\natexlab{}.
\newblock \bibinfo{title}{Octopus v2: On-device language model for super agent}.
\newblock
\newblock
\showeprint[arxiv]{2404.01744}~[cs.CL]


\bibitem[Clark et~al\mbox{.}(2018)]%
        {allenai:arc}
\bibfield{author}{\bibinfo{person}{Peter Clark}, \bibinfo{person}{Isaac Cowhey}, \bibinfo{person}{Oren Etzioni}, \bibinfo{person}{Tushar Khot}, \bibinfo{person}{Ashish Sabharwal}, \bibinfo{person}{Carissa Schoenick}, {and} \bibinfo{person}{Oyvind Tafjord}.} \bibinfo{year}{2018}\natexlab{}.
\newblock \showarticletitle{Think you have Solved Question Answering? Try ARC, the AI2 Reasoning Challenge}.
\newblock \bibinfo{journal}{\emph{arXiv:1803.05457v1}} (\bibinfo{year}{2018}).
\newblock


\bibitem[Devlin et~al\mbox{.}(2019)]%
        {devlin2019bertpretrainingdeepbidirectional}
\bibfield{author}{\bibinfo{person}{Jacob Devlin}, \bibinfo{person}{Ming-Wei Chang}, \bibinfo{person}{Kenton Lee}, {and} \bibinfo{person}{Kristina Toutanova}.} \bibinfo{year}{2019}\natexlab{}.
\newblock \bibinfo{title}{BERT: Pre-training of Deep Bidirectional Transformers for Language Understanding}.
\newblock
\newblock
\showeprint[arxiv]{1810.04805}~[cs.CL]
\urldef\tempurl%
\url{https://arxiv.org/abs/1810.04805}
\showURL{%
\tempurl}


\bibitem[Dong et~al\mbox{.}(2024)]%
        {dong2024surveyincontextlearning}
\bibfield{author}{\bibinfo{person}{Qingxiu Dong}, \bibinfo{person}{Lei Li}, \bibinfo{person}{Damai Dai}, \bibinfo{person}{Ce Zheng}, \bibinfo{person}{Jingyuan Ma}, \bibinfo{person}{Rui Li}, \bibinfo{person}{Heming Xia}, \bibinfo{person}{Jingjing Xu}, \bibinfo{person}{Zhiyong Wu}, \bibinfo{person}{Baobao Chang}, \bibinfo{person}{Xu Sun}, \bibinfo{person}{Lei Li}, {and} \bibinfo{person}{Zhifang Sui}.} \bibinfo{year}{2024}\natexlab{}.
\newblock \bibinfo{title}{A Survey on In-context Learning}.
\newblock
\newblock
\showeprint[arxiv]{2301.00234}~[cs.CL]
\urldef\tempurl%
\url{https://arxiv.org/abs/2301.00234}
\showURL{%
\tempurl}


\bibitem[Dubey et~al\mbox{.}(2024)]%
        {dubey2024llama3herdmodels}
\bibfield{author}{\bibinfo{person}{Abhimanyu Dubey}, \bibinfo{person}{Abhinav Jauhri}, {et~al\mbox{.}}} \bibinfo{year}{2024}\natexlab{}.
\newblock \bibinfo{title}{The Llama 3 Herd of Models}.
\newblock
\newblock
\showeprint[arxiv]{2407.21783}~[cs.AI]
\urldef\tempurl%
\url{https://arxiv.org/abs/2407.21783}
\showURL{%
\tempurl}


\bibitem[Fang et~al\mbox{.}(2018)]%
        {Fang_2018}
\bibfield{author}{\bibinfo{person}{Biyi Fang}, \bibinfo{person}{Xiao Zeng}, {and} \bibinfo{person}{Mi Zhang}.} \bibinfo{year}{2018}\natexlab{}.
\newblock \showarticletitle{NestDNN: Resource-Aware Multi-Tenant On-Device Deep Learning for Continuous Mobile Vision}. In \bibinfo{booktitle}{\emph{Proceedings of the 24th Annual International Conference on Mobile Computing and Networking}} \emph{(\bibinfo{series}{MobiCom ’18})}. \bibinfo{publisher}{ACM}.
\newblock
\urldef\tempurl%
\url{https://doi.org/10.1145/3241539.3241559}
\showDOI{\tempurl}


\bibitem[Fang et~al\mbox{.}(2023)]%
        {fang2023depgraph}
\bibfield{author}{\bibinfo{person}{Gongfan Fang}, \bibinfo{person}{Xinyin Ma}, \bibinfo{person}{Mingli Song}, \bibinfo{person}{Michael~Bi Mi}, {and} \bibinfo{person}{Xinchao Wang}.} \bibinfo{year}{2023}\natexlab{}.
\newblock \showarticletitle{Depgraph: Towards any structural pruning}. In \bibinfo{booktitle}{\emph{Proceedings of the IEEE/CVF Conference on Computer Vision and Pattern Recognition}}. \bibinfo{pages}{16091--16101}.
\newblock


\bibitem[Ge et~al\mbox{.}(2023)]%
        {ge2023llmosagentsapps}
\bibfield{author}{\bibinfo{person}{Yingqiang Ge}, \bibinfo{person}{Yujie Ren}, \bibinfo{person}{Wenyue Hua}, \bibinfo{person}{Shuyuan Xu}, \bibinfo{person}{Juntao Tan}, {and} \bibinfo{person}{Yongfeng Zhang}.} \bibinfo{year}{2023}\natexlab{}.
\newblock \bibinfo{title}{LLM as OS, Agents as Apps: Envisioning AIOS, Agents and the AIOS-Agent Ecosystem}.
\newblock
\newblock
\showeprint[arxiv]{2312.03815}~[cs.OS]
\urldef\tempurl%
\url{https://arxiv.org/abs/2312.03815}
\showURL{%
\tempurl}


\bibitem[Han et~al\mbox{.}(2021)]%
        {Han_2021}
\bibfield{author}{\bibinfo{person}{Rui Han}, \bibinfo{person}{Qinglong Zhang}, \bibinfo{person}{Chi~Harold Liu}, \bibinfo{person}{Guoren Wang}, \bibinfo{person}{Jian Tang}, {and} \bibinfo{person}{Lydia~Y. Chen}.} \bibinfo{year}{2021}\natexlab{}.
\newblock \showarticletitle{LegoDNN: block-grained scaling of deep neural networks for mobile vision}. In \bibinfo{booktitle}{\emph{Proceedings of the 27th Annual International Conference on Mobile Computing and Networking}} \emph{(\bibinfo{series}{ACM MobiCom ’21})}. \bibinfo{publisher}{ACM}.
\newblock
\urldef\tempurl%
\url{https://doi.org/10.1145/3447993.3483249}
\showDOI{\tempurl}


\bibitem[Han et~al\mbox{.}(2015)]%
        {han2015learningweightsconnectionsefficient}
\bibfield{author}{\bibinfo{person}{Song Han}, \bibinfo{person}{Jeff Pool}, \bibinfo{person}{John Tran}, {and} \bibinfo{person}{William~J. Dally}.} \bibinfo{year}{2015}\natexlab{}.
\newblock \bibinfo{title}{Learning both Weights and Connections for Efficient Neural Networks}.
\newblock
\newblock
\showeprint[arxiv]{1506.02626}~[cs.NE]
\urldef\tempurl%
\url{https://arxiv.org/abs/1506.02626}
\showURL{%
\tempurl}


\bibitem[He et~al\mbox{.}(2024)]%
        {he2024matterstransformersattentionneeded}
\bibfield{author}{\bibinfo{person}{Shwai He}, \bibinfo{person}{Guoheng Sun}, \bibinfo{person}{Zheyu Shen}, {and} \bibinfo{person}{Ang Li}.} \bibinfo{year}{2024}\natexlab{}.
\newblock \bibinfo{title}{What Matters in Transformers? Not All Attention is Needed}.
\newblock
\newblock
\showeprint[arxiv]{2406.15786}~[cs.LG]
\urldef\tempurl%
\url{https://arxiv.org/abs/2406.15786}
\showURL{%
\tempurl}


\bibitem[Hendrycks et~al\mbox{.}(2021)]%
        {hendryckstest2021}
\bibfield{author}{\bibinfo{person}{Dan Hendrycks}, \bibinfo{person}{Collin Burns}, \bibinfo{person}{Steven Basart}, \bibinfo{person}{Andy Zou}, \bibinfo{person}{Mantas Mazeika}, \bibinfo{person}{Dawn Song}, {and} \bibinfo{person}{Jacob Steinhardt}.} \bibinfo{year}{2021}\natexlab{}.
\newblock \showarticletitle{Measuring Massive Multitask Language Understanding}.
\newblock \bibinfo{journal}{\emph{Proceedings of the International Conference on Learning Representations (ICLR)}} (\bibinfo{year}{2021}).
\newblock


\bibitem[Hu et~al\mbox{.}(2021)]%
        {hu2021loralowrankadaptationlarge}
\bibfield{author}{\bibinfo{person}{Edward~J. Hu}, \bibinfo{person}{Yelong Shen}, \bibinfo{person}{Phillip Wallis}, \bibinfo{person}{Zeyuan Allen-Zhu}, \bibinfo{person}{Yuanzhi Li}, \bibinfo{person}{Shean Wang}, \bibinfo{person}{Lu Wang}, {and} \bibinfo{person}{Weizhu Chen}.} \bibinfo{year}{2021}\natexlab{}.
\newblock \bibinfo{title}{LoRA: Low-Rank Adaptation of Large Language Models}.
\newblock
\newblock
\showeprint[arxiv]{2106.09685}~[cs.CL]
\urldef\tempurl%
\url{https://arxiv.org/abs/2106.09685}
\showURL{%
\tempurl}


\bibitem[Huang et~al\mbox{.}(2023)]%
        {huang2023elastictrainer}
\bibfield{author}{\bibinfo{person}{Kai Huang}, \bibinfo{person}{Boyuan Yang}, {and} \bibinfo{person}{Wei Gao}.} \bibinfo{year}{2023}\natexlab{}.
\newblock \showarticletitle{ElasticTrainer: Speeding Up On-Device Training with Runtime Elastic Tensor Selection}. In \bibinfo{booktitle}{\emph{Proceedings of the 21st Annual International Conference on Mobile Systems, Applications and Services}}. \bibinfo{pages}{56--69}.
\newblock


\bibitem[Janowsky(1989)]%
        {PhysRevA.39.6600}
\bibfield{author}{\bibinfo{person}{Steven~A. Janowsky}.} \bibinfo{year}{1989}\natexlab{}.
\newblock \showarticletitle{Pruning versus clipping in neural networks}.
\newblock \bibinfo{journal}{\emph{Phys. Rev. A}}  \bibinfo{volume}{39} (\bibinfo{date}{Jun} \bibinfo{year}{1989}), \bibinfo{pages}{6600--6603}.
\newblock
Issue 12.
\urldef\tempurl%
\url{https://doi.org/10.1103/PhysRevA.39.6600}
\showDOI{\tempurl}


\bibitem[Jiang et~al\mbox{.}(2023)]%
        {jiang2023llmlinguacompressingpromptsaccelerated}
\bibfield{author}{\bibinfo{person}{Huiqiang Jiang}, \bibinfo{person}{Qianhui Wu}, \bibinfo{person}{Chin-Yew Lin}, \bibinfo{person}{Yuqing Yang}, {and} \bibinfo{person}{Lili Qiu}.} \bibinfo{year}{2023}\natexlab{}.
\newblock \bibinfo{title}{LLMLingua: Compressing Prompts for Accelerated Inference of Large Language Models}.
\newblock
\newblock
\showeprint[arxiv]{2310.05736}~[cs.CL]
\urldef\tempurl%
\url{https://arxiv.org/abs/2310.05736}
\showURL{%
\tempurl}


\bibitem[Jiang et~al\mbox{.}(2024)]%
        {jiang-etal-2024-longllmlingua}
\bibfield{author}{\bibinfo{person}{Huiqiang Jiang}, \bibinfo{person}{Qianhui Wu}, \bibinfo{person}{Xufang Luo}, \bibinfo{person}{Dongsheng Li}, \bibinfo{person}{Chin-Yew Lin}, \bibinfo{person}{Yuqing Yang}, {and} \bibinfo{person}{Lili Qiu}.} \bibinfo{year}{2024}\natexlab{}.
\newblock \showarticletitle{{L}ong{LLML}ingua: Accelerating and Enhancing {LLM}s in Long Context Scenarios via Prompt Compression}. In \bibinfo{booktitle}{\emph{Proceedings of the 62nd Annual Meeting of the Association for Computational Linguistics (Volume 1: Long Papers)}}, \bibfield{editor}{\bibinfo{person}{Lun-Wei Ku}, \bibinfo{person}{Andre Martins}, {and} \bibinfo{person}{Vivek Srikumar}} (Eds.). \bibinfo{publisher}{Association for Computational Linguistics}, \bibinfo{address}{Bangkok, Thailand}, \bibinfo{pages}{1658--1677}.
\newblock
\urldef\tempurl%
\url{https://aclanthology.org/2024.acl-long.91}
\showURL{%
\tempurl}


\bibitem[Jiang et~al\mbox{.}(2020)]%
        {alibaba2020mnn}
\bibfield{author}{\bibinfo{person}{Xiaotang Jiang}, \bibinfo{person}{Huan Wang}, \bibinfo{person}{Yiliu Chen}, \bibinfo{person}{Ziqi Wu}, \bibinfo{person}{Lichuan Wang}, \bibinfo{person}{Bin Zou}, \bibinfo{person}{Yafeng Yang}, \bibinfo{person}{Zongyang Cui}, \bibinfo{person}{Yu Cai}, \bibinfo{person}{Tianhang Yu}, \bibinfo{person}{Chengfei Lv}, {and} \bibinfo{person}{Zhihua Wu}.} \bibinfo{year}{2020}\natexlab{}.
\newblock \showarticletitle{MNN: A Universal and Efficient Inference Engine}. In \bibinfo{booktitle}{\emph{MLSys}}.
\newblock


\bibitem[Johannes~Welbl(2017)]%
        {SciQ}
\bibfield{author}{\bibinfo{person}{Matt~Gardner Johannes~Welbl, Nelson F.~Liu}.} \bibinfo{year}{2017}\natexlab{}.
\newblock \showarticletitle{Crowdsourcing Multiple Choice Science Questions}.
\newblock \bibinfo{journal}{\emph{arXiv:1707.06209v1}}.
\newblock


\bibitem[Kong et~al\mbox{.}(2023)]%
        {convreluplusplus}
\bibfield{author}{\bibinfo{person}{Rui Kong}, \bibinfo{person}{Yuanchun Li}, \bibinfo{person}{Yizhen Yuan}, {and} \bibinfo{person}{Linghe Kong}.} \bibinfo{year}{2023}\natexlab{}.
\newblock \showarticletitle{ConvReLU++: Reference-based Lossless Acceleration of Conv-ReLU Operations on Mobile CPU} \emph{(\bibinfo{series}{MobiSys '23})}. \bibinfo{publisher}{Association for Computing Machinery}, \bibinfo{address}{New York, NY, USA}, \bibinfo{pages}{503–515}.
\newblock
\showISBNx{9798400701108}
\urldef\tempurl%
\url{https://doi.org/10.1145/3581791.3596831}
\showDOI{\tempurl}


\bibitem[Kwon et~al\mbox{.}(2023)]%
        {kwon2023efficientmemorymanagementlarge}
\bibfield{author}{\bibinfo{person}{Woosuk Kwon}, \bibinfo{person}{Zhuohan Li}, \bibinfo{person}{Siyuan Zhuang}, \bibinfo{person}{Ying Sheng}, \bibinfo{person}{Lianmin Zheng}, \bibinfo{person}{Cody~Hao Yu}, \bibinfo{person}{Joseph~E. Gonzalez}, \bibinfo{person}{Hao Zhang}, {and} \bibinfo{person}{Ion Stoica}.} \bibinfo{year}{2023}\natexlab{}.
\newblock \bibinfo{title}{Efficient Memory Management for Large Language Model Serving with PagedAttention}.
\newblock
\newblock
\showeprint[arxiv]{2309.06180}~[cs.LG]
\urldef\tempurl%
\url{https://arxiv.org/abs/2309.06180}
\showURL{%
\tempurl}


\bibitem[Li et~al\mbox{.}(2021)]%
        {hermes}
\bibfield{author}{\bibinfo{person}{Ang Li}, \bibinfo{person}{Jingwei Sun}, \bibinfo{person}{Pengcheng Li}, \bibinfo{person}{Yu Pu}, \bibinfo{person}{Hai Li}, {and} \bibinfo{person}{Yiran Chen}.} \bibinfo{year}{2021}\natexlab{}.
\newblock \showarticletitle{Hermes: an efficient federated learning framework for heterogeneous mobile clients}. In \bibinfo{booktitle}{\emph{Proceedings of the 27th Annual International Conference on Mobile Computing and Networking}} (New Orleans, Louisiana) \emph{(\bibinfo{series}{MobiCom '21})}. \bibinfo{publisher}{Association for Computing Machinery}, \bibinfo{address}{New York, NY, USA}, \bibinfo{pages}{420–437}.
\newblock
\showISBNx{9781450383424}
\urldef\tempurl%
\url{https://doi.org/10.1145/3447993.3483278}
\showDOI{\tempurl}


\bibitem[Li et~al\mbox{.}(2017)]%
        {li2017pruningfiltersefficientconvnets}
\bibfield{author}{\bibinfo{person}{Hao Li}, \bibinfo{person}{Asim Kadav}, \bibinfo{person}{Igor Durdanovic}, \bibinfo{person}{Hanan Samet}, {and} \bibinfo{person}{Hans~Peter Graf}.} \bibinfo{year}{2017}\natexlab{}.
\newblock \bibinfo{title}{Pruning Filters for Efficient ConvNets}.
\newblock
\newblock
\showeprint[arxiv]{1608.08710}~[cs.CV]
\urldef\tempurl%
\url{https://arxiv.org/abs/1608.08710}
\showURL{%
\tempurl}


\bibitem[Li et~al\mbox{.}(2024)]%
        {li2024personal_llm_agents}
\bibfield{author}{\bibinfo{person}{Yuanchun Li}, \bibinfo{person}{Hao Wen}, \bibinfo{person}{Weijun Wang}, \bibinfo{person}{Xiangyu Li}, \bibinfo{person}{Yizhen Yuan}, \bibinfo{person}{Guohong Liu}, \bibinfo{person}{Jiacheng Liu}, \bibinfo{person}{Wenxing Xu}, \bibinfo{person}{Xiang Wang}, \bibinfo{person}{Yi Sun}, \bibinfo{person}{Rui Kong}, \bibinfo{person}{Yile Wang}, \bibinfo{person}{Hanfei Geng}, \bibinfo{person}{Jian Luan}, \bibinfo{person}{Xuefeng Jin}, \bibinfo{person}{Zilong Ye}, \bibinfo{person}{Guanjing Xiong}, \bibinfo{person}{Fan Zhang}, \bibinfo{person}{Xiang Li}, \bibinfo{person}{Mengwei Xu}, \bibinfo{person}{Zhijun Li}, \bibinfo{person}{Peng Li}, \bibinfo{person}{Yang Liu}, \bibinfo{person}{Ya-Qin Zhang}, {and} \bibinfo{person}{Yunxin Liu}.} \bibinfo{year}{2024}\natexlab{}.
\newblock \showarticletitle{Personal LLM Agents: Insights and Survey about the Capability, Efficiency and Security}.
\newblock \bibinfo{journal}{\emph{arXiv preprint arXiv:2401.05459}} (\bibinfo{year}{2024}).
\newblock


\bibitem[Liu et~al\mbox{.}(2023)]%
        {pmlr-v202-liu23am}
\bibfield{author}{\bibinfo{person}{Zichang Liu}, \bibinfo{person}{Jue Wang}, \bibinfo{person}{Tri Dao}, \bibinfo{person}{Tianyi Zhou}, \bibinfo{person}{Binhang Yuan}, \bibinfo{person}{Zhao Song}, \bibinfo{person}{Anshumali Shrivastava}, \bibinfo{person}{Ce Zhang}, \bibinfo{person}{Yuandong Tian}, \bibinfo{person}{Christopher Re}, {and} \bibinfo{person}{Beidi Chen}.} \bibinfo{year}{2023}\natexlab{}.
\newblock \showarticletitle{Deja Vu: Contextual Sparsity for Efficient {LLM}s at Inference Time}. In \bibinfo{booktitle}{\emph{Proceedings of the 40th International Conference on Machine Learning}} \emph{(\bibinfo{series}{Proceedings of Machine Learning Research}, Vol.~\bibinfo{volume}{202})}, \bibfield{editor}{\bibinfo{person}{Andreas Krause}, \bibinfo{person}{Emma Brunskill}, \bibinfo{person}{Kyunghyun Cho}, \bibinfo{person}{Barbara Engelhardt}, \bibinfo{person}{Sivan Sabato}, {and} \bibinfo{person}{Jonathan Scarlett}} (Eds.). \bibinfo{publisher}{PMLR}, \bibinfo{pages}{22137--22176}.
\newblock
\urldef\tempurl%
\url{https://proceedings.mlr.press/v202/liu23am.html}
\showURL{%
\tempurl}


\bibitem[Ma et~al\mbox{.}(2023)]%
        {ma2023llmpruner}
\bibfield{author}{\bibinfo{person}{Xinyin Ma}, \bibinfo{person}{Gongfan Fang}, {and} \bibinfo{person}{Xinchao Wang}.} \bibinfo{year}{2023}\natexlab{}.
\newblock \showarticletitle{LLM-Pruner: On the Structural Pruning of Large Language Models}. In \bibinfo{booktitle}{\emph{Advances in Neural Information Processing Systems}}.
\newblock


\bibitem[Mei et~al\mbox{.}(2024)]%
        {mei2024aiosllmagentoperating}
\bibfield{author}{\bibinfo{person}{Kai Mei}, \bibinfo{person}{Zelong Li}, \bibinfo{person}{Shuyuan Xu}, \bibinfo{person}{Ruosong Ye}, \bibinfo{person}{Yingqiang Ge}, {and} \bibinfo{person}{Yongfeng Zhang}.} \bibinfo{year}{2024}\natexlab{}.
\newblock \bibinfo{title}{AIOS: LLM Agent Operating System}.
\newblock
\newblock
\showeprint[arxiv]{2403.16971}~[cs.OS]
\urldef\tempurl%
\url{https://arxiv.org/abs/2403.16971}
\showURL{%
\tempurl}


\bibitem[Men et~al\mbox{.}(2024)]%
        {men2024shortgptlayerslargelanguage}
\bibfield{author}{\bibinfo{person}{Xin Men}, \bibinfo{person}{Mingyu Xu}, \bibinfo{person}{Qingyu Zhang}, \bibinfo{person}{Bingning Wang}, \bibinfo{person}{Hongyu Lin}, \bibinfo{person}{Yaojie Lu}, \bibinfo{person}{Xianpei Han}, {and} \bibinfo{person}{Weipeng Chen}.} \bibinfo{year}{2024}\natexlab{}.
\newblock \bibinfo{title}{ShortGPT: Layers in Large Language Models are More Redundant Than You Expect}.
\newblock
\newblock
\showeprint[arxiv]{2403.03853}~[cs.CL]
\urldef\tempurl%
\url{https://arxiv.org/abs/2403.03853}
\showURL{%
\tempurl}


\bibitem[Miao et~al\mbox{.}(2024)]%
        {Miao_2024}
\bibfield{author}{\bibinfo{person}{Xupeng Miao}, \bibinfo{person}{Gabriele Oliaro}, \bibinfo{person}{Zhihao Zhang}, \bibinfo{person}{Xinhao Cheng}, \bibinfo{person}{Zeyu Wang}, \bibinfo{person}{Zhengxin Zhang}, \bibinfo{person}{Rae Ying~Yee Wong}, \bibinfo{person}{Alan Zhu}, \bibinfo{person}{Lijie Yang}, \bibinfo{person}{Xiaoxiang Shi}, \bibinfo{person}{Chunan Shi}, \bibinfo{person}{Zhuoming Chen}, \bibinfo{person}{Daiyaan Arfeen}, \bibinfo{person}{Reyna Abhyankar}, {and} \bibinfo{person}{Zhihao Jia}.} \bibinfo{year}{2024}\natexlab{}.
\newblock \showarticletitle{SpecInfer: Accelerating Large Language Model Serving with Tree-based Speculative Inference and Verification}. In \bibinfo{booktitle}{\emph{Proceedings of the 29th ACM International Conference on Architectural Support for Programming Languages and Operating Systems, Volume 3}} \emph{(\bibinfo{series}{ASPLOS ’24})}. \bibinfo{publisher}{ACM}.
\newblock
\urldef\tempurl%
\url{https://doi.org/10.1145/3620666.3651335}
\showDOI{\tempurl}


\bibitem[Mihaylov et~al\mbox{.}(2018)]%
        {OpenBookQA2018}
\bibfield{author}{\bibinfo{person}{Todor Mihaylov}, \bibinfo{person}{Peter Clark}, \bibinfo{person}{Tushar Khot}, {and} \bibinfo{person}{Ashish Sabharwal}.} \bibinfo{year}{2018}\natexlab{}.
\newblock \showarticletitle{Can a Suit of Armor Conduct Electricity? A New Dataset for Open Book Question Answering}. In \bibinfo{booktitle}{\emph{EMNLP}}.
\newblock


\bibitem[Modarressi et~al\mbox{.}(2022)]%
        {modarressi-etal-2022-adapler}
\bibfield{author}{\bibinfo{person}{Ali Modarressi}, \bibinfo{person}{Hosein Mohebbi}, {and} \bibinfo{person}{Mohammad~Taher Pilehvar}.} \bibinfo{year}{2022}\natexlab{}.
\newblock \showarticletitle{{A}dap{L}e{R}: Speeding up Inference by Adaptive Length Reduction}. In \bibinfo{booktitle}{\emph{Proceedings of the 60th Annual Meeting of the Association for Computational Linguistics (Volume 1: Long Papers)}}, \bibfield{editor}{\bibinfo{person}{Smaranda Muresan}, \bibinfo{person}{Preslav Nakov}, {and} \bibinfo{person}{Aline Villavicencio}} (Eds.). \bibinfo{publisher}{Association for Computational Linguistics}, \bibinfo{address}{Dublin, Ireland}, \bibinfo{pages}{1--15}.
\newblock
\urldef\tempurl%
\url{https://doi.org/10.18653/v1/2022.acl-long.1}
\showDOI{\tempurl}


\bibitem[Mukherjee et~al\mbox{.}(2023)]%
        {mukherjee2023orca}
\bibfield{author}{\bibinfo{person}{Subhabrata Mukherjee}, \bibinfo{person}{Arindam Mitra}, \bibinfo{person}{Ganesh Jawahar}, \bibinfo{person}{Sahaj Agarwal}, \bibinfo{person}{Hamid Palangi}, {and} \bibinfo{person}{Ahmed Awadallah}.} \bibinfo{year}{2023}\natexlab{}.
\newblock \bibinfo{title}{Orca: Progressive Learning from Complex Explanation Traces of GPT-4}.
\newblock
\newblock
\showeprint[arxiv]{2306.02707}~[cs.CL]


\bibitem[Ouyang et~al\mbox{.}(2022)]%
        {ouyang2022traininglanguagemodelsfollow}
\bibfield{author}{\bibinfo{person}{Long Ouyang}, \bibinfo{person}{Jeff Wu}, \bibinfo{person}{Xu Jiang}, \bibinfo{person}{Diogo Almeida}, \bibinfo{person}{Carroll~L. Wainwright}, \bibinfo{person}{Pamela Mishkin}, \bibinfo{person}{Chong Zhang}, \bibinfo{person}{Sandhini Agarwal}, \bibinfo{person}{Katarina Slama}, \bibinfo{person}{Alex Ray}, \bibinfo{person}{John Schulman}, \bibinfo{person}{Jacob Hilton}, \bibinfo{person}{Fraser Kelton}, \bibinfo{person}{Luke Miller}, \bibinfo{person}{Maddie Simens}, \bibinfo{person}{Amanda Askell}, \bibinfo{person}{Peter Welinder}, \bibinfo{person}{Paul Christiano}, \bibinfo{person}{Jan Leike}, {and} \bibinfo{person}{Ryan Lowe}.} \bibinfo{year}{2022}\natexlab{}.
\newblock \bibinfo{title}{Training language models to follow instructions with human feedback}.
\newblock
\newblock
\showeprint[arxiv]{2203.02155}~[cs.CL]
\urldef\tempurl%
\url{https://arxiv.org/abs/2203.02155}
\showURL{%
\tempurl}


\bibitem[Packer et~al\mbox{.}(2024)]%
        {packer2024memgptllmsoperatingsystems}
\bibfield{author}{\bibinfo{person}{Charles Packer}, \bibinfo{person}{Sarah Wooders}, \bibinfo{person}{Kevin Lin}, \bibinfo{person}{Vivian Fang}, \bibinfo{person}{Shishir~G. Patil}, \bibinfo{person}{Ion Stoica}, {and} \bibinfo{person}{Joseph~E. Gonzalez}.} \bibinfo{year}{2024}\natexlab{}.
\newblock \bibinfo{title}{MemGPT: Towards LLMs as Operating Systems}.
\newblock
\newblock
\showeprint[arxiv]{2310.08560}~[cs.AI]
\urldef\tempurl%
\url{https://arxiv.org/abs/2310.08560}
\showURL{%
\tempurl}


\bibitem[Pan et~al\mbox{.}(2024)]%
        {wu2024llmlingua2}
\bibfield{author}{\bibinfo{person}{Zhuoshi Pan}, \bibinfo{person}{Qianhui Wu}, \bibinfo{person}{Huiqiang Jiang}, \bibinfo{person}{Menglin Xia}, \bibinfo{person}{Xufang Luo}, \bibinfo{person}{Jue Zhang}, \bibinfo{person}{Qingwei Lin}, \bibinfo{person}{Victor Ruhle}, \bibinfo{person}{Yuqing Yang}, \bibinfo{person}{Chin-Yew Lin}, \bibinfo{person}{H.~Vicky Zhao}, \bibinfo{person}{Lili Qiu}, {and} \bibinfo{person}{Dongmei Zhang}.} \bibinfo{year}{2024}\natexlab{}.
\newblock \showarticletitle{{LLML}ingua-2: Data Distillation for Efficient and Faithful Task-Agnostic Prompt Compression}.
\newblock \bibinfo{journal}{\emph{ArXiv preprint}}  \bibinfo{volume}{abs/2403.12968} (\bibinfo{year}{2024}).
\newblock
\urldef\tempurl%
\url{https://arxiv.org/abs/2403.12968}
\showURL{%
\tempurl}


\bibitem[Paszke et~al\mbox{.}(2019)]%
        {paszke2019pytorchimperativestylehighperformance}
\bibfield{author}{\bibinfo{person}{Adam Paszke}, \bibinfo{person}{Sam Gross}, \bibinfo{person}{Francisco Massa}, {et~al\mbox{.}}} \bibinfo{year}{2019}\natexlab{}.
\newblock \bibinfo{title}{PyTorch: An Imperative Style, High-Performance Deep Learning Library}.
\newblock
\newblock
\showeprint[arxiv]{1912.01703}~[cs.LG]
\urldef\tempurl%
\url{https://arxiv.org/abs/1912.01703}
\showURL{%
\tempurl}


\bibitem[Raffel et~al\mbox{.}(2023)]%
        {raffel2023exploringlimitstransferlearning}
\bibfield{author}{\bibinfo{person}{Colin Raffel}, \bibinfo{person}{Noam Shazeer}, \bibinfo{person}{Adam Roberts}, \bibinfo{person}{Katherine Lee}, \bibinfo{person}{Sharan Narang}, \bibinfo{person}{Michael Matena}, \bibinfo{person}{Yanqi Zhou}, \bibinfo{person}{Wei Li}, {and} \bibinfo{person}{Peter~J. Liu}.} \bibinfo{year}{2023}\natexlab{}.
\newblock \bibinfo{title}{Exploring the Limits of Transfer Learning with a Unified Text-to-Text Transformer}.
\newblock
\newblock
\showeprint[arxiv]{1910.10683}~[cs.LG]
\urldef\tempurl%
\url{https://arxiv.org/abs/1910.10683}
\showURL{%
\tempurl}


\bibitem[Selvaraju et~al\mbox{.}(2019)]%
        {Selvaraju_2019}
\bibfield{author}{\bibinfo{person}{Ramprasaath~R. Selvaraju}, \bibinfo{person}{Michael Cogswell}, \bibinfo{person}{Abhishek Das}, \bibinfo{person}{Ramakrishna Vedantam}, \bibinfo{person}{Devi Parikh}, {and} \bibinfo{person}{Dhruv Batra}.} \bibinfo{year}{2019}\natexlab{}.
\newblock \showarticletitle{Grad-CAM: Visual Explanations from Deep Networks via Gradient-Based Localization}.
\newblock \bibinfo{journal}{\emph{International Journal of Computer Vision}} \bibinfo{volume}{128}, \bibinfo{number}{2} (\bibinfo{date}{Oct.} \bibinfo{year}{2019}), \bibinfo{pages}{336–359}.
\newblock
\showISSN{1573-1405}
\urldef\tempurl%
\url{https://doi.org/10.1007/s11263-019-01228-7}
\showDOI{\tempurl}


\bibitem[Song et~al\mbox{.}(2023)]%
        {song2023powerinfer}
\bibfield{author}{\bibinfo{person}{Yixin Song}, \bibinfo{person}{Zeyu Mi}, \bibinfo{person}{Haotong Xie}, {and} \bibinfo{person}{Haibo Chen}.} \bibinfo{year}{2023}\natexlab{}.
\newblock \bibinfo{title}{PowerInfer: Fast Large Language Model Serving with a Consumer-grade GPU}.
\newblock
\newblock
\showeprint[arxiv]{2312.12456}~[cs.LG]


\bibitem[Su et~al\mbox{.}(2023)]%
        {su2023roformerenhancedtransformerrotary}
\bibfield{author}{\bibinfo{person}{Jianlin Su}, \bibinfo{person}{Yu Lu}, \bibinfo{person}{Shengfeng Pan}, \bibinfo{person}{Ahmed Murtadha}, \bibinfo{person}{Bo Wen}, {and} \bibinfo{person}{Yunfeng Liu}.} \bibinfo{year}{2023}\natexlab{}.
\newblock \bibinfo{title}{RoFormer: Enhanced Transformer with Rotary Position Embedding}.
\newblock
\newblock
\showeprint[arxiv]{2104.09864}~[cs.CL]
\urldef\tempurl%
\url{https://arxiv.org/abs/2104.09864}
\showURL{%
\tempurl}


\bibitem[Sun et~al\mbox{.}(2024)]%
        {sun2024simpleeffectivepruningapproach}
\bibfield{author}{\bibinfo{person}{Mingjie Sun}, \bibinfo{person}{Zhuang Liu}, \bibinfo{person}{Anna Bair}, {and} \bibinfo{person}{J.~Zico Kolter}.} \bibinfo{year}{2024}\natexlab{}.
\newblock \bibinfo{title}{A Simple and Effective Pruning Approach for Large Language Models}.
\newblock
\newblock
\showeprint[arxiv]{2306.11695}~[cs.CL]
\urldef\tempurl%
\url{https://arxiv.org/abs/2306.11695}
\showURL{%
\tempurl}


\bibitem[Sun et~al\mbox{.}(2020)]%
        {sun2020mobilebertcompacttaskagnosticbert}
\bibfield{author}{\bibinfo{person}{Zhiqing Sun}, \bibinfo{person}{Hongkun Yu}, \bibinfo{person}{Xiaodan Song}, \bibinfo{person}{Renjie Liu}, \bibinfo{person}{Yiming Yang}, {and} \bibinfo{person}{Denny Zhou}.} \bibinfo{year}{2020}\natexlab{}.
\newblock \bibinfo{title}{MobileBERT: a Compact Task-Agnostic BERT for Resource-Limited Devices}.
\newblock
\newblock
\showeprint[arxiv]{2004.02984}~[cs.CL]
\urldef\tempurl%
\url{https://arxiv.org/abs/2004.02984}
\showURL{%
\tempurl}


\bibitem[Sundararajan et~al\mbox{.}(2017)]%
        {sundararajan2017axiomaticattributiondeepnetworks}
\bibfield{author}{\bibinfo{person}{Mukund Sundararajan}, \bibinfo{person}{Ankur Taly}, {and} \bibinfo{person}{Qiqi Yan}.} \bibinfo{year}{2017}\natexlab{}.
\newblock \bibinfo{title}{Axiomatic Attribution for Deep Networks}.
\newblock
\newblock
\showeprint[arxiv]{1703.01365}~[cs.LG]
\urldef\tempurl%
\url{https://arxiv.org/abs/1703.01365}
\showURL{%
\tempurl}


\bibitem[Swokowski(1979)]%
        {swokowski1979calculus}
\bibfield{author}{\bibinfo{person}{Earl~William Swokowski}.} \bibinfo{year}{1979}\natexlab{}.
\newblock \bibinfo{booktitle}{\emph{Calculus with analytic geometry}}.
\newblock \bibinfo{publisher}{Taylor \& Francis}.
\newblock


\bibitem[Tambe et~al\mbox{.}(2021)]%
        {tambe2021edgebertsentencelevelenergyoptimizations}
\bibfield{author}{\bibinfo{person}{Thierry Tambe}, \bibinfo{person}{Coleman Hooper}, \bibinfo{person}{Lillian Pentecost}, \bibinfo{person}{Tianyu Jia}, \bibinfo{person}{En-Yu Yang}, \bibinfo{person}{Marco Donato}, \bibinfo{person}{Victor Sanh}, \bibinfo{person}{Paul~N. Whatmough}, \bibinfo{person}{Alexander~M. Rush}, \bibinfo{person}{David Brooks}, {and} \bibinfo{person}{Gu-Yeon Wei}.} \bibinfo{year}{2021}\natexlab{}.
\newblock \bibinfo{title}{EdgeBERT: Sentence-Level Energy Optimizations for Latency-Aware Multi-Task NLP Inference}.
\newblock
\newblock
\showeprint[arxiv]{2011.14203}~[cs.AR]
\urldef\tempurl%
\url{https://arxiv.org/abs/2011.14203}
\showURL{%
\tempurl}


\bibitem[Taori et~al\mbox{.}(2023)]%
        {alpaca}
\bibfield{author}{\bibinfo{person}{Rohan Taori}, \bibinfo{person}{Ishaan Gulrajani}, \bibinfo{person}{Tianyi Zhang}, \bibinfo{person}{Yann Dubois}, \bibinfo{person}{Xuechen Li}, \bibinfo{person}{Carlos Guestrin}, \bibinfo{person}{Percy Liang}, {and} \bibinfo{person}{Tatsunori~B. Hashimoto}.} \bibinfo{year}{2023}\natexlab{}.
\newblock \bibinfo{title}{Stanford Alpaca: An Instruction-following LLaMA model}.
\newblock \bibinfo{howpublished}{\url{https://github.com/tatsu-lab/stanford_alpaca}}.
\newblock


\bibitem[team(2023)]%
        {mlc-llm}
\bibfield{author}{\bibinfo{person}{MLC team}.} \bibinfo{year}{2023}\natexlab{}.
\newblock \bibinfo{booktitle}{\emph{{MLC-LLM}}}.
\newblock
\urldef\tempurl%
\url{https://github.com/mlc-ai/mlc-llm}
\showURL{%
\tempurl}


\bibitem[Teerapittayanon et~al\mbox{.}(2017)]%
        {teerapittayanon2017branchynetfastinferenceearly}
\bibfield{author}{\bibinfo{person}{Surat Teerapittayanon}, \bibinfo{person}{Bradley McDanel}, {and} \bibinfo{person}{H.~T. Kung}.} \bibinfo{year}{2017}\natexlab{}.
\newblock \bibinfo{title}{BranchyNet: Fast Inference via Early Exiting from Deep Neural Networks}.
\newblock
\newblock
\showeprint[arxiv]{1709.01686}~[cs.NE]
\urldef\tempurl%
\url{https://arxiv.org/abs/1709.01686}
\showURL{%
\tempurl}


\bibitem[Touvron et~al\mbox{.}(2023)]%
        {touvron2023llamaopenefficientfoundation}
\bibfield{author}{\bibinfo{person}{Hugo Touvron}, \bibinfo{person}{Thibaut Lavril}, \bibinfo{person}{Gautier Izacard}, \bibinfo{person}{Xavier Martinet}, \bibinfo{person}{Marie-Anne Lachaux}, \bibinfo{person}{Timothée Lacroix}, \bibinfo{person}{Baptiste Rozière}, \bibinfo{person}{Naman Goyal}, \bibinfo{person}{Eric Hambro}, \bibinfo{person}{Faisal Azhar}, \bibinfo{person}{Aurelien Rodriguez}, \bibinfo{person}{Armand Joulin}, \bibinfo{person}{Edouard Grave}, {and} \bibinfo{person}{Guillaume Lample}.} \bibinfo{year}{2023}\natexlab{}.
\newblock \bibinfo{title}{LLaMA: Open and Efficient Foundation Language Models}.
\newblock
\newblock
\showeprint[arxiv]{2302.13971}~[cs.CL]
\urldef\tempurl%
\url{https://arxiv.org/abs/2302.13971}
\showURL{%
\tempurl}


\bibitem[Wang et~al\mbox{.}(2019)]%
        {wang2019gluemultitaskbenchmarkanalysis}
\bibfield{author}{\bibinfo{person}{Alex Wang}, \bibinfo{person}{Amanpreet Singh}, \bibinfo{person}{Julian Michael}, \bibinfo{person}{Felix Hill}, \bibinfo{person}{Omer Levy}, {and} \bibinfo{person}{Samuel~R. Bowman}.} \bibinfo{year}{2019}\natexlab{}.
\newblock \bibinfo{title}{GLUE: A Multi-Task Benchmark and Analysis Platform for Natural Language Understanding}.
\newblock
\newblock
\showeprint[arxiv]{1804.07461}~[cs.CL]
\urldef\tempurl%
\url{https://arxiv.org/abs/1804.07461}
\showURL{%
\tempurl}


\bibitem[Wang et~al\mbox{.}(2024)]%
        {wang2024mmluprorobustchallengingmultitask}
\bibfield{author}{\bibinfo{person}{Yubo Wang}, \bibinfo{person}{Xueguang Ma}, {et~al\mbox{.}}} \bibinfo{year}{2024}\natexlab{}.
\newblock \bibinfo{title}{MMLU-Pro: A More Robust and Challenging Multi-Task Language Understanding Benchmark}.
\newblock
\newblock
\showeprint[arxiv]{2406.01574}~[cs.CL]
\urldef\tempurl%
\url{https://arxiv.org/abs/2406.01574}
\showURL{%
\tempurl}


\bibitem[Wen et~al\mbox{.}(2024)]%
        {autodriod}
\bibfield{author}{\bibinfo{person}{Hao Wen}, \bibinfo{person}{Yuanchun Li}, \bibinfo{person}{Guohong Liu}, \bibinfo{person}{Shanhui Zhao}, \bibinfo{person}{Tao Yu}, \bibinfo{person}{Toby Jia-Jun Li}, \bibinfo{person}{Shiqi Jiang}, \bibinfo{person}{Yunhao Liu}, \bibinfo{person}{Yaqin Zhang}, {and} \bibinfo{person}{Yunxin Liu}.} \bibinfo{year}{2024}\natexlab{}.
\newblock \showarticletitle{AutoDroid: LLM-powered Task Automation in Android}. In \bibinfo{booktitle}{\emph{Proceedings of the 30th Annual International Conference on Mobile Computing and Networking}} (Washington D.C., DC, USA) \emph{(\bibinfo{series}{ACM MobiCom '24})}. \bibinfo{publisher}{Association for Computing Machinery}, \bibinfo{address}{New York, NY, USA}, \bibinfo{pages}{543–557}.
\newblock
\showISBNx{9798400704895}
\urldef\tempurl%
\url{https://doi.org/10.1145/3636534.3649379}
\showDOI{\tempurl}


\bibitem[Wen et~al\mbox{.}(2023)]%
        {wen2023adaptivenetpostdeploymentneuralarchitecture}
\bibfield{author}{\bibinfo{person}{Hao Wen}, \bibinfo{person}{Yuanchun Li}, \bibinfo{person}{Zunshuai Zhang}, \bibinfo{person}{Shiqi Jiang}, \bibinfo{person}{Xiaozhou Ye}, \bibinfo{person}{Ye Ouyang}, \bibinfo{person}{Ya-Qin Zhang}, {and} \bibinfo{person}{Yunxin Liu}.} \bibinfo{year}{2023}\natexlab{}.
\newblock \bibinfo{title}{AdaptiveNet: Post-deployment Neural Architecture Adaptation for Diverse Edge Environments}.
\newblock
\newblock
\showeprint[arxiv]{2303.07129}~[cs.LG]
\urldef\tempurl%
\url{https://arxiv.org/abs/2303.07129}
\showURL{%
\tempurl}


\bibitem[Wies et~al\mbox{.}(2023)]%
        {wies2023learnabilityincontextlearning}
\bibfield{author}{\bibinfo{person}{Noam Wies}, \bibinfo{person}{Yoav Levine}, {and} \bibinfo{person}{Amnon Shashua}.} \bibinfo{year}{2023}\natexlab{}.
\newblock \bibinfo{title}{The Learnability of In-Context Learning}.
\newblock
\newblock
\showeprint[arxiv]{2303.07895}~[cs.CL]
\urldef\tempurl%
\url{https://arxiv.org/abs/2303.07895}
\showURL{%
\tempurl}


\bibitem[Wolf et~al\mbox{.}(2020)]%
        {wolf-etal-2020-transformers}
\bibfield{author}{\bibinfo{person}{Thomas Wolf}, \bibinfo{person}{Lysandre Debut}, \bibinfo{person}{Victor Sanh}, \bibinfo{person}{Julien Chaumond}, \bibinfo{person}{Clement Delangue}, {et~al\mbox{.}}} \bibinfo{year}{2020}\natexlab{}.
\newblock \showarticletitle{Transformers: State-of-the-Art Natural Language Processing}. In \bibinfo{booktitle}{\emph{Proceedings of the 2020 Conference on Empirical Methods in Natural Language Processing: System Demonstrations}}. \bibinfo{publisher}{Association for Computational Linguistics}, \bibinfo{address}{Online}, \bibinfo{pages}{38--45}.
\newblock
\urldef\tempurl%
\url{https://www.aclweb.org/anthology/2020.emnlp-demos.6}
\showURL{%
\tempurl}


\bibitem[Wu et~al\mbox{.}(2023)]%
        {lamini-lm}
\bibfield{author}{\bibinfo{person}{Minghao Wu}, \bibinfo{person}{Abdul Waheed}, \bibinfo{person}{Chiyu Zhang}, \bibinfo{person}{Muhammad Abdul-Mageed}, {and} \bibinfo{person}{Alham~Fikri Aji}.} \bibinfo{year}{2023}\natexlab{}.
\newblock \showarticletitle{LaMini-LM: A Diverse Herd of Distilled Models from Large-Scale Instructions}.
\newblock \bibinfo{journal}{\emph{CoRR}}  \bibinfo{volume}{abs/2304.14402} (\bibinfo{year}{2023}).
\newblock
\showeprint[arXiv]{2304.14402}
\urldef\tempurl%
\url{https://arxiv.org/abs/2304.14402}
\showURL{%
\tempurl}


\bibitem[Xia et~al\mbox{.}(2024)]%
        {xia2024shearedllamaacceleratinglanguage}
\bibfield{author}{\bibinfo{person}{Mengzhou Xia}, \bibinfo{person}{Tianyu Gao}, \bibinfo{person}{Zhiyuan Zeng}, {and} \bibinfo{person}{Danqi Chen}.} \bibinfo{year}{2024}\natexlab{}.
\newblock \bibinfo{title}{Sheared LLaMA: Accelerating Language Model Pre-training via Structured Pruning}.
\newblock
\newblock
\showeprint[arxiv]{2310.06694}~[cs.CL]
\urldef\tempurl%
\url{https://arxiv.org/abs/2310.06694}
\showURL{%
\tempurl}


\bibitem[Xie et~al\mbox{.}(2024)]%
        {xie2024droidcall}
\bibfield{author}{\bibinfo{person}{Weikai Xie}, \bibinfo{person}{Li Zhang}, \bibinfo{person}{Shihe Wang}, \bibinfo{person}{Rongjie Yi}, {and} \bibinfo{person}{Mengwei Xu}.} \bibinfo{year}{2024}\natexlab{}.
\newblock \showarticletitle{DroidCall: A Dataset for LLM-powered Android Intent Invocation}.
\newblock \bibinfo{journal}{\emph{arXiv preprint arXiv:2412.00402}} (\bibinfo{year}{2024}).
\newblock


\bibitem[Xu et~al\mbox{.}(2023)]%
        {xu2023llmcadfastscalableondevice}
\bibfield{author}{\bibinfo{person}{Daliang Xu}, \bibinfo{person}{Wangsong Yin}, \bibinfo{person}{Xin Jin}, \bibinfo{person}{Ying Zhang}, \bibinfo{person}{Shiyun Wei}, \bibinfo{person}{Mengwei Xu}, {and} \bibinfo{person}{Xuanzhe Liu}.} \bibinfo{year}{2023}\natexlab{}.
\newblock \bibinfo{title}{LLMCad: Fast and Scalable On-device Large Language Model Inference}.
\newblock
\newblock
\showeprint[arxiv]{2309.04255}~[cs.NI]
\urldef\tempurl%
\url{https://arxiv.org/abs/2309.04255}
\showURL{%
\tempurl}


\bibitem[Xu et~al\mbox{.}(2024)]%
        {xu2024empowering1000tokenssecondondevice}
\bibfield{author}{\bibinfo{person}{Daliang Xu}, \bibinfo{person}{Hao Zhang}, \bibinfo{person}{Liming Yang}, \bibinfo{person}{Ruiqi Liu}, \bibinfo{person}{Gang Huang}, \bibinfo{person}{Mengwei Xu}, {and} \bibinfo{person}{Xuanzhe Liu}.} \bibinfo{year}{2024}\natexlab{}.
\newblock \bibinfo{title}{Empowering 1000 tokens/second on-device LLM prefilling with mllm-NPU}.
\newblock
\newblock
\showeprint[arxiv]{2407.05858}~[cs.AI]
\urldef\tempurl%
\url{https://arxiv.org/abs/2407.05858}
\showURL{%
\tempurl}


\bibitem[Xue et~al\mbox{.}(2024)]%
        {xue2024powerinfer2fastlargelanguage}
\bibfield{author}{\bibinfo{person}{Zhenliang Xue}, \bibinfo{person}{Yixin Song}, \bibinfo{person}{Zeyu Mi}, \bibinfo{person}{Le Chen}, \bibinfo{person}{Yubin Xia}, {and} \bibinfo{person}{Haibo Chen}.} \bibinfo{year}{2024}\natexlab{}.
\newblock \bibinfo{title}{PowerInfer-2: Fast Large Language Model Inference on a Smartphone}.
\newblock
\newblock
\showeprint[arxiv]{2406.06282}~[cs.LG]
\urldef\tempurl%
\url{https://arxiv.org/abs/2406.06282}
\showURL{%
\tempurl}


\bibitem[Yang et~al\mbox{.}(2024b)]%
        {PIQA}
\bibfield{author}{\bibinfo{person}{Sohee Yang}, \bibinfo{person}{Jonghyeon Kim}, \bibinfo{person}{Joel Jang}, \bibinfo{person}{Seonghyeon Ye}, \bibinfo{person}{Hyunji Lee}, {and} \bibinfo{person}{Minjoon Seo}.} \bibinfo{year}{2024}\natexlab{b}.
\newblock \bibinfo{title}{Improving Probability-based Prompt Selection Through Unified Evaluation and Analysis}.
\newblock
\newblock
\showeprint[arxiv]{2305.14877}~[cs.CL]
\urldef\tempurl%
\url{https://arxiv.org/abs/2305.14877}
\showURL{%
\tempurl}


\bibitem[Yang et~al\mbox{.}(2024a)]%
        {yang2024laco}
\bibfield{author}{\bibinfo{person}{Yifei Yang}, \bibinfo{person}{Zouying Cao}, {and} \bibinfo{person}{Hai Zhao}.} \bibinfo{year}{2024}\natexlab{a}.
\newblock \showarticletitle{Laco: Large language model pruning via layer collapse}.
\newblock \bibinfo{journal}{\emph{arXiv preprint arXiv:2402.11187}} (\bibinfo{year}{2024}).
\newblock


\bibitem[Yin et~al\mbox{.}(2024)]%
        {yin2024llmservicemobiledevices}
\bibfield{author}{\bibinfo{person}{Wangsong Yin}, \bibinfo{person}{Mengwei Xu}, \bibinfo{person}{Yuanchun Li}, {and} \bibinfo{person}{Xuanzhe Liu}.} \bibinfo{year}{2024}\natexlab{}.
\newblock \bibinfo{title}{LLM as a System Service on Mobile Devices}.
\newblock
\newblock
\showeprint[arxiv]{2403.11805}~[cs.OS]
\urldef\tempurl%
\url{https://arxiv.org/abs/2403.11805}
\showURL{%
\tempurl}


\bibitem[Yu et~al\mbox{.}(2022)]%
        {Orca_osdi}
\bibfield{author}{\bibinfo{person}{Gyeong-In Yu}, \bibinfo{person}{Joo~Seong Jeong}, \bibinfo{person}{Geon-Woo Kim}, \bibinfo{person}{Soojeong Kim}, {and} \bibinfo{person}{Byung-Gon Chun}.} \bibinfo{year}{2022}\natexlab{}.
\newblock \showarticletitle{Orca: A Distributed Serving System for {Transformer-Based} Generative Models}. In \bibinfo{booktitle}{\emph{16th USENIX Symposium on Operating Systems Design and Implementation (OSDI 22)}}. \bibinfo{publisher}{USENIX Association}, \bibinfo{address}{Carlsbad, CA}, \bibinfo{pages}{521--538}.
\newblock
\showISBNx{978-1-939133-28-1}
\urldef\tempurl%
\url{https://www.usenix.org/conference/osdi22/presentation/yu}
\showURL{%
\tempurl}


\bibitem[Yuan et~al\mbox{.}(2024)]%
        {yuanrethinking}
\bibfield{author}{\bibinfo{person}{Jinliang Yuan}, \bibinfo{person}{Chen Yang}, \bibinfo{person}{Dongqi Cai}, \bibinfo{person}{Shihe Wang}, \bibinfo{person}{Xin Yuan}, \bibinfo{person}{Zeling Zhang}, \bibinfo{person}{Xiang Li}, \bibinfo{person}{Dingge Zhang}, \bibinfo{person}{Hanzi Mei}, \bibinfo{person}{Xianqing Jia}, \bibinfo{person}{Shangguang Wang}, {and} \bibinfo{person}{Mengwei Xu}.} \bibinfo{year}{2024}\natexlab{}.
\newblock \showarticletitle{Mobile Foundation Model as Firmware}. In \bibinfo{booktitle}{\emph{Proceedings of the 30th Annual International Conference on Mobile Computing and Networking}} (Washington D.C., DC, USA) \emph{(\bibinfo{series}{ACM MobiCom '24})}. \bibinfo{publisher}{Association for Computing Machinery}, \bibinfo{address}{New York, NY, USA}, \bibinfo{pages}{279–295}.
\newblock
\showISBNx{9798400704895}
\urldef\tempurl%
\url{https://doi.org/10.1145/3636534.3649361}
\showDOI{\tempurl}


\bibitem[Zhang et~al\mbox{.}(2024)]%
        {zhang2024llamatouch}
\bibfield{author}{\bibinfo{person}{Li Zhang}, \bibinfo{person}{Shihe Wang}, \bibinfo{person}{Xianqing Jia}, \bibinfo{person}{Zhihan Zheng}, \bibinfo{person}{Yunhe Yan}, \bibinfo{person}{Longxi Gao}, \bibinfo{person}{Yuanchun Li}, {and} \bibinfo{person}{Mengwei Xu}.} \bibinfo{year}{2024}\natexlab{}.
\newblock \bibinfo{title}{LlamaTouch: A Faithful and Scalable Testbed for Mobile UI Task Automation}.
\newblock
\newblock
\showeprint[arxiv]{2404.16054}~[cs.HC]
\urldef\tempurl%
\url{https://arxiv.org/abs/2404.16054}
\showURL{%
\tempurl}


\bibitem[Zhang et~al\mbox{.}(2022)]%
        {zhang2022optopenpretrainedtransformer}
\bibfield{author}{\bibinfo{person}{Susan Zhang}, \bibinfo{person}{Stephen Roller}, {et~al\mbox{.}}} \bibinfo{year}{2022}\natexlab{}.
\newblock \bibinfo{title}{OPT: Open Pre-trained Transformer Language Models}.
\newblock
\newblock
\showeprint[arxiv]{2205.01068}~[cs.CL]
\urldef\tempurl%
\url{https://arxiv.org/abs/2205.01068}
\showURL{%
\tempurl}


\bibitem[Zheng et~al\mbox{.}(2023a)]%
        {zheng2023judging}
\bibfield{author}{\bibinfo{person}{Lianmin Zheng}, \bibinfo{person}{Wei-Lin Chiang}, \bibinfo{person}{Ying Sheng}, \bibinfo{person}{Siyuan Zhuang}, \bibinfo{person}{Zhanghao Wu}, \bibinfo{person}{Yonghao Zhuang}, \bibinfo{person}{Zi Lin}, \bibinfo{person}{Zhuohan Li}, \bibinfo{person}{Dacheng Li}, \bibinfo{person}{Eric.~P Xing}, \bibinfo{person}{Hao Zhang}, \bibinfo{person}{Joseph~E. Gonzalez}, {and} \bibinfo{person}{Ion Stoica}.} \bibinfo{year}{2023}\natexlab{a}.
\newblock \bibinfo{title}{Judging LLM-as-a-judge with MT-Bench and Chatbot Arena}.
\newblock
\newblock
\showeprint[arxiv]{2306.05685}~[cs.CL]


\bibitem[Zheng et~al\mbox{.}(2023b)]%
        {PIT}
\bibfield{author}{\bibinfo{person}{Ningxin Zheng}, \bibinfo{person}{Huiqiang Jiang}, \bibinfo{person}{Quanlu Zhang}, \bibinfo{person}{Zhenhua Han}, \bibinfo{person}{Lingxiao Ma}, \bibinfo{person}{Yuqing Yang}, \bibinfo{person}{Fan Yang}, \bibinfo{person}{Chengruidong Zhang}, \bibinfo{person}{Lili Qiu}, \bibinfo{person}{Mao Yang}, {and} \bibinfo{person}{Lidong Zhou}.} \bibinfo{year}{2023}\natexlab{b}.
\newblock \showarticletitle{PIT: Optimization of Dynamic Sparse Deep Learning Models via Permutation Invariant Transformation}. In \bibinfo{booktitle}{\emph{Proceedings of the 29th Symposium on Operating Systems Principles}} (Koblenz, Germany) \emph{(\bibinfo{series}{SOSP '23})}. \bibinfo{publisher}{Association for Computing Machinery}, \bibinfo{address}{New York, NY, USA}, \bibinfo{pages}{331–347}.
\newblock
\showISBNx{9798400702297}
\urldef\tempurl%
\url{https://doi.org/10.1145/3600006.3613139}
\showDOI{\tempurl}


\bibitem[Zhou et~al\mbox{.}(2023)]%
        {zhou2023instructionfollowingevaluationlargelanguage}
\bibfield{author}{\bibinfo{person}{Jeffrey Zhou}, \bibinfo{person}{Tianjian Lu}, \bibinfo{person}{Swaroop Mishra}, \bibinfo{person}{Siddhartha Brahma}, \bibinfo{person}{Sujoy Basu}, \bibinfo{person}{Yi Luan}, \bibinfo{person}{Denny Zhou}, {and} \bibinfo{person}{Le Hou}.} \bibinfo{year}{2023}\natexlab{}.
\newblock \bibinfo{title}{Instruction-Following Evaluation for Large Language Models}.
\newblock
\newblock
\showeprint[arxiv]{2311.07911}~[cs.CL]
\urldef\tempurl%
\url{https://arxiv.org/abs/2311.07911}
\showURL{%
\tempurl}


\bibitem[Zhu et~al\mbox{.}(2015)]%
        {Zhu_2015_ICCV}
\bibfield{author}{\bibinfo{person}{Yukun Zhu}, \bibinfo{person}{Ryan Kiros}, \bibinfo{person}{Rich Zemel}, \bibinfo{person}{Ruslan Salakhutdinov}, \bibinfo{person}{Raquel Urtasun}, \bibinfo{person}{Antonio Torralba}, {and} \bibinfo{person}{Sanja Fidler}.} \bibinfo{year}{2015}\natexlab{}.
\newblock \showarticletitle{Aligning Books and Movies: Towards Story-Like Visual Explanations by Watching Movies and Reading Books}. In \bibinfo{booktitle}{\emph{The IEEE International Conference on Computer Vision (ICCV)}}.
\newblock


\end{thebibliography}

\end{document}